\documentclass[a4paper,10pt]{article}
\pdfoutput=1
\usepackage{jheppub}
\usepackage[T1]{fontenc}
\usepackage{lmodern}
\usepackage{xcolor}
\usepackage{amsmath,amssymb}
\usepackage{mathrsfs,wasysym}
\usepackage{booktabs}
\usepackage{multirow}
\usepackage{slashed}
\usepackage{cancel}
\usepackage{xspace}
\usepackage{epsfig}
\usepackage{amssymb}
\usepackage{slashed}
\usepackage{amsmath}
\usepackage{pifont}
\usepackage{mathtools}
\usepackage{booktabs,multirow,tabularx}
\usepackage{lscape}

\allowdisplaybreaks

\def\({\left(}
\def\){\right)}
\def\[{\left[}
\def\]{\right]}
\def\beq{\begin{equation}}
\def\beqn{\begin{eqnarray}}
\def\eeq{\end{equation}}
\def\eeqn{\end{eqnarray}}
\def\abs#1{\left|#1\right|}

\def\binomial#1#2{
\left(\!\!
\begin{array}{c}
#1\\
#2
\end{array}
\!\!\right)
}

\DeclarePairedDelimiter\floor{\lfloor}{\rfloor}
\newcommand\sss{\scriptscriptstyle}
\newcommand\mydot{\!\cdot\!}
\newcommand\ep{\epsilon}
\newcommand\half{\frac{1}{2}}
\newcommand\as{\alpha_s}

\newcommand\asb{\bar\alpha_s}

\newcommand\gE{\gamma_{\sss\rm E}}
\newcommand{\nfac}{\lambda}
\newcommand{\Mmin}{\gamma_{\rm min}}

\newcommand{\epem}{e^+e^-}

\newcommand{\ord}{{\cal O}}

\newcommand\TR{T_R}
\newcommand\CA{C_A}

\newcommand\hsig{\hat{\sigma}}

\newcommand\MSb{\overline{\rm MS}}

\newcommand\stepf{\Theta}

\newcommand{\Eof}{{\rm Ei}}
\newcommand{\rhob}{\bar{\rho}}

\newcommand{\hth}{\hat{\theta}}

\newcommand{\homega}{\hat{\omega}}

\newcommand{\hh}{\Omega}
\newcommand{\tilh}{\tilde{h}}

\newcommand{\MSbar}{\overline{\rm MS}}
\newcommand\bbas{\alpha_N}

\newcommand{\gggf}{\gamma_s}
\newcommand{\hgggf}{\hat{\gamma}_s}
\newcommand{\G}{{\cal G}}
\newcommand{\Ord}{{\cal O}}
\newcommand{\hchi}{\hat\chi}
\renewcommand{\Re}{{\rm Re}}
\renewcommand{\Im}{{\rm Im}}
\newcommand{\gam}{\gamma}
\newcommand{\gamz}{z}
\newcommand{\cc}{\varphi}

\newcommand{\LLp}{{\rm LL}^\prime}
\newcommand{\NLLp}{{\rm NLL}^\prime}
\newcommand{\NorLLp}{{\rm (N)LL}^\prime}
\newcommand{\LL}{{\rm LL}}
\newcommand{\NLL}{{\rm NLL}}

\newcommand{\HELL}{\href{https://www.roma1.infn.it/~bonvini/hell/}{\texttt{HELL}}\xspace}

\title{New results on small-$x$ resummation for splitting functions}

\author[a]{Marco Bonvini,}
\affiliation[a]{INFN, Sezione di Roma, Piazzale Aldo Moro~5,
I-00185, Rome, Italy}

\author[b,c]{Stefano Frixione,}
\affiliation[b]{INFN, Sezione di Genova, Via Dodecaneso 33,
I-16146, Genoa, Italy}
\affiliation[c]{CERN, Theoretical Physics Department,
CH-1211 Geneva, Switzerland}

\author[b,d]{Giovanni Stagnitto}
\affiliation[d]{Universit\`a degli Studi di Genova, Via Dodecaneso 33,
I-16146 Genoa, Italy}

\emailAdd{Marco.Bonvini@roma1.infn.it}
\emailAdd{Stefano.Frixione@cern.ch}
\emailAdd{Giovanni.Stagnitto@ge.infn.it}

\abstract{
We revisit the basic steps necessary to obtain next-to-leading-logarithmic
accurate small-$x$ results for the DGLAP splitting functions, and their
implementations within the \HELL framework. We derive new analytical
all-order results for the leading-logarithmic $gg$ anomalous dimension,
the $qg$ and $gg$ finite Green functions, and most importantly for the
$qg$ anomalous dimension, which allows us to arrive for the first time
at a properly resummed $qg$ splitting kernel. We use these results as
cornerstones of a new implementation of small-$x$ splitting-function
resummation which is more solid and numerically better behaved with
respect to those available thus far. All of these
novelties are included in the upcoming 4.0 version of \HELL.
}

\keywords{Parton Distribution Functions, Resummation, Small-$x$}

\begin{document}
\maketitle
\flushbottom

\section{Introduction}
\label{sec:intro}

Predictions for high-energy hadron colliders rely on the collinear
factorization formula
\begin{equation}\label{factfor}
  d\sigma_{h_1h_2 \to X}(Q^2,s) = \sum_{ij}
  \int dx_1 dx_2\,f_{i/h_1}(x_1,\mu_F^2) f_{j/h_2}(x_2,\mu_F^2)\,
  d\hsig_{ij\to X}(x_1, x_2, Q^2, s,\mu_F^2)\,,
\end{equation}
where $s$ denotes the centre-of-mass energy squared, $f_{i/h_1}$ and
$f_{j/h_2}$ are non-perturbative Parton Distribution Functions (PDFs), that
encode the longitudinal momentum distribution of partons $i$ and $j$ inside
the incoming hadrons $h_1$ and $h_2$, respectively, and $d\hsig_{ij\to X}$ is
the short-distance cross section (a.k.a.~coefficient function) that models the
hard scattering of the incoming partons for the sufficiently inclusive
production of a generic final state $X$. The symbol $Q^2$ denotes a hard scale
squared, typical of the scattering process, e.g.~the invariant mass squared
$M^2$ of the final state $X$.  The factorization in eq.~\eqref{factfor},
namely the separation of long- and short-distance effects, is associated with
a scale $\mu_F$, thus called factorisation scale: the dependence of PDFs on
such a scale is dictated by the DGLAP equations~\cite{Gribov:1972ri,
Dokshitzer:1977sg,Altarelli:1977zs}, whose kernels (also called splitting
functions in configuration space and anomalous dimensions in a space
Mellin-conjugated to the latter one) are known exactly up to
next-to-next-to-leading order (NNLO)~\cite{Moch:2004pa, Vogt:2004mw}.  At the
N$^3$LO, the currently available partial and approximate
results~\cite{Davies:2016jie,Moch:2017uml,Falcioni:2023vqq,Falcioni:2023luc,
Falcioni:2023tzp,Gehrmann:2023cqm,Gehrmann:2023iah,Falcioni:2024xyt,
Falcioni:2024qpd,Kniehl:2025ttz,Falcioni:2025hfz,Gehrmann:2026qbl} already
enable phenomenological applications.

While a factorization formula such as that of eq.~\eqref{factfor} originally
stems from a careful analysis of QCD colour dynamics\footnote{And it is
in fact unproved.}, a formally identical expression underpinned
by QED interactions and much better understood perturbatively can be employed
in the case of high-energy {\em lepton} colliders, like LEP or the proposed
\mbox{FCC-ee} or CEPC. There, it serves the primary purpose of resumming
potentially-large logarithmic terms of $m_\ell^2/Q^2$ ($m_\ell$ being the
mass of the colliding leptons) that are associated with initial-state collinear
radiation, and which emerge order-by-order in perturbative computations,
whose predictions might thus be spoiled by them. Technically, the resummation
is achieved by introducing lepton (antilepton) PDFs $f_{i/\ell^-}$
($f_{i/\ell^+}$), which are the strict analogues of their QCD counterparts;
indeed, as in QCD, such PDFs evolve perturbatively according to a DGLAP
equation, thereby achieving the sought resummation --- currently, the
state of the art is a next-to-leading logarithmic (NLL)
evolution~\cite{Bertone:2019hks,Bertone:2022ktl}.  Importantly, at
variance with the hadronic case, the initial conditions for the evolution
at $\mu_0^2\sim m_\ell^2$ are fully perturbative: at the leading-order they are
trivially given by $f_{\ell^\pm/\ell^\pm}(x,\mu_0^2) = \delta(1-x)$ for the
lepton themselves, and are equal to zero for all of the other partons.
At the next-to-leading order, the photon acquires a nonzero initial
condition~\cite{Frixione:2019lga}. The next-to-next-to-leading order initial
conditions have recently also become available (see
refs.~\cite{Stahlhofen:2025hqd,Schnubel:2025ejl}, and
refs.~\cite{Blumlein:2011mi,Blumlein:2020jrf,Ablinger:2020qvo}
for earlier works).

Among possible future lepton colliders, one interesting option is constituted
by a muon collider.  Thanks to its reduced synchrotron radiation, such a
machine could potentially reach energies of order \mbox{10-30~TeV}. Muon
PDFs relevant to muon-collider predictions have attracted some recent
interest~\cite{Han:2021kes,Garosi:2023bvq,Frixione:2023gmf}. As it will be
discussed more extensively later, owing to its much larger centre-of-mass
energy (w.r.t.~that of an $\epem$ collider), a muon collider will be sensitive
to much smaller values of parton momentum fractions. In that regime, two
things happen: firstly, the fixed-order perturbative approach of DGLAP
is liable to become unstable owing to the presence of $\log x$ terms
in the splitting kernels (small-$x$ effects, i.e.~the main argument of
this paper). And secondly, QCD dynamics inside the incoming leptons cannot
be neglected any longer. Indeed, although quark radiation first appears at
relative $\mathcal{O}(\alpha^2)$ through photon splitting, with gluon
radiation kicking in at $\mathcal{O}(\alpha^2\as)$, the resulting PDFs appear
to be non-negligible at the end of the evolution, with the gluon large in
size at small values of $x < 10^{-2}$--$10^{-3}$, second only to the
photon PDF\footnote{At large values of $x$, the muon-in-muon PDF
$f_{\mu^\pm/\mu^\pm}$ remains the only relevant contribution at any final
scale.}.  This is due to the interplay between the enhancement of QCD
radiation at small energies and the significant evolution range, which
spans at least three order of magnitudes from the muon mass
$m_\mu \sim 100$~MeV to the electroweak scale. Because of this,
refs.~\cite{Han:2021kes,Garosi:2023bvq,Frixione:2023gmf} all deal with
QCD radiation, although by means of vastly different approaches.

\begin{figure}[t]
  \begin{center}
    \includegraphics[scale=1.0,width=\textwidth]{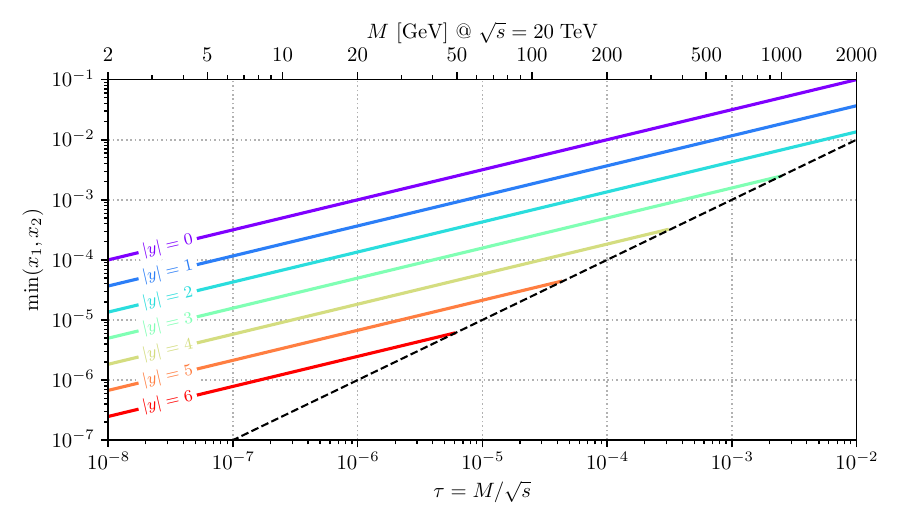}
    \caption{Momentum fractions probed at high-energy colliders as a
      function of \mbox{$\tau= M/\sqrt{s}$} for various rapidities.
      On the secondary axis on the top,
      the value of $M$ for $\sqrt{s} = 20$~TeV is indicated as reference. Note
      that the largest accessible rapidity is determined by $\tau$, with $|y| <
      1/2\log(1/\tau)$, as marked by the black dashed line.}
    \label{fig:kin}
  \end{center}
\end{figure}

We now return to the first item mentioned before, i.e.~the relevance
of small-$x$ effects in high-energy collisions. In fact, since such
effects are driven by kinematics, they are equally relevant to hadron
as well as to lepton collisions, and can thus be discussed by means of the
same language in these two cases. The starting point is the observation
that, in the context of a factorisation theorem, the production of a
final state with invariant mass $M$ entails that the PDFs will be
evaluated (in leading-order kinematics) at the following $x$ values
\begin{equation}\label{xvals}
  x_{1,2}=\frac{M}{\sqrt{s}}\,e^{\pm y}=\sqrt{\tau}\,e^{\pm y}\,,
\end{equation}
with $y$ the laboratory-frame rapidity of that final state, and
$\tau \equiv M^2/s$.  To illustrate the kinematical regions that emerge
from eq.~(\ref{xvals}), in fig.~\ref{fig:kin} we plot eq.~\eqref{xvals}
as a function of $\tau$ for given values of $y$.  The main message of
fig.~\ref{fig:kin} is that, over a large part of the phase space, small values
of $x$ are probed. For example, at a $\sqrt{s}=20$~TeV collider, assuming
the onset of the small-$x$ regime at \mbox{$x\sim10^{-3}$} (i.e.~the regime
in which there could possibly be a breakdown of a fixed-order-based DGLAP
PDF evolution), this corresponds to the central ($y=0$) production of a
system with $M=20$~GeV; at $|y|=2$ that value increases to about $M=180$~GeV,
while production of any final state with absolute rapidity larger than 3 will
involve the small-$x$ region.
%%%%%%%%%%%%%%%%%%%%%%%%%%%%%%%%%%%%%%%%%%%%%%%%%%%%%%%%%%%%%%%%%%%%
%%%%%%%%%%%%%%%%%%%%%%%%%%%%%%%%%%%%%%%%%%%%%%%%%%%%%%%%%%%%%%%%%%%%%%%%

The presence of potentially large $\log x$ terms in PDF evolution (and in
the coefficient functions) calls for their all-order resummation.
Small-$x$ (or high-energy) resummation is a longstanding topic in QCD,
and has been addressed with a variety of methods; for a pedagogical
introduction we refer the reader to textbooks~\cite{Kovchegov:2012mbw,
Forshaw:1997dc}.  The small-$x$ resummation of DGLAP evolution is known
to next-to-leading logarithmic accuracy (NLL) and it is based on the BFKL
equation~\cite{Lipatov:1976zz,Fadin:1975cb,Kuraev:1976ge,Kuraev:1977fs,
Balitsky:1978ic}. After the computation of the two-loop BKFL
kernel~\cite{Fadin:1998py,Ciafaloni:1998gs}%
\footnote{The calculation of the three-loop BFKL kernel is in progress~\cite
{Caola:2021izf,Falcioni:2021dgr,Falcioni:2021buo,Fadin:2023roz,
Buccioni:2024gzo,Abreu:2024xoh,Byrne:2022wzk,Byrne:2025phh}.},
it has been realised that the
BFKL evolution kernel is perturbative unstable, making its implementation in
phenomenological applications non trivial. The problem of constructing
resummed anomalous dimensions for PDF evolution has been tackled by several
groups, see
refs.~\cite{Ball:1995vc,Ball:1997vf,Altarelli:2001ji,Altarelli:2003hk,
Altarelli:2005ni,Altarelli:2008aj},
refs.~\cite{Salam:1998tj,Ciafaloni:1999yw,Ciafaloni:2003rd,Ciafaloni:2007gf},
and refs.~\cite{Thorne:1999sg,Thorne:1999rb,Thorne:2001nr,White:2006yh}.
More recently, further progress has been made in
refs.~\cite{Bonvini:2016wki,Bonvini:2017ogt}, where the
Altarelli-Ball-Forte (ABF henceforth) approach~\cite{Ball:1995vc,Ball:1997vf,
Altarelli:2001ji,Altarelli:2003hk,Altarelli:2005ni,Altarelli:2008aj}
has been exploited to obtain state-of-the-art resummed predictions for DGLAP
splitting functions, publicly available through the \HELL code.

We reiterate that fig.~\ref{fig:kin} implies that small-$x$ resummation is
also relevant for muon collider physics, and therefore one should include
it at least in (but not necessarily limited to) the QCD-evolution sector
of lepton densities. However, in doing so, one faces a problem whose analogue
is not relevant to hadronic-PDF evolution. Namely, muon-PDF evolution
starts at a scale of the order of the muon mass, i.e.~smaller than
$\Lambda_{\rm QCD}$, that is a regime where hadron PDFs are not evolved
by means of DGLAP equations. Therefore, with muon PDFs one is forced to
either artificially switch off QCD dynamics~\cite{Han:2021kes,Garosi:2023bvq}
without a proper dynamical mechanism to do so\footnote{This implies the
necessity of introducing a switch-off scale upon which PDF predictions
will depend in an inherently infrared-unstable manner.}, or to extend the
validity of $\as(\mu^2)$ down to $\mu^2=0$. The latter option has been
suggested by two of us in ref.~\cite{Frixione:2023gmf}, where it is
achieved by means of the adoption of a
simple analytical parametrisation of $\as$, inspired by a dispersive
approach~\cite{Webber:1998um}. With such a parametrisation,
at small scales the coupling behaves as\footnote{Expressions valid at
two-loop, as well as with quark-mass thresholds, can be found in
ref.~\cite{Frixione:2023gmf}.}
\begin{equation}
\as(\mu) \sim \frac{1}{\beta_0}
\[ \as^{\rm pert}\(\frac{\mu^2}{\Lambda_{\rm QCD}^2} \) + \as^{\rm np}
\(\frac{\mu^2}{\Lambda_{\rm QCD}^2} \) \],
\end{equation}
with $\beta_0$ the first coefficient of the QCD $\beta$ function and
\begin{equation}
\as^{\rm pert}(X) = \frac{1}{\log(X)}, \quad \as^{\rm np}(X)
= \frac{X+b}{(1-X)(1+b)}\left( \frac{1+c}{X+c} \right)^p.
\end{equation}
When $\mu^2 \to \Lambda_{\rm QCD}^2$, both $\as^{\rm pert}(X)$ and $\as^{\rm
np}(X)$ behave as $1/(1-X)$ with opposite signs, canceling the divergence.
Elsewhere, this functional form guarantees the smooth transition betweem
an RGE-driven evolution relevant to large and intermediate values of $\mu^2$,
and its low-energy parametrised small-scale counterpart. In general, it is
this smooth behaviour coupled with the requirement that RGE dynamics be
used as much as is sensible (i.e.~down to around the charm-quark mass)
which implies a rapid growth of $\as(\mu^2)$ for decreasing $\mu$ before
non-perturbative effects turn it around. In particular,
with the default values for the parameters of $\as^{\rm np}$ adopted in
ref.~\cite{Frixione:2023gmf} ($p=4$, $b=0.25$ and $c=4$) $\as(\mu^2)$
can be as large as $0.8$ (this happens for $\mu \sim 0.3$~GeV).

The bottom line is that, on top of the small-$x$ resummation procedure
which is common to both hadron and lepton PDFs, the physics of a muon
collider (and, in fact, that of an $\epem$ collider whose physics program
is based on the ability of reaching extremely large precision where all
ingredients must be controlled to the minutest details) forces one
to consider a large-$\as$ regime as well. This regime has been largely
ignored in the literature and in the resulting computer codes
such as \HELL, which will be prominent in this paper.

By using the large-$\as$ argument as a motivation, in this work we address
and solve the problem it poses. In doing so, we achieve what are nominally
by-product ingredients in such a solution, but which are in fact of general
greater importance than the ability to carry out small-$x$, large-$\as$
evolution. Specifically, we have obtained for the first time in the
literature analytical all-order results for, among other things, the so
called $\gamma_s$ and $h_{qg}$ functions, as well as for the complete
$\ord(\ep^0)$ $\G_{qg}$ Green function. We stress that the $h_{qg}$
result is of particular importance, since it leads to a properly
all-order resummed expression for the $P_{qg}$ kernel, which currently
is resummed to all orders only in an approximate manner that is, as we
shall show, liable to extremely large instabilities. We use these novel
results as a basis for improving the \HELL code, which thus becomes
more robust (importantly, also in the context of hadron-collision
applications), and capable of dealing with large-$\as$ evolution.
In view of this, the present paper has a broader and more technical
scope than muon-PDF small-$x$ evolution --- the results for the latter
topic are therefore postponed to a forthcoming
work~\cite{Bonvini:muonpdf}.

The paper is organised as follows: in sect.~\ref{sec:general} we give
a brief overview of the small-$x$ resummation procedure we follow
(sects.~\ref{sec:gengampp} and~\ref{sec:gengamqg}), and of the new
analytical results we have obtained in this work (sect.~\ref{sec:res}).
In sect.~\ref{sec:Pqg} we present our findings relevant to the $P_{qg}$
splitting kernel, in both its all-order and perturbative-expansion forms.
The implications of the new analytical results we have found, and of the
new approach for the inclusion of subleading-logarithmic effects, including
those stemming from strong-coupling running, are discussed in
sect.~\ref{sec:largeas}. Sect.~\ref{sec:Gqgres} reports the analytical
results for the $\ord(\ep^0)$ $qg$ and $gg$ Green functions, which do
not enter any of the numerical predictions we have obtained, but are
nevertheless useful to elucidate the all-order structure emerging
from high-energy factorisation. Finally, in sect.~\ref{sec:concl} we
give our conclusions. Additional technical material is reported in
the appendices: sect.~\ref{sec:gammagg} deals with the LL $gg$
anomalous dimension, sect.~\ref{sec:appqg} with the $qg$ anomalous
dimension and Green function, and sect.~\ref{sec:appHELL} with the
numerical implementation relevant to the new version of \HELL.

\section{Generalities on small-$x$ resummation}
\label{sec:general}

We begin by broadly summarizing the strategy used to perform
small-$x$ resummation of the DGLAP splitting functions. For pedagogical
purposes, in the present section we keep technical details to a minimum.

Leading small-$x$ logarithms arise only in the singlet sector\footnote{The
non-singlet sector is affected by small-$x$ logarithmic enhancements only
starting at the next-to-leading power. As such, it is not of relevance
to us for the current work.
This also implies that the impressive progress towards the computation of
N$^3$LO DGLAP splitting functions is currently of no help for improving the
resummed results, as so far only the non-singlet splitting functions are
known exactly~\cite{Gehrmann:2026qbl}, while for the singlet ones only some
integer Mellin moments are known, which do not allow one to constrain the
small-$x$ region.  Rather, any approximate N$^3$LO prediction of the singlet
splitting functions must supplement the constraints from the known Mellin
moments with the small-$x$ behaviour known from resummation (see
e.g.~\cite{Falcioni:2024qpd}).}, and we therefore consider a $2\times2$
evolution matrix that couples together the quark singlet and the gluon.
Small-$x$ resummation is typically performed in a conjugate (Mellin) space.
We define the Mellin transformation following the standard small-$x$
conventions, where the entries $\gamma_{ij}$ of the anomalous dimension
$2\times2$ matrix are defined as
\beq \label{gamma_def}
\gamma_{ij}(\as,N) = \int_0^1 dx\, x^N P_{ij}(\as,x),
\eeq
so that the leading-power small-$x$ logarithms that have the generic
form $\frac{1}{x}\log^nx$ correspond to poles at $N=0$ (rather than
at $N=1$, which one would obtain by using the standard Mellin-transform
conventions). Specifically, the leading logarithmic (LL) contributions
are defined to be all terms that behave as $\frac1x\as^n\log^{n-1}x$
for any $n$, which correspond to $(\as/N)^n$ structures in the Mellin space,
while next-to-leading logarithmic (NLL) terms are suppressed by an overall
power of $\as$ w.r.t.~the LL ones, and so on.  Once the all-order
functions of $\as/N$ are constructed in the Mellin space, the resummed
splitting functions $P_{ij}(x)$ are then obtained in the configuration
space by Mellin inversion.

The NLL resummation of anomalous dimensions in the singlet sector requires
just two resummed ingredients.  Historically, one of them is taken to be the
eigenvalue $\gamma_+(\as,N)$ of the matrix in eq.~\eqref{gamma_def}, that is
defined as the one containing {\em all} small-$N$ singularities, while the
other, $\gamma_-$, is nonsingular at $N=0$. In addition to that, one needs
to include information on the rotation back to flavour space, which is
carried out for instance by the entry $\gamma_{qg}$. At the NLL all of the
other matrix entries can be derived from these two ingredients
according to the following relations\footnote
{While formally eq.~\eqref{CCgq} is valid only at LL,
  we can still define $\gamma_{gq}^{\NLLp}$ as in eq.~\eqref{CCgq}
  since as far as PDF evolution is concerned
  the yet unknown NLL component of $\gamma_{gq}$ is irrelevant because that kernel is multiplied
  by the quark PDF, which is itself a NLL object.}
\begin{subequations}
\label{Delta4gamma}
\begin{align}
\gamma_{gg}^{\NLLp}(\as,N) &=
\gamma_+^{\NLLp}(\as,N) -\frac{C_F}{C_A}\gamma_{qg}^{\NLLp}(\as,N), \\
\gamma_{gq}^{\NLLp}(\as,N) &= \frac{C_F}{C_A}\gamma_{gg}^{\NLLp}(\as,N),
\label{CCgq}\\
\gamma_{qq}^{\NLLp}(\as,N) &= \frac{C_F}{C_A}\gamma_{qg}^{\NLLp}(\as,N).
\end{align}
\end{subequations}
We stress that the resummation procedure of
$\gamma_+^{\NLLp}$ and $\gamma_{qg}^{\NLLp}$ that we are about to review
actually contains a number of contributions in addition to the pure LL
and NLL small-$x$ terms, chiefly emerging from the running of the strong
coupling, that, despite being subleading w.r.t.~the indicated logarithmic
accuracy, are necessary for the perturbative stability of the result.
This is what informs the notation used in eq.~\eqref{Delta4gamma}, i.e.~the
fact that we have written $\NLLp$ rather than simply $\NLL$. This an
example of a notation which we shall systematically adopt in the paper,
by following the
\begin{itemize}
\item[]{\bf Notation rule:} for any given resummed quantity $\varsigma$,
the symbol $\varsigma$ itself denotes the all-order function that includes
all small-$x$ logarithmic towers. By $\varsigma|_{\LL}$ and
$\varsigma|_{\NLL}$ we denote the corresponding quantities which strictly
contain all and only the leading and the leading plus next-to-leading
small-$x$ logarithms, respectively. Finally, by $\varsigma^{\LLp}$ and
$\varsigma^{\NLLp}$ we denote
$\varsigma|_{\LL}$ and $\varsigma|_{\NLL}$
augmented by logarithmic contributions beyond the leading and the
next-to-leading ones, respectively, as well as by power-suppressed terms.
\end{itemize}
Clearly, there are infinite ways to include beyond-accuracy logarithmic
contributions into a strictly logarithmically-ordered quantity; the
precise definition of such extra contributions will be clear from the
context.

We also stress that the results that we are going to present are in a
factorization scheme which is a variant of the $\MSbar$ scheme usually called
$Q_0\MSbar$~\cite{Catani:1993ww,Catani:1994sq,Ciafaloni:2005cg,Marzani:2007gk}.
We also recall that all resummed results, however they are
obtained, must then be matched to the known exact fixed-order results. This is
usually performed by using an additive matching: from the resummed result its
$\as$ expansion up to the desired order is subtracted, and this ``resummed
contribution'' (after a suitable damping\footnote{We remind the reader that,
while not mandatory, damping is common practice in matched results whose
resummation components are potentially anomalously large in regions where
resummation should actually not be necessary.} at large $x$) is added to the
corresponding fixed-order complete result.

Small-$x$ resummation in Mellin space is performed differently for the
eigenvalue $\gamma_+$ and the $\gamma_{qg}$ entry: the former is resummed
through duality with BFKL (see sect.~\ref{sec:gengampp}), while the latter is
obtained through the collinear factorization of the quark Green function in
the high-energy limit (see sect.~\ref{sec:gengamqg}).

\subsection{Resummation of $\gamma_+$}\label{sec:gengampp}
The core of the resummation of the anomalous dimension matrix is the eigenvalue
$\gamma_+$.  At the LL, it is the only object needed for resummation, and it
thus coincides with $\gamma_{gg}$ and (up to a colour factor) $\gamma_{gq}$,
while the quark entries $\gamma_{qg}$ and $\gamma_{qq}$ are of NLL.
Its resummation is driven by the consistency between DGLAP and BFKL
equations, which leads to an all-order duality relation between
$\gamma_+(\as,N)$ and the BFKL kernel $\chi_{\rm BFKL}(\as,\gamma)$\footnote
{Note that $\chi_{\rm BFKL}(\as,\gamma)$ is sometimes called the eigenvalue or
characteristic function of the BFKL equation, obtained by Mellin
transformation from the BFKL kernel in momentum space, with $\gamma$ being the
Mellin conjugate variable.  The choice of using $\gamma$ as the generic name
for this variable, often adopted in the literature, is motivated by the fact
that by duality the kernel is computed with argument $\gamma$ coinciding with
the anomalous dimension.}.
By ignoring running coupling effects, the duality
takes generally the form
\beq\label{eq:duality}
\chi_{\rm BFKL}\(\as, \gamma_+(\as,N)\) = N,
\eeq
namely the two functions are the inverse of one another.
By using the LO expression of the kernel, namely
\mbox{$\chi_{\rm BFKL}(\as,\gamma) = \as \chi_0(\gamma) + \Ord(\as^2)$}, one
obtains the resummation of the pure LL contributions to $\gamma_+$.  Indeed,
eq.~\eqref{eq:duality} becomes
\beq\label{eq:dualityLO}
\as \chi_0\(\gamma_+(\as,N)\) + \Ord(\as^2) = N,
\eeq
which, neglecting the $\Ord(\as^2)$ contributions, clearly gives as
solution for $\gamma_+$ an all-order function of $\as/N$.
Explicitly, we can therefore write
\begin{align}
  \chi_0(\gam) &= \frac{\CA}\pi \,\chi(\gam), \label{eq:BFKLkerLO}\\*
  \chi(\gam) &\equiv 2\psi_0(1)-\psi_0(\gam)-\psi_0(1-\gam)
= \frac{1}{\gam}+\sum_{k=1}^\infty 2\zeta_{2k+1}\gam^{2k},\label{eq:chidef}
\end{align}
with $\psi_0$ the digamma function and $\zeta_{2k+1}$ the Riemann $\zeta$
constants.
We can thus rewrite the duality at this order,
eq.~\eqref{eq:dualityLO}, as
\begin{equation}
\label{dualLO}
  1 = \bbas\,\chi\left(\gamma_s(\bbas)\right),
\end{equation}
where for convenience we have introduced the symbol (in four dimensions ---
for a $d$-dimensional definition see eq.~\eqref{abarnorm2X})
\begin{equation}\label{bbasdef}
  \bbas \equiv \frac{\CA\as}{\pi N}.
\end{equation}
The anomalous dimension $\gamma_s$ emerging from eq.~\eqref{dualLO}
corresponds to the pure LL part of $\gamma_+$, and resums all and only
the $(\as/N)^k$ terms\footnote{Therefore, according to our conventions
it should be denoted by $\gamma_+|_{\LL}$; the use $\gamma_s$ instead
of that symbol is in keeping with the conventions of the literature.}.
Solving the implicit equation~\eqref{dualLO} is complicated
due to the functional form of $\chi(\gam)$, and indeed
so far in the literature only fairly limited analytical information
had been available.
Conversely, in this work, we have been able to solve
eq.~\eqref{dualLO} perturbatively to all orders, thus finding an explicit
series representation. While the former cannot be resummed in a closed
form directly in terms of elementary functions, we have managed to do so
through an integral representation; an all-order fixed-point solution
is reported later as well. A detailed discussion of these novel results
is given in appendix~\ref{sec:gggPert}.

Unfortunately, the resummation in terms of $\gamma_s$, despite being a useful
tool for perturbative analytic manipulations, is not suitable for obtaining
phenomenological results numerically.  Indeed, when moving to the NLL, which
in turns requires the usage of the NLO BFKL kernel, the solution of the duality
equation has a completely different behaviour at small $N$ w.r.t.~the pure LL
one, thus exposing a problem of perturbative stability of the resummation
procedure. Such an instability in the resummed
perturbative progression is induced by a severe perturbative instability of
the fixed-order BFKL kernel.  In particular, in the ``physical region'' of the
kernel, $0<\gamma<1$, NLO corrections change completely the behaviour near
$\gamma=0$ and $\gamma=1$, owing to the novel presence of higher poles with
opposite signs.  Therefore, in order to achieve a sensible resummation,
the instability of the fixed-order BFKL kernel has to be addressed first.

The achievement of perturbatively-stable resummed results took several years
in the late 90s and early 2000s. The strategy towards this goal, used by most
groups with some differences in the practical implementations, is based on the
following steps.
\begin{itemize}
\item Firstly, the duality eq.~\eqref{eq:duality} is employed ``in reverse'',
  namely by using the knowledge of the fixed-order anomalous dimension to
  solve the duality equation for the BFKL kernel, thereby leading to the
  resummation of the poles of such kernel located at $\gamma=0$.
\item This procedure does not cure the instability at $\gamma=1$. In order to
  achieve this, the symmetry property of the BFKL kernel for
  $\gamma\to1-\gamma$ is exploited to extend the stabilisation of the pole at
  $\gamma=0$ to $\gamma=1$ as well.
\item At this point the fixed-order BFKL kernel has been stabilised by the
  resummation of its leading singularities.  After further imposing a
  constraint due to momentum conservation, the duality can be used again to
  obtain a resummed anomalous dimension, usually dubbed double-leading,
  because \emph{both} (hence the ``double'') the kernel and its dual anomalous
  dimension are accurate to (N)LO+(N)LL accuracy.
\item Finally, the resummation of a class of subleading terms originating from
  the running of the strong coupling is also included. Despite being
  subleading, these terms are very important because they change the kind of
  the leading small-$N$ singularity (from a branch point to a simple pole),
  thus impacting the leading small-$x$ behaviour of the splitting functions.
\end{itemize}
The concrete realisation of this procedure as proposed by the ABF group is
what has been implemented in the \HELL code, with various improvements
detailed in refs~\cite{Bonvini:2016wki,Bonvini:2017ogt,Bonvini:2018xvt}.
As a result, the perturbatively stable resummed anomalous dimensions
contain, in addition to the purely LL and NLL contributions,
several subleading logarithmic and subleading power terms.
Thus, we can write
\begin{align}
  \gamma_+^{\LLp} (\as,N) &= \gamma_s(\bbas) + \text{subleading log} + \text{subleading power} \\
  \gamma_+^{\NLLp} (\as,N) &= \gamma_s(\bbas) + \as \gamma_{ss}(\bbas) + \text{subleading log} + \text{subleading power},
\end{align}
where $\gamma_{ss}(\bbas)$ is another all-order function of $\as/N$
representing all purely NLL contributions to $\gamma_+$.
The subleading contributions, despite formally being beyond accuracy, are a
crucial part of the resummed result, and cannot be neglected.

Unfortunately, explicit expressions for the subleading terms (or equivalently
for the full $\gamma_+^{\NorLLp}$) are very difficult (if not impossible) to
obtain, given that part of the anomalous dimension comes from the solution of
the duality equation with a kernel which is much more complicated than the LO
one of eq.~\eqref{eq:BFKLkerLO} and \eqref{eq:chidef}.  The standard approach,
also adopted in the \HELL code, is to compute the solution to the duality
relation numerically, a procedure that has some criticalities (we discuss some
of them in sect.~\ref{sec:HELL-as}).  Our findings on the explicit forms of
$\gamma_s$ presented in appendix~\ref{sec:gggPert} may provide some insight on
this procedure, possibly paving the way for an alternative, more robust
determination of the resummed anomalous dimension $\gamma_+^{\NorLLp}$.

We conclude this section by noting that eq.~\eqref{dualLO} allows one to
trivially express $\bbas$ in terms of $\gamma_s$ as follows\footnote{The full
perturbative expansion of $\bbas$ in terms of $\gamma_s$ is provided in
eq.~\eqref{atom}.}
\begin{equation}
\label{dualLOaN} \bbas =
\frac{1}{\chi(\gamma_s)} = \gamma_s + \Ord(\gamma_s^2)\,.
\end{equation}
We stress that this is an exact relationship between two given quantities
($\bbas$ and $\gamma_s$), which has nothing to do with the logarithmic
accuracy at which one is working.  This property allows one to recast any
perturbative expansion in $\bbas$ as an expansion in $\gamma_s$;
it will be extensively exploited in sect.~\ref{sec:Pqg}, since $\gamma_s$
enters as a building block in the resummation of $\gamma_{qg}$.  Given the
intricate nature of $\gamma_s$ as is outlined above and detailed in
appendix~\ref{sec:gammagg}, being able to express results in terms of
$\gamma_s$ rather than $\bbas$ simplifies the analytical expressions.

\subsection{Resummation of $\gamma_{qg}$}\label{sec:gengamqg}

The resummation of the $\gamma_{qg}$ element of the splitting matrix relies on
the collinear factorization of the quark Green function in the high-energy
limit~\cite{Catani:1994sq}.  As we shall come back to this factorization later
in sect.~\ref{sec:Pqg} we refrain from giving additional details here, and focus
instead on available results and on how they have been used.  We limit
ourselves to stressing that this factorisation procedure has been employed
so far in the literature only to extract the perturbative expansion of
$\gamma_{qg}$ in an iterative manner (thus in practice for a limited
number of them).

The $qg$ anomalous dimension is enhanced at small $x$ starting at NLL, which
means that its leading small-$x$ dependence can be written as a simple
$\as$ prefactor, times an all-order function of $\as/N$. More explicitly,
from ref.~\cite{Catani:1994sq}, we have
\beq\label{gammaqgCH}
\left.\gamma_{qg}(\as,N)\right|_{\NLL}=
\frac{\as}{3\pi}\,\TR
\[1+\frac{5}{3}\bbas+\frac{14}{9}\bbas^2
 +\(\frac{82}{81}+2\zeta_3\)\bbas^3+\ldots\].
\eeq
A complete all-order expression for $\gamma_{qg}$ has never been fully
achieved so far. Available implementations of the resulting resummation
therefore must rely on a finite number of coefficients of the $\gamma_{qg}$
expansion in $\as$: ref.~\cite{Catani:1994sq} presented the
analytical results for the first six coefficients of such an expansion,
and later in the context of ref.~\cite{Altarelli:2008aj}
about thirty-five coefficients have been computed numerically.
As we shall show, the numerical extraction leads one to significant deviations
w.r.t.~to the actual analytical result, which we shall present later.
The authors of ref.~\cite{Catani:1994sq} have also been able
to derive an all-order expression for the rational part of the coefficients,
which is however insufficient to obtain an improved numerical approximation,
as the trascendental $\zeta$ terms are as important as (if not more than)
the rational ones.

Even though any implementation of the resummation of $\gamma_{qg}$ based on a
finite number of coefficients is necessarily approximate, in the literature
some additional manipulations have been introduced in order to obtain a result
which is as good as possible.  These include on the one hand methods to try and
improve the convergence properties of the series, in order to make the most out
of the few coefficients which were known, and on the other hand to include in
the resummed formula additional contributions due to the running of the
coupling, which are important\footnote {Of course, the improvement given by
the resummation of subleading running coupling effects may be irrelevant in
comparison with the irreducible uncertainty due to the missing higher order
coefficients.  However, as we shall see in sect.~\ref{sec:results}, there are
kinematical regimes in which the approximation obtained from the limited
number of coefficients is good enough to justify the inclusion of the running
coupling corrections.}  for a more reliable description of the small-$N$
regime (as is the case for $\gamma_+$).

The first of these manipulations stems from
introducing~\cite{Altarelli:2008aj} a new function, $h_{qg}$, such that
the anomalous dimension $\gamma_{qg}$ at the NLL can be rewritten as
\beq\label{hqgdef}
\left.\gamma_{qg}(\as,N)\right|_{\NLL}=
\frac{\as}{3\pi}\,\TR\, h_{qg}(\gamma_s(\bbas)),
\eeq
in terms of the pure LL anomalous dimension $\gamma_s(\bbas)$, defined in
eq.~\eqref{dualLO}. In writing thus its resummed expression, $\gamma_{qg}$
has the same form as that of the resummed expressions of the coefficient
functions~\cite{Catani:1994sq}; this form is indeed underpinned by the idea
that $\gamma_{qg}$ represents the collinear divergent piece of a coefficient
function, and it is therefore a part of it.
The advantage of eq.~\eqref{hqgdef} is that the anomalous dimension $\gamma_s$
captures the essence of the irreducible complexity of the resummed result,
so that the function $h_{qg}$ is expected to be simpler, and with better
convergence properties, w.r.t.~the original function $\gamma_{qg}$.
In particular, by making use of eq.~\eqref{dualLOaN}, it is easy to extract
the coefficients of the expansion of $h_{qg}$ in powers of $\gamma_s$ from
those of $\gamma_{qg}$ in powers of $\bbas$: truncating the series expansion
of the function $h_{qg}$ at a given order $\gamma_s^k$ is likely less
impactful than truncating the expansion of $\gamma_{qg}$ at the corresponding
order $\bbas^k$, owing to the expected better convergence properties of
the former w.r.t.~the latter.

The second manipulation consists in introducing the resummation of subleading
running coupling contributions.  This is more easily achieved in terms of the
function $h_{qg}$.  One simple (but insufficient) step is to compute the
function $h_{qg}$ in terms of the full resummed $\gamma_+$ rather than simply
in terms of $\gamma_s$, so that the resummation of running coupling effects in
$\gamma_+$ is automatically included.  Technically, at this order, it is
sufficient to use $\gamma_+^{\LLp}$; however, ABF~\cite{Altarelli:2008aj}
suggest to use the
NLL anomalous dimension $\gamma_+^{\NLLp}$ so that the leading $N$ pole is
in the same position for all entries of the singlet anomalous dimension
matrix. For the rest of the section we simply write $\gamma_+$
without specifying the order, but keeping in mind that in practical
implementations (like in \HELL) it is to be intended as $\gamma_+^{\NLLp}$.

In order to actually resum all leading running-coupling contributions (those
proportional to powers of $\beta_0$), it is also necessary to change the way
the function $h_{qg}$ depends on $\gamma_+$~\cite{Ball:2007ra,Altarelli:2008aj}.
By introducing the series expansion of $h_{qg}$,
\beq
\label{eq:hqgser}
h_{qg}(\gamz)=\sum_{k=0}^\infty h_k \gamz^k\,,
\eeq
the resummed $\NLLp$ anomalous dimension appearing in eq.~\eqref{Delta4gamma}
is computed according to the formula
\beq\label{hqgABF}
\gamma_{qg}^{\NLLp}(\as,N) =
\frac{\as}{3\pi}\,\TR\, \sum_{k=0}^\infty h_k \[\gamma_+^k(\as,N)\],
\eeq
where we have introduced the notation $\[\gamma_+^k\]$ defined by
the following recursive procedure~\cite{Altarelli:2008aj}
\beq\label{eq:recursion[]}
\[\gamma_+^{k+1}\] = \(\gamma_+-k\,r(\as,N)\)\[\gamma_+^k\],
\qquad
\[\gamma_+^0\] = 1.
\eeq
The quantity $r(\as,N)$, in its simplest realisation,
is given by $r(\as,N)=\as\beta_0$.
This expression stems from approximating the dependence on $\as$ of $\gamma_+$
as if it were linear (like at the LO).  ABF also suggest an alternative form of
$r(\as,N)$ such that the coefficient of $\as$ is the actual derivative of
$\gamma_+$.  We shall come back to this point shortly.

Note that eq.~\eqref{hqgABF} cannot straightforwardly be written as an
all-order closed expression (a non-trivial integral expression will be
presented in sect.~\ref{sec:HELLimplementation}).  Paradoxically, this
has not posed a significant problem, since in any case the number of known
$h_k$ coefficients was quite limited.  Indeed, at this stage, in order to
obtain a resummed expression one had to decide what to do with the limited
knowledge available.
This problem is even more severe after the inclusion of running coupling
effects, because the convergence of the series in eq.~\eqref{hqgABF} is worse
than the original one of eq.~\eqref{eq:hqgser}, due to the presence of the
factor $k$ in the recursion eq.~\eqref{eq:recursion[]}, which reduces the radius
of convergence\footnote{We point out that eq.~\eqref{hqgABF} is an
asymptotic expansion~\cite{Altarelli:2008aj} of an all-order result which was
unknown at the time of the publication of ref.~\cite{Altarelli:2008aj}; as
such, it is affected by an intrinsic, spurious (i.e.~non-physical)
ambiguity.}.
In order to tame
this problem, in ref.~\cite{Altarelli:2008aj} the Borel summation of the
series was considered, supplementing the asymptotic behaviour of the Borel
transform with a guessed functional form.  Later, in
ref.~\cite{Bonvini:2016wki} this procedure has been modified by
computing a Pad\'e approximant of the Borel transform and using it in its place,
which leads to a more robust
result. The \HELL implementation prior to this paper
is based on the latter approach, and
employs just the first 16 coefficients of the expansion of
$h_{qg}$\footnote{The
choice of using 16 coefficients was based on an analysis of the stability of
the result upon variations of several parameters (the number of coefficients
used, the order of the Pad\'e approximant, the order of the Borel
transform).\label{foot:16}}.  The quality of such an approximation can be
assessed indirectly by comparing it with the exact resummed result presented
for the first time in this paper (see sect.~\ref{sec:results}).

Before concluding this section, we give some extra detail about this
resummed formula that will be useful later in this work.  We start by
reporting an alternative form of eq.~\eqref{hqgABF}, derived in
ref.~\cite{Bonvini:2016wki}, which reads
\beq\label{gammaqgResHELLv2}
\gamma_{qg}^{\NLLp}(\as(Q^2),N) = \frac{\as(Q^2)}{3\pi}\,\TR\,
\sum_{k=0}^\infty h_k \[\partial_\nu^k U(N,Q^2e^\nu,Q^2)\]_{\nu=0}.
\eeq
This equation coincides with eq.~\eqref{hqgABF} for a specific form of
the function $U$, but it is more general than that for a generic $U$,
and in particular it allows one to incorporate the $\beta_0$ terms to
all orders without making any approximation.
The function $U$ represents the leading DGLAP evolution at small $x$, given by
\beq\label{eq:Udef}
U(N,\mu^2,\mu_0^2) = \exp\left[\int_{\mu_0^2}^{\mu^2} \frac{dq^2}{q^2}\,
\gamma_+\left(\as(q^2), N \right)\right].
\eeq
In order to be able to compute the integral analytically, some
approximation must be introduced. The most drastic one consists in
neglecting the running of the strong coupling altogether, namely to use
\beq\label{eq:Ufc}
U(N,\mu^2,\mu_0^2) \overset{\rm f.c.}=
\(\frac{\mu^2}{\mu_0^2}\)^{\gamma_+(\as,N)},
\eeq
so that the derivatives in this fixed-coupling (f.c.) case become
\beq
\[\partial_\nu^k U(N,Q^2e^\nu,Q^2)\]_{\nu=0} \overset{\rm f.c.}=
\gamma_+^k(\as,N),
\eeq
thus reproducing the power series of $h_{qg}$ computed in $\gamma_+(\as,N)$
(to fully recover the fixed-coupling result eq.~\eqref{hqgdef}, one should
also replace $\gamma_+(\as,N)$ with its fixed-coupling expression
$\gamma_s(\bbas)$).
A less drastic approximation leading to the ABF result of eq.~\eqref{hqgABF}
is
\beq\label{gamPABF}
\gamma_+(\as(q^2),N) \overset{\rm ABF}=
\frac{\gamma_+(\as(\mu_0^2),N)}{1+r(\as(\mu_0^2),N)\log\frac{q^2}{\mu_0^2}},
\eeq
whence
\beq\label{eq:UABF}
U(N,\mu^2,\mu_0^2) \overset{\rm ABF}=
U_{\rm ABF}(N,\mu^2,\mu_0^2) \equiv \(1+r(\as(\mu_0^2),N)\log\frac{\mu^2}{\mu_0^2}\)^{\gamma_+(\as(\mu_0^2),N)/r(\as(\mu_0^2),N)}.
\eeq
Note that the ABF-approximate evolution operator in eq.~\eqref{eq:UABF}
depends only on $\as(\mu_0^2)$, while the $\mu$ dependence is explicit.
It is easy to see~\cite{Bonvini:2016wki} that the derivatives in
eq.~\eqref{gammaqgResHELLv2} applied to this function reproduce the
same recursion pattern as in eq.~\eqref{eq:recursion[]}, namely
\beq
\[\partial_\nu^k U_{\rm ABF}(N,Q^2e^\nu,Q^2)\]_{\nu=0} =
\[\gamma_+^k(\as(Q^2),N)\].
\eeq
The dependence on $\log(q^2/\mu_0^2)$ of eq.~\eqref{gamPABF} resembles
that of a function which depends linearly on $\as$ and with one-loop
running. In this case, the function $r$ would simply be $r(\as,N) =
\as\beta_0$, which is what we already discussed.  Since in this approximation
the anomalous dimension has the right value but the wrong slope at
$q^2=\mu_0^2$, ABF suggest a possibly improved version that reproduces the
correct derivative.
This is achieved by choosing $r(\as,N) = \as^2\beta_0
\frac{d}{d\as}\log\gamma_+(\as,N)$. This variant, which has been used to obtain
the default resummed predictions in previous \HELL works, is clearly more
complicated as it requires the $\as$-derivative of the resummed anomalous
dimension.  We shall come back to this point in sect.~\ref{sec:HELL-as}.

Another interesting result of ref.~\cite{Bonvini:2016wki}
is that, if a function (or a distribution) ${\cal P}_{qg}(\xi)$ exists
such that
\beq\label{eq:hqg_ktfact}
h_{qg}(\gam) = \gam \int_0^\infty d\xi\, \xi^{\gam-1} {\cal P}_{qg}(\xi),
\eeq
then it is possible to rewrite eq.~\eqref{gammaqgResHELLv2} as follows:
\beq\label{gammaqgResHELLv3}
\gamma_{qg}^{\NLLp} (\as(Q^2),N)
= \frac{\as(Q^2)}{3\pi}\,\TR\, \int_0^\infty d\xi\,
\frac{d}{d\xi}U(N,Q^2\xi,Q^2)\, {\cal P}_{qg}(\xi).
\eeq
This expression reminds one of the $k_t$ factorization of
ref.~\cite{Catani:1994sq}\footnote {In particular, it reproduces the $k_t$
factorization expressions of ref.~\cite{Catani:1994sq} in the LL
fixed-coupling limit, where the evolution function is replaced by its
fixed-coupling expression eq.~\eqref{eq:Ufc} and the anomalous dimension
$\gamma_+$ by $\gamma_s$.}, and indeed a form of this kind has been used
extensively for the resummation of coefficient functions in the \HELL
literature.  The actual function ${\cal P}_{qg}(\xi)$ has never been introduced
before in the literature because it is impossible to obtain it without the
full knowledge of the function $h_{qg}(\gamz)$.

\subsection{Synopsis of new analytical results\label{sec:res}}

Before closing this section, we briefly summarize the main new analytical
results of this paper for the reader's convenience. These results are detailed
in sects.~\ref{sec:Pqg} and~\ref{sec:Gqgres} and in
appendices~\ref{sec:gammagg} and~\ref{sec:appqg}, and here we limit ourselves to
refer to the relevant equations. They constitute the basic ingredients
through which we revisit and improve several aspects of what has been
outlined in sects.~\ref{sec:gengampp} and~\ref{sec:gengamqg}.

A first set of results give different expressions for $\gamma_s$, the pure
LL part of the $\gamma_+$ eigenvalue, introduced in eq.~\eqref{dualLO}.
Firstly, in eq.~\eqref{hgggex2}, we write $\gamma_s(a)$ as a perturbative
expansion in its argument $a$, with coefficients in closed analytical form;
its inverse relationship, namely the argument $a$ written as a series in
$\gamma_s$, is presented in eq.~\eqref{atom}, which thus reports the
exact expressions for the terms collectively denoted by $\Ord(\gamma_s^2)$
in eq.~\eqref{dualLOaN}.
In eq.~\eqref{fk0LNLres}, we provide a closed-form non-integral implicit
expression for $\gamma_s$. In eq.~\eqref{hgggex4} and eq.~\eqref{hgggex10},
$\gamma_s$ is written through equivalent integral representations featuring
both a Borel and a Hankel integral. It is possible to perform the Borel
integrals analytically, and in this way one obtains pure-Hankel integral
representations, as in eqs.~\eqref{hgggex4b} and~\eqref{hgggex10c}.
Finally, in  eq.~\eqref{FPggg}, $\gamma_s(a)$ is expressed as a fixed point
of a function, parametrically dependent on $a$, directly related to the
LO BFKL kernel.

Concerning the resummation of the anomalous dimension $\gamma_{qg}$, the main
result of the paper presented in eq.~\eqref{hqgres10} is an all-order
closed-form expression for the function $h_{qg}$ introduced in
eq.~\eqref{hqgdef}.
A closed-form non-recursive expression for the coefficients $h_k$ of the
expansion of $h_{qg}$ in terms of $\gamma$, eq.~\eqref{eq:hqgser}, is provided
in eq.~\eqref{hncoeff}.
It is also possible to find expressions for the coefficients of the series when
$h_{qg}$ is expanded in terms of $\bbas$: two alternative analytical expressions
for such coefficients are given in eq.~\eqref{tkcoeff} and
eq.~\eqref{Gqgresm02}, with the former more compact than the latter.

Further ancillary results related to the resummation of the anomalous dimension
$\gamma_{qg}$ are presented in appendix~\ref{sec:appqg}: in particular, they
concern closed-form expressions for the perturbative expansions of some objects
introduced in sect.~\ref{sec:Pqg}.
The coefficients of the expansion of the bare quark Green function
$\G_{qg}^{(0)}$ (introduced in eq.~\eqref{Gqgfact}) in terms of
$\bbas/\ep$, see eq.~\eqref{Gzqgdef}, are given in
eqs.~\eqref{rho0res}--\eqref{rhokres} or alternatively in eq.~\eqref{rhokres2}.

Finally, we present new results for the finite Green function $\G_{qg}$ and
its $gg$ counterpart, $\G_{gg}$.  In eq.~\eqref{Fzfunres2}, we provide a
closed and all-order form for the $\ep=0$ part of $\G_{qg}$.
Eq.~\eqref{Fzfunres4} gives a relationship between the $\ord(\ep^0)$
contributions to $\G_{qg}$ and $\G_{gg}$.
Finally, the analytical non-recursive coefficients of the expansion of
$\G_{qg}$ and $\G_{gg}$ in terms of $\gamma_s$ are reported in
eqs.~\eqref{rncoeff} and~\eqref{sncoeff}, respectively.

As a concluding comment for this section, we point out that many of the
all-order results we have just introduced stem from a methodology which entails
some guesswork; typically, by computing exactly the first few (up to twenty,
in some cases) elements of a sequence, and by using them to establish
a parametric pattern, eventually employed to resum the series formed by the
elements of that sequence. Therefore, while in number theory such results
would be referred to as ``conjectures'', as physicists we feel confident
that this method has allowed us to find the true all-order functions
we needed for our small-$x$ studies; in all cases, this has meant
comparing the re-expansion of the all-order results with as many
perturbative coefficients as we could evaluate, whose numbers were always
larger than those upon which the original pattern-extrapolations
had been based.

\section{All-order resummation of $P_{qg}$}
\label{sec:Pqg}

The goal of this section is to obtain a resummed expression for the $qg$
splitting function $P_{qg}$.  Specifically, in sect.~\ref{sec:hqgAO} we first
determine the function $h_{qg}$ of eq.~\eqref{hqgdef} to all orders.
In sect.~\ref{sec:hqgPert} we then provide closed-form expressions for the
coefficients of the power series expansion of $h_{qg}$ in terms of
both $\gamma_s$ and $\bbas$,
and we compare our results with the coefficients
previously known in the literature.  Finally, in
sect.~\ref{sec:HELLimplementation} we discuss how to use the new $h_{qg}$
result to obtain a complete resummation of $P_{qg}$ that includes subleading
running-coupling effects.

\subsection{All-order result in conjugate Mellin space}
\label{sec:hqgAO}

The way to determine the function $h_{qg}$ is through the collinear
factorization of the quark Green function in the high-energy
limit~\cite{Catani:1994sq}. The bare (i.e.~divergent in four dimensions)
quark Green function $\G_{qg}^{(0)}$ can be factorized as follows,
\beq\label{Gqgfact}
\G_{qg}^{(0)} = \G_{qg}\,\Gamma_{gg} + \Gamma_{qg},
\eeq
where $\G_{qg}$ represent the $\ep\to 0$ finite (factorized) Green function,
and $\Gamma_{gg}$ and $\Gamma_{qg}$ are the transition functions (or collinear
counterterms) in the high-energy limit.
Eq.~\eqref{Gqgfact} is valid in $d=4+2\ep$ dimensions, and the functions in
eq.~\eqref{Gqgfact} depend on $N$, $\as$ and $\ep$.
As we are working with $d$-dimensional quantities, we find it convenient to
extend the definition of $\bbas$ in eq.~\eqref{bbasdef} to\footnote{In case
of $4$-dimensional quantities, eq.~\eqref{abarnorm2X} coincides with
eq.~\eqref{bbasdef} since $S_\ep = 1$ for $\ep = 0$.}
\beq
\label{abarnorm2X}
\bbas\equiv\frac{\CA\as S_\ep}{\pi N},
\qquad
S_\ep=\(\frac{e^{\gE}}{4\pi}\)^\ep.
\eeq
For the purpose of the algebraic manipulations that we shall perform,
we factor out the term $\frac{\TR N}{3\CA}$ from $\gamma_{qg}$,
the Green functions and the $qg$ transition function\footnote{In order to
avoid any possible misunderstanding, we have explicitly indicated on the
r.h.s.~of eqs.~\eqref{Gqg0hatdef}-\eqref{Gamqgdef}, the dependence of the
functions which appear there on two different arguments in order to make
explicit the fact that the dependence upon $\ep$ is not entirely captured
by $\bbas$. In the following this dependence will often be understood.}
\begin{subequations}
\begin{align}
\left.\gamma_{qg}\right|_{\NLL} &=
\frac{\TR N}{3\CA} \hat\gamma_{qg}(\bbas),
\label{hatgamqgdef}\\
\left.\G_{qg}^{(0)}\right|_{\NLL} &=
\frac{\TR N}{3\CA} \hat\G_{qg}^{(0)} (\bbas,\ep),
\label{Gqg0hatdef}\\
\left.\G_{qg}\right|_{\NLL} &=
\frac{\TR N}{3\CA} \hat\G_{qg}(\bbas,\ep),
\\
\left.\Gamma_{qg}\right|_{\NLL} &=
\frac{\TR N}{3\CA} \hat\Gamma_{qg}(\bbas,\ep),
\label{Gamqgdef}
\end{align}
\end{subequations}
so that the functions with a hat depend on $\as$ and $N$ only through $\bbas$.
Note that $\gamma_{qg}$ lives in 4 dimensions, whereas the functions
$\G_{qg}^{(0)}$, $\G_{qg}$ and $\Gamma_{qg}$ live in $d$ dimensions, and so
they also depend on $\ep$, both explicitly and implicitly (through $\bbas$,
per eq.~\eqref{abarnorm2X}). We also observe that, according to
eq.~\eqref{hatgamqgdef} and eq.~\eqref{hqgdef},
the functions $\hat\gamma_{qg}$ and $h_{qg}$ are related by
\beq\label{eq:gammaqghat_hqg}
\hat\gamma_{qg}(\bbas) = \bbas\, h_{qg}(\gamma_s(\bbas)).
\eeq
With these definitions, we can rewrite eq.~\eqref{Gqgfact} as
\beq\label{Gqgfact2}
\hat\G_{qg}^{(0)} = \hat\G_{qg}\,\Gamma_{gg} + \hat\Gamma_{qg},
\eeq
with\footnote{Note that there is a typo in the definition of the
function $\Gamma_{qg}$ in ref.~\cite{Catani:1994sq}: in equation (4.13)
of that paper, the first argument of $\Gamma_{gg,N}$ should read
$\alpha/S_\ep$ rather than $\alpha$.}
\begin{align}
\hat\Gamma_{qg}(\bbas,\ep)&=
\frac{1}{\ep} \int_0^{\bbas}\frac{da}{a}\,
\hat\gamma_{qg}(a)\,\left.\Gamma_{gg}(a,\ep)\right|_{\LL}
\nonumber\\*
&= \frac{1}{\ep} \int_0^{\bbas}da\, h_{qg}(\gamma_s(a))\,
\left.\Gamma_{gg}(a,\ep)\right|_{\LL},
\label{Gqgdef}
\\*
\left.\Gamma_{gg}(\bbas,\ep)\right|_{\LL} &=
\exp\left[\frac{1}{\ep}\int_0^{\bbas}\frac{da}{a}\,\gamma_s(a)\right],
\label{Gggdef}
\end{align}
For later convenience, we introduce the function
\begin{align}
\hh\(z\) &\equiv \int_0^{1/\chi(z)}\frac{da}{a}\,\gamma_s(a),
\label{hhdef}
\end{align}
where $\chi(z)$ is the LO BFKL kernel, eq.~\eqref{eq:chidef}.
We observe that when $\hh$ is computed in $\gamma_s(\bbas)$
we can use eq.~\eqref{dualLO} to rewrite the upper limit of the
integral as $\bbas$, thus leading to
\beq
\hh\(\gamma_s(\bbas)\) = \int_0^{\bbas}\frac{da}{a}\,\gamma_s(a).
\eeq
This observation allows us to rewrite eq.~\eqref{Gggdef} as
\beq
\left.\Gamma_{gg}(\bbas,\ep)\right|_{\LL}=
\exp\left[\frac{1}{\ep}\,\hh\(\gamma_s(\bbas)\)\right].
\eeq
We stress that we have made the choice of defining $\hh$ such that it depends
on $\bbas$ through the function $\gamma_s(\bbas)$\footnote {This is always
possible, as we commented already in sect.~\ref{sec:gengampp}, thanks to
eq.~\eqref{dualLOaN}.}  for a good reason.  Indeed, this choice will play a
crucial role in the following, as working in terms of $\gamma_s$ instead of
$\bbas$ simplifies analytical derivations.  Moreover, this function will
also appear in our all-order results for $h_{qg}$ and $\hat\G_{qg}$.  Further
results on the function $\hh(z)$, and in particular explicit expressions
depending only on the BFKL kernel $\chi$, are collected in
appendix~\ref{sec:hhexpr}.

In eq.~\eqref{Gqgfact2} both the factorized Green function $\hat\G_{qg}$ and
the anomalous dimension $\hat\gamma_{qg}$ appearing implicitly through
eq.~\eqref{Gqgdef} are unknowns.  The way to extract both functions from a
single equation relies on the fact that $\hat\G_{qg}$ is finite in the
limit $\ep\to0$.  Indeed, we can write eq.~\eqref{Gqgfact2} in the form
\beq\label{masterqg}
\hat\G_{qg} = \[\hat\G_{qg}^{(0)} - \hat\Gamma_{qg}\]\Gamma_{gg}^{-1},
\eeq
so that it is clear that the constraint that the left-hand side be finite in
the $\ep\to0$ limit imposes the cancellation of the $\ep$ poles on the
right-hand side, allowing an order-by-order extraction of $\hat\gamma_{qg}$
and thus of $\hat\G_{qg}$\footnote{In particular, at each order in $\as$ the
largest power of the $\ep$ pole determines the corresponding order of
$\hat\gamma_{qg}$, and the lower powers provide consistency conditions with
lower orders.}.  A fundamental ingredient in this computation is
$\hat\G_{qg}^{(0)}$, whose form is straightforward but extremely
involved~\cite{Catani:1994sq}.
Here in the main text we therefore assume that this function is
known, and proceed to computing the quantities of primary interest. The
details of the computation of $\hat\G_{qg}^{(0)}$, and its results, are
reported in appendix~\ref{sec:G0qg}.

We now turn to the determination of $h_{qg}$ to all orders.
We begin by rewriting eq.~\eqref{masterqg} as
\beq
\hat\G_{qg}\,\Gamma_{gg} = F(\hh)\,\Gamma_{gg}-\hat\Gamma_{qg},
\qquad
F(\hh)\equiv \hat\G_{qg}^{(0)}\,\Gamma_{gg}^{-1},
\label{masterqg2}
\eeq
where we have introduced the quantity $F$, which we assume to
be a function of $\hh$, eq.~\eqref{hhdef}. This assumption is
actually innocuous, since it is always possible to trade a dependence on
$\bbas$ for a dependence on $\hh$: one first writes $\bbas$ as a function
of $\gamma_s$ by exploiting eq.~\eqref{dualLOaN}, and then one writes
$\gamma_s$ as a function of $\hh$ by exploiting eq.~\eqref{gamvshh2XX}.
More generally, regardless of the details of the specific linear
transformation, any perturbative series in $\bbas$ can equivalently be given
as a series in either $\gamma_s$ or (importantly) $\hh$. Thus, the dependence
of $F$ upon $\hh$ is not meant to be exclusive of other possible dependences
(e.g., $\ep$), which are understood, but it merely indicates that one
assumes $\hh$ to be an appropriate expansion parameter.

By deriving both sides of eq.~(\ref{masterqg2}) with respect to $\bbas$
and by solving for $h_{qg}$ the resulting expression, one obtains
\beq
h_{qg}(\gamma_s) = \gamma_s\chi(\gamma_s)\left(\ep\,\frac{dF}{d\hh}+F(\hh)\right)-
\left(\ep\,\frac{d\hat\G_{qg}}{d\bbas}+\gamma_s\chi(\gamma_s) \hat\G_{qg}\right),
\label{hqgmasterqg2}
\eeq
where we have used eq.~\eqref{dualLOaN} and
\beq \label{dhhda}
\frac{d\hh}{d\bbas}=\gamma_s\, \chi(\gamma_s)\,.
\eeq
Note that the right-hand-sides of eq.~\eqref{hqgmasterqg2} and eq.~\eqref{dhhda}
are meant to be evaluated at
$\gamma_s = \gamma_s(\bbas)$ and $\hh = \hh(\gamma_s(\bbas))$.
In order to proceed, we note that owing to eqs.~(\ref{Gqgdef}),
(\ref{Gggdef}), and to the general characteristics of the coefficients
of $\hat\G_{qg}^{(0)}$ (whose details are reported in appendix~\ref{sec:G0qg}),
the structure of eq.~(\ref{masterqg2}), as far as the $\ep$ dependence
is concerned, is the following,
\beq
\hat\G_{qg}
=
F(\hh)-\hat\Gamma_{qg}\Gamma_{gg}^{-1}
\qquad\Longleftrightarrow\qquad
\sum_{k=0}^\infty A_k\ep^k=
\sum_{k=-\infty}^\infty B_k\ep^k-
\sum_{k=-\infty}^{-1} C_k\ep^k\,,
\label{G0master}
\eeq
where $A_k$, $B_k$, and $C_k$ are $\ep$-independent parameters
(except for the implicit dependence through $\bbas$)\footnote{The fact
that the $S_\ep$ dependence is implicit through $\bbas$ is crucial to
simplify the expressions. Indeed, if we had expanded also that dependence,
the $\hat\Gamma_{qg}\Gamma_{gg}^{-1}$ would produce also positive powers
of $\ep$, hampering the separation between the various terms.},
whose precise definition is unimportant here. Therefore,
\begin{align}
B_k&=C_k\,, \qquad k<0\,,
\label{BkCkneg}
\\
B_k&=A_k\,, \qquad k\ge 0\,.
\label{BkCkpos}
\end{align}
This immediately shows that the $C_k$ coefficients, containing the
information on $\hat\gamma_{qg}$, are fully determined by the singular
part of $F$, while they are completely independent of $\hat\G_{qg}$,
which is in turn determined by the non-singular part of $F$.
Bearing this in mind, we separate the $\ep$ dependence of $F(\hh)$
into three contributions, respectively divergent, finite, and vanishing
in the $\ep\to 0$ limit, as follows:
\beq
F=F^{\rm div}+F^0+F^{\rm reg}\equiv
\sum_{k=-\infty}^{-1} B_k\ep^k+B_0+\sum_{k=1}^\infty B_k\ep^k\,.
\label{Fhsplit}
\eeq
Since $h_{qg}$ is IR-finite, and we are interested in its form in
four dimensions, eqs.~(\ref{hqgmasterqg2}) and~(\ref{Fhsplit}) imply
\begin{align}
  h_{qg}(\gamma_s)
  &= \gamma_s\chi(\gamma_s)
    \left(\ep\,\frac{dF^{\rm div}}{d\hh}+F^{\rm div}(\hh)
    +
    B_0-A_0\right)
    +\Ord(\ep)
    \nonumber
  \\
  &= \gamma_s\chi(\gamma_s)
    \left(\ep\,\frac{dF^{\rm div}}{d\hh}+F^{\rm div}(\hh)\right)
    +\Ord(\ep),
\label{hqgmasterqg3}
\end{align}
which is consistent with the known fact that $\hat\gamma_{qg}$ can be obtained
by only considering the divergent part of the eq.~(\ref{masterqg}).
Crucially in eq.~\eqref{hqgmasterqg3} the explicit dependence on $\bbas$
has disappeared, with $h_{qg}$ depending only on $\gamma_s$, possibly
through $\hh$.

With an explicit computation which uses eqs.~(\ref{Gqgzex}), (\ref{rhokidef}),
and~(\ref{rho0res})--(\ref{rhokres}) (or~(\ref{rhokres2})) we can infer
the all-order behaviour of the coefficients in the expansion of
$F^{\rm div}$ in $\hh$, which reads thus:
\begin{align}
F^{\rm div}(\hh)&=
\frac{3}{4}\sum_{k=1}^\infty\sum_{j=k}^\infty
\frac{(-)^{k-1}}{j!}2^{j-k}\frac{\hh^j}{\ep^k}+
\frac{1}{4}\sum_{k=1}^\infty\sum_{j=k}^\infty
\frac{(-)^{k-1}}{j!}\frac{2^{j-k}}{3^{j-k}}\frac{\hh^j}{\ep^k}
\label{Fdivtmp}
\\* &=
\sum_{k=1}^\infty\frac{(-)^{k+1}3}{2^{k+2}}\frac{e^{2\hh}}{\ep^k}+
\sum_{k=1}^\infty\sum_{j=0}^{k-1}
\frac{(-)^{k}3}{2^{k+1}}\frac{2^{j-1}\hh^j}{j!\ep^k}
\nonumber\\* &
+\sum_{k=1}^\infty\frac{(-)^{k+1}3^k}{2^{k+2}}\frac{e^{2/3\hh}}{\ep^k}
+\sum_{k=1}^\infty\sum_{j=0}^{k-1}\frac{(-)^k 3^k}{2^{k+1}}\frac{2^{j-1}\hh^j}{3^jj!\ep^k}\,.
\end{align}
By summing the series we finally obtain
\beq
F^{\rm div}(\hh) = \frac{3e^{2\hh}}{4(1+2\ep)}-
\frac{3e^{-\hh/\ep}}{4(1+2\ep)}+
\frac{3e^{\frac23\hh}}{4(3+2\ep)}-
\frac{3e^{-\hh/\ep}}{4(3+2\ep)}\,,
\label{Fdivres}
\eeq
whence
\beq\label{dFdiv}
\ep\,\frac{dF^{\rm div}}{d\hh}+F^{\rm div}(\hh)=
\frac{1}{4}\left(3e^{2\hh}+e^{\frac23\hh}\right).
\eeq
We note that, as a function of $\hh$, $F^{\rm div}$ turns out to only have
rational coefficients, which is remarkable given that trascendental
coefficients proportional to odd Riemann $\zeta$ constants appear when expanding
in $\bbas$. Moreover, we observe that eq.~\eqref{dFdiv} is independent of
$\ep$ and finite, despite the fact that the function $F^{\rm div}$ contains
only divergent contributions.  By plugging eq.~(\ref{dFdiv}) into
eq.~(\ref{hqgmasterqg3}), we obtain the sought result in four dimensions,
\beq
h_{qg}(\gamz) = \frac{\gamz\,\chi(\gamz)}{4}\left(3e^{2\hh(\gamz)}+
e^{\frac{2}{3}\hh(\gamz)}\right)\,,
\label{hqgres10}
\eeq
whence, using eqs.~\eqref{hatgamqgdef}, \eqref{eq:gammaqghat_hqg}, and
\eqref{dualLOaN},
\beq\label{eq:gammaqgallorder}
\left.\gamma_{qg}(\as,N)\right|_{\NLL}= \frac{\TR N}{3\CA}
\frac{\gamma_s(\bbas)}{4}\left(3e^{2\hh(\gamma_s(\bbas))}+
e^{\frac{2}{3}\hh(\gamma_s(\bbas))}\right)\,.
\eeq
Eq.~\eqref{hqgres10}, or equivalently eq.~\eqref{eq:gammaqgallorder},
is our main new result.
Eq.~\eqref{hqgres10} is explicitly written as a function of a generic
$\gamz$ to stress the fact that it is valid for any (complex) value of $\gamz$
and not just for $\gamz=\gamma_s(\bbas)$.
We note that by replacing the functions with their expansions to lowest order,
$\gamz\,\chi(\gamz)=1+\Ord(\gamz)$ and $\hh(\gamz)=\gamz+\Ord(\gamz^2)$,
we reproduce the result of ref.~\cite{Catani:1994sq} for the rational
coefficients of $\hat\gamma_{qg}$\footnote{The rational part of the
coefficients of the expansion of $\hat\gamma_{qg}$ are the same of those of
$h_{qg}$, because the relation between the two is given by the relation between
$\bbas$ and $\gamma_s$, which differs from an identity only by terms
proportional to (odd) $\zeta$ constants --- see eq.~\eqref{dualLOaN} and
eq.~\eqref{eq:chidef}.}.

As we can see, the all-order expression for $h_{qg}(\gamma)$ depends on the
function $\hh(\gamma)$.  Since its definition, eq.~\eqref{hhdef}, is based on
an integral with complex variables, its evaluation is not straightforward.
Unfortunately, we have not been able to compute the integral in terms of
elementary functions.  Nevertheless, in appendix~\ref{sec:hhexpr} we have been
able to manipulate it such that its form is easier to handle numerically,
e.g.\ by expressing it in terms of the BFKL kernel $\chi$ without involvement
of the function $\gamma_s$, and by recasting all integrals in the real domain.

\subsection{Perturbative coefficients}
\label{sec:hqgPert}

Thanks to eq.~(\ref{hqgres10}), the computation of the coefficients of the
expansion of $h_{qg}$ in terms of $\gamz$ is straightforward. We rewrite
that equation as follows
\beq
h_{qg}(\gamz)=
\frac{1}{4}\left[3\exp\left(2\tilh_{\frac{1}{2}}(\gamz)\right)+
\exp\left(\frac{2}{3}\tilh_{\frac{3}{2}}(\gamz)\right)\right],
\label{hqgres1}
\eeq
where, from eq.~\eqref{eq:hhv2},
\begin{align}
  \tilh_b(\gamz)
  &\equiv \hh(\gamz) +b\log(\gamz\chi(\gamz)) \nonumber\\
  &= \gamz-(\gamz-b)\log\left(1+\hchi(\gamz)\right) +\int_0^1 dy\,\log\left(1+\hchi(y\gamz)\right)\,,
\label{tilhdef}
\end{align}
with $\hchi$ related to the BFKL LO kernel, see eq.~\eqref{Gfun} and
eq.~\eqref{FPggg0}. Then, by using the expression of the generating
function of the complete exponential Bell polynomials $B_n$, we obtain
\beq
\exp\left(b\tilh_\frac{1}{b}(\gamz)\right)=
\sum_{n=0}^\infty\frac{\gamz^n}{n!}B_n\left(\widehat{Z}_b\right)\,,
\label{exhtb}
\eeq
where the $k^{\rm th}$ element of the vector $\widehat{Z}_b$ is
\beq
\widehat{Z}_b^{[k]}
=
b\Bigg(\delta_{k,1}+
\frac{1}{b}\sum_{j=1}^k(-)^{j+1}(j-1)!B_{k,j}(Z)
-(k-1)\sum_{j=1}^{k-1}(-)^{j+1}(j-1)!B_{k-1,j}(Z)\Bigg),
\label{hZvecdef}
\eeq
with $Z$ being the vector defined in eq.~\eqref{Zvec}
and by $B_{k,j}$ we denote the incomplete exponential Bell polynomials.
By re-expanding eq.~\eqref{hqgres1} in powers of $\gamz$ through
eq.~\eqref{exhtb}, and by equating the resulting expression to that
of a generic perturbative expansion of $h_{qg}(\gamz)$ as in
eq.~\eqref{eq:hqgser},
one obtains a closed-form, non-recursive expression for the coefficients
$h_n$. Explicitly, that is:
\beq
h_n=\frac{1}{n!}\left[\frac{3}{4}\,B_n\left(\widehat{Z}_2\right)
+\frac{1}{4}\,B_n\left(\widehat{Z}_{\frac{2}{3}}\right)\right].
\label{hncoeff}
\eeq
For convenience, we report the explicit form of the first ten coefficients
which emerge from eq.~\eqref{hncoeff}
\begin{subequations}
\begin{align}
  h_0 &= 1,\\
  h_1 &= \frac53,\\
  h_2 &= \frac{14}9,\\
  h_3 &= \frac{82}{81}+2\zeta_3,\\
  h_4 &= \frac{122}{243}+\frac{5}6\zeta_3,\\
  h_5 &= \frac{146}{729}-\frac{14}9\zeta_3+2\zeta_5,\\
  h_6 &= \frac{2188}{32805}-\frac{205}{81}\zeta_3+\frac{5}9\zeta_5,\\
  h_7 &= \frac{13124}{688905}-\frac{488}{243}\zeta_3-\frac{15}{7}\zeta_3^2-\frac{56}{27}\zeta_5+2\zeta_7,\\
  h_8 &= \frac{1406}{295245}-\frac{803}{729}\zeta_3-\frac1{2}\zeta_3^2-\frac{82}{27}\zeta_5+\frac{5}{12}\zeta_7,\\
  h_9 &= \frac{11810}{11160261}-\frac{15316}{32805}\zeta_3+\frac{41}{14}\zeta_3^2-\frac{1708}{729}\zeta_5-\frac{125}{27}\zeta_3\zeta_5-\frac{7}3\zeta_7+2\zeta_9.
\end{align}
\end{subequations}
With the result of eq.~(\ref{hncoeff}), we can readily obtain its analogue
for the expansion of $\hat\gamma_{qg}$ itself, which reads
(see appendix~\ref{sec:exprtn2} for an explicit derivation)
\begin{align}
  \hat\gamma_{qg}(\bbas)
  &= \frac{\gamma_s(\bbas)}{4}\left(3e^{2\hh(\gamma_s(\bbas))}+e^{\frac{2}{3}\hh(\gamma_s(\bbas))}\right) \label{gqgvsggg2fun}\\*
  &=\sum_{k=0}^\infty t_k \bbas^{k+1}\,,
\label{gqgvsggg}
\\
t_k&=\frac{1}{k!}\sum_{n=0}^k h_n n!B_{k,n}\big(\hat{G}\big)
=\frac{1}{k!}\sum_{n=0}^k \left(\frac{3}{4}\,B_n\left(\widehat{Z}_2\right)
+\frac{1}{4}\,B_n\left(\widehat{Z}_{\frac{2}{3}}\right)\right)
B_{k,n}\big(\hat{G}\big)\,,
\label{tkcoeff}
\end{align}
where $\hat G$ is defined in eq.~\eqref{hGvec}.
The first ten coefficients of $\gamma_{qg}$ stemming from
eq.~(\ref{tkcoeff}) read thus:
\begin{subequations}
\begin{align}
  t_0 &= 1,\\
  t_1 &= \frac53,\\
  t_2 &= \frac{14}9,\\
  t_3 &= \frac{82}{81}+2\zeta_3,\\
  t_4 &= \frac{122}{243}+\frac{25}6\zeta_3,\\
  t_5 &= \frac{146}{729}+\frac{14}3\zeta_3+2\zeta_5,\\
  t_6 &= \frac{2188}{32805}+\frac{287}{81}\zeta_3+12\zeta_3^2+\frac{35}9\zeta_5,\\
  t_7 &= \frac{13124}{688905}+\frac{488}{243}\zeta_3+\frac{515}{21}\zeta_3^2+\frac{112}{27}\zeta_5+2\zeta_7,\\
  t_8 &= \frac{1406}{295245}+\frac{73}{81}\zeta_3+\frac{55}{2}\zeta_3^2+\frac{82}{27}\zeta_5+32\zeta_3\zeta_5+\frac{15}4\zeta_7,\\
  t_9 &= \frac{11810}{11160261}+\frac{2188}{6561}\zeta_3+\frac{2665}{126}\zeta_3^2+96\zeta_3^3+\frac{1220}{729}\zeta_5+\frac{1675}{27}\zeta_3\zeta_5+\frac{35}9\zeta_7+2\zeta_9.
\end{align}
\end{subequations}
The coefficients up to $t_5$ were already known from
ref.~\cite{Catani:1994sq}.  We have been able to obtain a large number of
coefficients (more than thirty) in an analytical way by solving directly
eq.~\eqref{masterqg} order by order, and they coincide with the expansion of
the closed form found above, thus providing us with a further confirmation
of our all-order result eq.~\eqref{hqgres10}, and of the self-consistency
of the procedures we have followed.

As a marginal comment, we have observed that if we try to expand the function
$\hat\gamma_{qg}$ in powers of $\gamma_s$ rather than $\bbas$ we obtain a
direct expression for the expansion coefficients that is simpler than the two
above. We have
\begin{align}
  \hat\gamma_{qg}(\bbas)
  &=\sum_{n=0}^\infty u_n \gamma_s^{n+1}(\bbas)\,,
\label{gqgvsggg2}
\\*
u_n&=\frac{1}{n!}\left[
\frac{3}{4}\,B_{n}\left(\widehat{Z}_2^\prime\right)
+\frac{1}{4}\,B_{n}\left(\widehat{Z}_{\frac{2}{3}}^\prime\right)\right],
\label{qncoeff}
\end{align}
with
\beqn
\widehat{Z}_b^{\prime[k]}&=&b\Bigg(\delta_{k,1}
-(k-1)\sum_{j=1}^{k-1}(-)^{j+1}(j-1)!B_{k-1,j}(Z)\Bigg).
\label{hZpvecdef}
\eeqn
This form, simpler with respect to that of the $h_n$ coefficients
of eq.~\eqref{hncoeff}, is likely due to the absence of the factor
$\chi(\gamma_s(\bbas))$ in eq.~\eqref{gqgvsggg2fun}, which is compensated by
the factor $\bbas$ relating $\hat\gamma_{qg}$ and $h_{qg}$, eq.~\eqref{hqgdef},
thanks to eq.~\eqref{dualLO}.  As these coefficients $u_n$ are not used
anywhere in the text or in the literature, we do not report any of them here.

As we mentioned above, in the context of ref.~\cite{Altarelli:2008aj} 36
coefficients have been extracted by solving eq.~\eqref{masterqg} order by
order numerically, due to the limited computing resources available at
that time that could not allow analytic manipulations up to that order.  It is
interesting to compare the results of such an extraction with the exact
coefficients. We do this at the level of the expansion of $h_{qg}$, which
is better behaved.
%%%%%%%%%%%%%%%%%%%%%%%%%%%%%%%%%%%%%%%%%%%%%%%%%%%%%%%%%%%%%%%%%%%%%%%
\begin{figure}[t]
  \centering
  \includegraphics[width=0.6\textwidth]{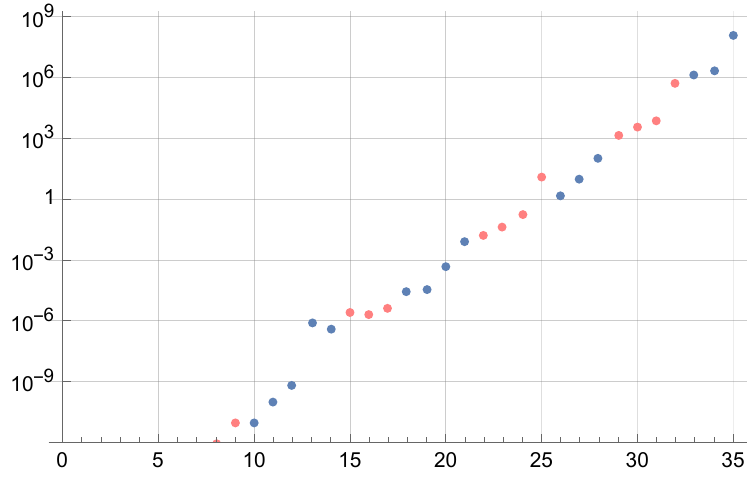}
  \caption{Plot of the relative difference $R_n$, eq.~\eqref{RABFcoeff},
    between exact and numerical $h_k$ coefficients in logarithmic scale as
    a function of $n$. The blue (red) points feature $h_n(\text{exact}) > 0$
    ($h_n(\text{exact}) < 0$).}
  \label{fig:ABFcoeff}
\end{figure}
%%%%%%%%%%%%%%%%%%%%%%%%%%%%%%%%%%%%%%%%%%%%%%%%%%%%%%%%%%%%%%%%%%%%%%%
In fig.~\ref{fig:ABFcoeff} we display the ratio of the coefficients $h_n$
as computed in ref.~\cite{Altarelli:2008aj} over the exact ones computed here,
by plotting the deviation from unity of that ratio, in absolute value and
with a logarithmic scale:
\beq\label{RABFcoeff}
R_n = \left| 1 - \frac{h_n(\text{ref.~\cite{Altarelli:2008aj}})}{h_n(\text{exact})} \right|.
\eeq
We observe that above $n\sim25$ the numerically computed coefficients
are completely wrong. This is due to large numerical cancellations among
the various terms contributing to these coefficients, cancellations which
are responsible for increasingly large errors --- while the size of these
errors is a function of the specific numerical accuracy employed in the
calculation, the fact that they increase with $n$ (the coefficient index)
is inherently due to the numerical, as opposed to analytical, approach.
Having said that, it is reassuring that the lowest-order numerical
coefficients are in good agreement with their analytical counterparts; in
particular, the agreement is better than permille for all coefficients
$h_n$ with $n\leq20$.
The approximate implementation of the resummation of $P_{qg}$
in all of the \HELL versions prior to that of this paper
is based on the first 16 coefficients, and is
therefore unaffected by the inaccuracies which plague large-$n$
coefficients\footnote{As was already explained in footnote~\ref{foot:16},
the choice of this number of coefficients  was based on numerical stability
checks upon variations of several parameters. At the time of the publication
of ref.~\cite{Bonvini:2016wki} it was not possible to know that some
coefficients were very different from the exact ones.}.
The actual accuracy of the old approximate implementation of resummation
will be studied in sect.~\ref{sec:results}.

\subsection{Resummation of subleading running coupling effects}
\label{sec:HELLimplementation}

Having computed the function $h_{qg}$ to all orders in sect.~\ref{sec:hqgAO},
we now need to use it to obtain a resummed expression for the splitting
function $P_{qg}$, which must include the resummation of subleading
running coupling effects as well.

As we discussed in sect.~\ref{sec:gengamqg}, the resummation of subleading
running coupling effects can be obtained in a number of alternative
(equivalent) ways.  The most powerful one is given by
eq.~\eqref{gammaqgResHELLv3}, as it allows one to achieve the resummation
through the computation of an integral, in a way that would be fully
equivalent to what is done in the \HELL framework for the resummation
of coefficient functions.  The splitting function would also be
straightforwardly obtained as follows
\beq\label{eq:PqgNLL-ktfact}
P_{qg}^{\NLLp} (\as(Q^2),x) = \frac{\as(Q^2)}{3\pi}\TR
\int_0^\infty d\xi\, \frac{d}{d\xi}U(x,Q^2\xi,Q^2)\, {\cal P}_{qg}(\xi),
\eeq
by means of the evolution function $U(x,\mu^2,Q^2)$ in momentum space,
which is simply given as the inverse Mellin transform of
eq.~\eqref{eq:Udef}\footnote{In \HELL, the evolution function in $x$ space
in the ABF approximation (the inverse Mellin transform of eq.~\eqref{eq:UABF})
is precomputed in grids of values of the arguments (as it is needed for
various applications) and readily available for its use in
eq.~\eqref{eq:PqgNLL-ktfact}}.
This formulation of resummation requires the generalised function
(i.e.~a distribution) ${\cal P}_{qg}(\xi)$, which is the inverse Mellin
transform of $h_{qg}(\gamma)/\gamma$ (see eq.~\eqref{eq:hqg_ktfact}).
Unfortunately, this Mellin inversion is highly nontrivial.
The function is too complicated to allow the analytical computation
of its inverse, and even numerically there are issues due to the asymptotic
behaviour of the function $\hh$, which makes the numerical convergence
very challenging\footnote{See in particular eq.~(\ref{sFsasyE1}). The
presence of exponentials of the exponential integral implies that the sought
inverse Mellin transform approaches zero (starting at about $\xi\simeq 0.9$)
in an incredibly fast way (approximately, by normalising its value to $1$
at $\xi=0.9$, it is equal to $10^{-40}$ at $\xi=0.995$). Unfortunately,
the full control of the $\xi\to 1$ region is essential, since the
inverse Mellin transform we are talking about is actually a generalised
plus distribution, with the function mentioned above as its kernel.}.
In addition to that, a numerical inversion alone is not able to
reproduce the nature of ${\cal P}_{qg}(\xi)$ as a distribution.
We have therefore decided to adopt an alternative approach.

Specifically, we consider the equivalent formulation of the resummation
in terms of the series expansion of $h_{gq}$, eq.~\eqref{gammaqgResHELLv2}.
Also in this case the splitting function can be obtained straightforwardly
by means of the evolution function in $x$ space,
\beq\label{PqgResHELLv2}
P_{qg}^{\NLLp} (\as(Q^2),x) = \frac{\as(Q^2)}{3\pi}\TR
\sum_{k=0}^\infty h_k \[\partial_\nu^k U(x,Q^2e^\nu,Q^2)\]_{\nu=0}.
\eeq
Such a formula depends directly on the coefficients $h_k$ of the expansion
of the function $h_{qg}$, eq.~\eqref{eq:hqgser}. We remind the reader that
the current formulation of the (approximate) resummation of the $qg$
anomalous dimension in \HELL is based on this formula, by using just a finite
number of coefficients and by extrapolating the others through a Borel-Pad\'e
procedure~\cite{Bonvini:2016wki}.  Here, instead, we manipulate the series
in order to be able to sum it exactly to the now-known all-order result of
eq.~\eqref{hqgres10}. In doing so, what we obtain is free of the spurious
ambiguities induced by the asymptotic nature of the series in
eq.~\eqref{PqgResHELLv2}.

\begin{figure}[t]
  \centering
  \includegraphics[width=0.3\textwidth]{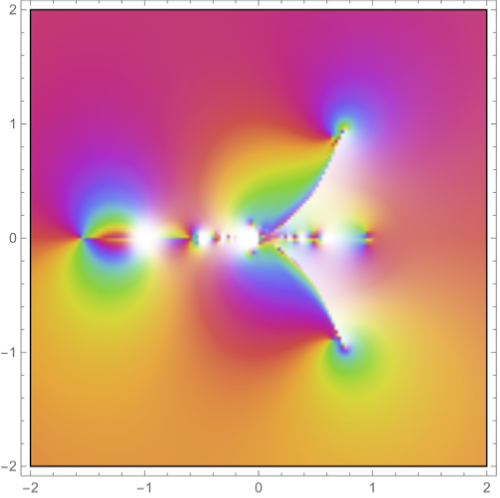}\qquad
  \includegraphics[width=0.31\textwidth]{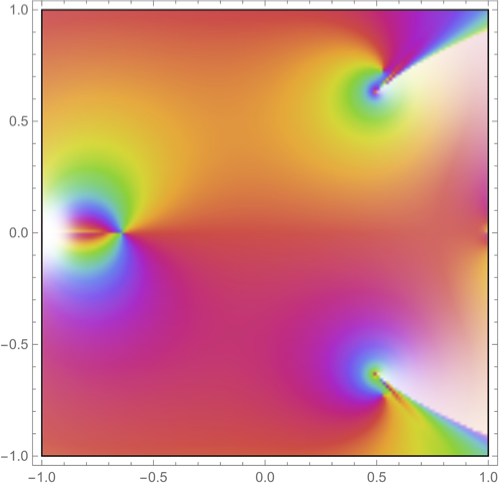}
  \caption{The function $h_{qg}(1/s)$ in the complex $s$ plane (left) and
    $h_{qg}(z)$ in the complex $z$ plane (right).
    The plot is obtained by means of the \texttt{ComplexPlot} function of
 \texttt{Mathematica}~\cite{Mathematica}, which uses different colours for the phase
 of the function, and varying intensity for its absolute value, so as to
 maximise visibility. Although both colours and intensity can in principle
 be controlled by the user,
 %(e.g.~through {\tt ColorFunction}, {\tt ColorFunctionScaling}, and {\tt Opacity}),
 we have refrained
 from doing so, since these plots (and those of similar nature which
 will follow) are meant to only convey qualitative information.}
  \label{fig:hqg1s}
\end{figure}

The first step consists in writing the derivatives that appear in
eq.~(\ref{PqgResHELLv2}) by using the Cauchy formula, and in expressing
the factorial there with the integral representation
of the $\Gamma$ function~\cite{Bonvini:2014joa},
\begin{align}
  f^{(k)}(0)
  &= \frac{k!}{2\pi i} \oint \frac{ds}s \, \(\frac 1s\)^k f(s)
    \nonumber\\
  &= \frac1{2\pi i} \int_0^\infty dw\, e^{-w} w^k \oint \frac{ds}s \, \(\frac 1s\)^k f(s)
    \nonumber\\
  &= \frac1{2\pi i} \int_0^\infty dw\, e^{-w} \oint \frac{ds}s \, \(\frac1s\)^k f(ws),
\end{align}
which is valid for any function $f$ that is analytic around the origin.
In the last step we have rescaled the variable $s$ by $w$, which is
safe as $w$ is positive. The integration contour in the variable $s$
is a closed counterclockwise one that must encircle the origin and avoid
other possible singularities of the function $f$ far from $s=0$.
The advantage of this expression is that it converts a $k^{\rm th}$ derivative
into a $k^{\rm th}$ power, at the price of introducing two integrals.
We can thus rewrite eq.~\eqref{PqgResHELLv2} as follows
\begin{align}\label{PqgResHELLv3}
  P_{qg}^{\NLLp} (\as(Q^2),x)
  &= \frac{\as(Q^2)}{3\pi}\TR
    \sum_{k=0}^\infty h_k \frac1{2\pi i} \int_0^\infty dw\, e^{-w} \oint \frac{ds}s \, U(x,Q^2e^{ws},Q^2) \(\frac1s\)^k \nonumber\\
  &= \frac{\as(Q^2)}{3\pi}\TR\,
    \frac1{2\pi i} \int_0^\infty dw\, e^{-w} \oint \frac{ds}s \, U(x,Q^2e^{ws},Q^2)\, h_{qg}\(\frac1s\) \nonumber\\
  &= \frac{\as(Q^2)}{3\pi}\TR\,
    \frac1{2\pi i} \int_0^\infty dw\, e^{-w} \oint \frac{dz}z \, U(x,Q^2e^{w/z},Q^2)\, h_{qg}(z),
\end{align}
where in the last step we have changed integration variable from
$s$ to $z=1/s$, as this will turn out to be convenient.
In Mellin space the result above simply reads
\begin{align}\label{gammaqgResHELLv4}
  \gamma_{qg}^{\NLLp} (\as(Q^2),N)
  &= \frac{\as(Q^2)}{3\pi}\TR\,
    \frac1{2\pi i} \int_0^\infty dw\, e^{-w} \oint \frac{dz}z \, U(N,Q^2e^{w/z},Q^2)\, h_{qg}(z).
\end{align}
Note that, after summing the series, the singularity structure in $s$ changes:
while there were only singularities in $s=0$ for each term of the sum,
after summing the series more singularities and cuts appear in the complex
$s$ plane. Owing to this new structure, the integration contour in the
variable $s$ must be suitably adapted. We observe that all singularities of
$h_{qg}(1/s)$ lie in a region around the origin, and get denser as $s=0$
is approached. Therefore, most of these singularities must lie within the
integration contour, lest the contour should cross a branch cut.
The simplest option is to encircle them all, which can be easily realized
by using for the contour a circle with a large radius --- a radius equal
to $2$ turns out to be sufficient, as one can see from fig.~\ref{fig:hqg1s}.
Equivalently, in the variable $z$ one can choose a circle with radius
smaller than $1/2$. We stress that inside such a circle in the complex
$z$ plane, the function $h_{qg}(z)$ has no singularities (see
fig.~\ref{fig:hqg1s}). This makes the formulation that employs the
last line of eq.~\eqref{PqgResHELLv3} convenient for further manipulations,
in which the only singularities that contribute to the $z$ integral are
those from the function $U$ and the explicit pole at $z=0$.
%%%%%%%%%%%%%%%%%%%%%%%%%%%%%%%%%%%%%%%%%%%%%%%%%%%%%%%%%%%%%%%%%%%%
%%%%%%%%%%%%%%%%%%%%%%%%%%%%%%%%%%%%%%%%%%%%%%%%%%%%%%%%%%%%%%%%%%%%%

In order to assess the goodness of eq.~\eqref{PqgResHELLv3} and of
the choice of the contour, we consider the fixed-coupling limit:
in this case, the evolution function is given by eq.~\eqref{eq:Ufc}.
By plugging eq.~\eqref{eq:Ufc} into eq.~\eqref{gammaqgResHELLv4} we obtain
\begin{align}
  \gamma_{qg}^{\NLLp} (\as,N)
  &\overset{\rm f.c.}= \frac{\as}{3\pi}\TR\,
    \frac1{2\pi i} \int_0^\infty dw\, e^{-w} \oint \frac{dz}z \, e^{w\gamma_+(\as,N)/z}\, h_{qg}(z) \nonumber\\
  &= \frac{\as}{3\pi}\TR\,
    \frac1{2\pi i} \oint \frac{dz}{z-\gamma_+(\as,N)}\, h_{qg}(z)  \label{gqgFCint2} \\
  &= \frac{\as}{3\pi}\TR\, h_{qg}\(\gamma_+(\as,N)\), \label{gqgFC}
\end{align}
which coincides, as is expected, with the result of eq.~\eqref{hqgdef}
when $\gamma_+(\as,N)\to\gamma_s(\bbas)$.
Here, in the first step we have computed the $w$ integral which converges
only if $\Re(1-\gamma_+/z)>0$, and extended the result analytically
elsewhere\footnote{A similar procedure is used in appendix~\ref{sec:gggIR}
for the manipulations of an integral representation of
$\gamma_s$.}. Then, we could use in the last step the fact that the
only singularity present inside the integration contour is the explicit $z$
pole. This is strictly speaking true only if $\abs{\gamma_+(\as,N)}$ is
sufficiently small to be inside the circle, but the result can be
analytically continued elsewhere if that is not the case.

Although this example is not a proof of our assumption on the
integration contour, it gives us confidence that our choice is reasonable.
In practice though, in order to implement eq.~\eqref{PqgResHELLv3},
\eqref{gammaqgResHELLv4} numerically, we need the evolution operator
$U$ with full access to complex values of its second argument.
As this is related to the scale at which $\as$ is computed, it is not
straightforward to treat it in a general way. In the following, we thus
consider an approximation of the evolution operator that allows us to obtain
numerical results, consistently with previous \HELL implementation.

Specifically, in order to avoid computing the strong coupling with
complex argument, we consider the ABF approximation of the evolution
operator that is customarily used in \HELL, eq.~\eqref{eq:UABF}.
Plugging it into eq.~\eqref{gammaqgResHELLv4} we obtain
\begin{align}\label{gqgABFint2}
  \gamma_{qg}^{\NLLp} (\as,N)
  &\overset{\rm ABF}= \frac{\as}{3\pi}\TR\,
  \frac1{2\pi i} \int_0^\infty dw\, e^{-w} \oint \frac{dz}z \, \(1+\frac{r (\as,N) w}z\)^{\frac{\gamma_+(\as,N)}{r (\as,N)}}\, h_{qg}(z) \\
  &= \frac{\as}{3\pi}\TR\,
    \frac1{2\pi i} \oint \frac{dz}z \, e^{\frac z{r(\as,N)}}\(\frac{r (\as,N)}z\)^{\frac{\gamma_+(\as,N)}{r (\as,N)}}\Gamma\(1+\frac{\gamma_+(\as,N)}{r (\as,N)},\frac z{r (\as,N)}\) h_{qg}(z)\nonumber
\end{align}
in terms of the incomplete $\Gamma$ function. In deriving this expression
we have performed an analytic continuation in $z$ of the $w$ integral thanks
to the fact that we have been able to compute the $w$ integral analytically.
In order to obtain the splitting function $P_{qg}$, the inverse Mellin of
this result must be computed. However, the integrand of the $z$ integral
computed along the inverse Mellin contour has a branch-cut for real
negative $z$ values. This means that the original choice of the $z$
integration contour, a circle around the origin, is not suitable anymore,
and it must be modified. The most obvious option is to ``stretch'' the
circle to enclose the cut, thereby leading to a Hankel-like contour along
the negative real $z$ axis.

It is useful to observe that we can simplify the expression of
eq.~\eqref{gqgABFint2} by noting that, for small $z$, we can expand
\begin{align}
  \Gamma\(1+\frac{\gamma_+}r,\frac zr\)
  &= \Gamma\(1+\frac{\gamma_+}r\) - \int_0^{\frac zr}dt\,e^{-t}t^{\frac{\gamma_+}r} \nonumber\\
  &= \Gamma\(1+\frac{\gamma_+}r\) - \Ord\(z^{\frac{\gamma_+}r+1}\).
\end{align}
Plugging this into eq.~\eqref{gqgABFint2} we obtain
\begin{align}\label{gqgABFint3}
  \gamma_{qg}^{\NLLp} (\as,N)
  &\overset{\rm ABF}= \frac{\as}{3\pi}\TR\, \frac1{2\pi i} \oint \frac{dz}z \,
  e^{\frac z{r(\as,N)}}\[\(\frac{r(\as,N)}z\)^{\frac{\gamma_+(\as,N)}{r(\as,N)}}\Gamma\(1+\frac{\gamma_+(\as,N)}{r(\as,N)}\) + \Ord(z)\] h_{qg}(z)  \nonumber\\
  &= \frac{\as}{3\pi}\TR\, \Gamma\(1+\frac{\gamma_+(\as,N)}{r(\as,N)}\) \frac1{2\pi i} \oint \frac{dz}z \, e^{\frac z{r(\as,N)}}\(\frac{r(\as,N)}z\)^{\frac{\gamma_+(\as,N)}{r(\as,N)}} h_{qg}(z),
\end{align}
where we have used the residue theorem to deduce that the $\Ord(z)$ term
has a null integral.  Eq.~\eqref{gqgABFint3} represents a significant
simplification of eq.~\eqref{gqgABFint2}.  In particular, the integration
along the cut on the real negarive $z$ axis converges well for
eq.~\eqref{gqgABFint3}.  We use this representation for the new numerical
implementation of the resummation of $P_{qg}$ in \HELL.  Additional details,
including the choice of the integration contours in $z$ and $N$, are given in
appendix~\ref{sec:appHELL-Pqg}.

\section{Resummed DGLAP evolution to large $\as$}
\label{sec:largeas}

Having obtained an all-order expression for $P_{qg}$, we report in this
section new numerical results for the singlet DGLAP evolution matrix,
which allow one to handle large values of $\as$. For this to be possible,
we had to improve some of the core steps of the \HELL code, which were
optimised in the past only for $\as\lesssim0.3$.
We thus first briefly report on these technical changes.

\subsection{Improvements to employ large $\as$ values in $\gamma_+$}
\label{sec:HELL-as}

The first part of the \HELL workflow consists in resumming the anomalous
dimension $\gamma_+$ along the Mellin inversion contour.  This procedure is
based on the ingredients discussed in sect.~\ref{sec:gengampp}.  One of the
steps involved is the resummation of collinear poles in the BFKL kernel,
obtained through the duality with DGLAP.  For this step, \HELL (from version
2.0 onwards) uses an approximate fixed-order anomalous dimension with the
correct small-$N$ structure but with a modified asymptotic behaviour at large
$N$ (see appendix~B.3 of ref.~\cite{Bonvini:2017ogt}). The reason for this is
twofold.  Firstly, it gives one the possibility of computing the
inverse function analytically, which leads to numerical performances
which are significantly more stable.  Secondly, when using it to resum
the collinear ($\gamma=0$) pole of the BFKL kernel, one does not hit the
subleading singularity at $\gamma=-1$ of such a kernel, and thus avoids a
problem in the subsequent matching of the resummed anomalous dimension
with the fixed order.

However, such approximate fixed-order anomalous dimension does avoid hitting
the subleading singularity at $\gamma=-1$ only for sufficiently small values
of the coupling constant, namely $\as\lesssim0.5$.  As we now want to employ
larger values of $\as$ this approximation will have to be improved.  We have
done this in a minimal way, which retains the advantages of the previous
formulation, but also allows us to push the procedure safely to larger $\as$
values.  Moreover, we have introduced our modification in a smooth way, so
that for small values of $\as$ the result is basically unchanged w.r.t.~to the
original approximation of ref.~\cite{Bonvini:2017ogt}.  Technical details of
this procedure are reported in appendix~\ref{sec:appHELL-as}.

An additional issue we have faced when employing large values of $\as$ is
related to the approximation used for implementing the resummation of the
subleading running-coupling effects discussed in sect.~\ref{sec:gengampp}.
We have noticed that with such an approximation the resummed
splitting function behaves at small $x$ in a
unexpected way; namely, it decreases, rather than increases, with decreasing
$x$'s. This behaviour is driven by a spurious singularity
generated by the aforementioned approximation, which turns out to have a
negative residue with large absolute value, and which thus dominates over the
leading singularity that has a positive but small residue.  We explain this
issue in greater detail in appendix~\ref{sec:appHELL-RC}.

In order to solve this problem, we propose a very simple modification of the
approximation introduced in ref.~\cite{Bonvini:2017ogt}, which is consistent
with the original idea, but adds suitable subleading terms that make the
spurious pole innocuous.  After this modification, which we also argue to lead
to a more natural approximation than the original one, the behaviours of the
resummed splitting functions are as expected, namely at small $x$ they grow
with decreasing $x$'s.  All of the details about these changes are collected in
appendix~\ref{sec:appHELL-RC}, and their impact is investigated in
sect.~\ref{sec:results-comparison}.

Finally, another issue which shows up in the computation of the resummed
$\gamma_+$ at large $\as$ is a change of branch in the solution of the duality
condition with the BFKL kernel.  Such a solution is implemented numerically
with a zero-finding algorithm in the complex $N$ plane, for values of $N$
along the Mellin inversion contour (we give some details on how this is
performed in practice in appendix~\ref{sec:appHELL-branch}).  The functions
involved are very complicated, with rich singularity structures, so it is very
natural to expect that when changing the value of $\as$ the branch selected by
the algorithm may change.  Luckily, this happens only at large $|\Im(N)|$
values, which impacts the Mellin inversion only at large $x$, leaving the
small-$x$ region, which is the one we are interested in, unaffected by this
problem.

In practice, however, having a jump from one branch to another is undesirable
because it creates a sudden change in the large-$x$ behaviour of the relevant
Mellin inverse (here, the splitting function $P_+$) when changing $\as$.  We
have therefore decided to adopt a numerical procedure\footnote{Note that, in
principle, it could be possible to avoid the branch change problem by
modifying the BFKL kernel with suitable subleading contributions. The problem
of this approach is that without a clear understanding of the analytic
structure of the kernel and of the origin of the branch cuts it is very
difficult to guess which kind of subleading contributions could fix the
problem.  Given the complexity of functions involved, we did not pursue this
approach.}  in which the large-$x$ behaviour is extrapolated from a lower-$x$
region where the change of branch does not produce any effect.  This procedure
destroys momentum conservation, which we reimpose manually.  In order to do
so, we find it convenient to project the splitting function $P_+$ on a basis
of Chebyshev polynomials of the first kind,
which we can use to compute the Mellin transform
analytically in terms of a small number of coefficients.  This approach, that
we describe in detail in appendix~\ref{sec:appHELL-branch}, works very well,
and it also allows us to speed up the \HELL code, thanks to the simpler access
to $\gamma_+$.

The only limitation of this procedure is related to the computation of the
derivative of $\gamma_+$ with respect to $\as$, needed for one of the
incarnations of the ABF evolution operator given in eq.~\eqref{eq:UABF} (see the
discussion in sect.~\ref{sec:gengamqg}).  Indeed, the numerical computation of
the $\as$ derivative of the anomalous dimension obviously fails for values of
$\as$ in proximity of the branch-change region.  There are different solutions
which we may adopt.  For instance, we may consider an asymmetric difference
quotient on the side where the branch does not change, but it is difficult to
implement this approach in an automated way.  Given also that we have noticed
that the numerical convergence of the integral in eq.~\eqref{gqgABFint3} is
problematic when using
$r(\as,N)=\as^2\beta_0\frac{d}{d\as}\log\gamma_+(\as,N)$, we have decided to
limit ourselves to using the incarnation of the ABF evolution operator of
eq.~\eqref{eq:UABF} which does not require any $\as$ derivative, namely
we set $r(\as,N)=\as\beta_0$. Thus, we work exclusively with the latter
choice of $r(\as,N)$, and we postpone further studies of
this problem to future work.

\subsection{Numerical results for $P_{qg}$ and comparison with previous
predictions}
\label{sec:results-comparison}

We now present some numerical results for the splitting functions that emerge
from the procedures discussed thus far. We start by focussing on $P_{qg}$, and
we discuss the comparison with its previous implementation as is included in
version 3 of \HELL.

For the sake of this comparison, we find it more convenient to plot the
resummed result after subtracting its expansion at the
NNLO~\cite{Bonvini:2017ogt,Bonvini:2018xvt},
\beq
\Delta_3P_{qg}(\as,x) \equiv P_{qg}^{\NLLp}(\as,x) -
\sum_{k=0}^2 \as^{k+1} P_{qg}^{{\NLLp} (k)}(x)
\eeq
since this is the default output of \HELL. We also apply a large-$x$ damping of
the form $(1-x)^2(1-\sqrt x)^4$, as is used in previous versions of \HELL, to
kill the effect of resummation in the large-$x$ region where it is not
expected to be accurate.
%%%%%%%%%%%%%%%%%%%%%%%%%%%%%%%%%%%%%%%%%%%%%%%%%%%%%%%%%%%%%%%%%%%%%
\begin{figure}[t]
  \centering
  \includegraphics[width=0.495\textwidth,page=1]{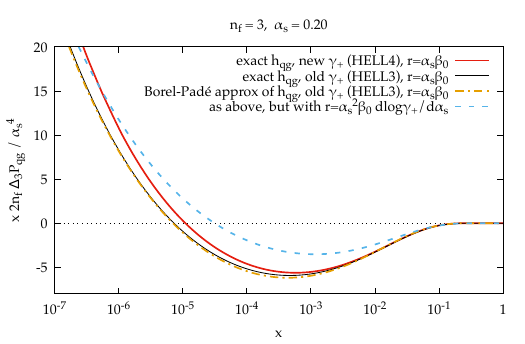}
  \includegraphics[width=0.495\textwidth,page=1]{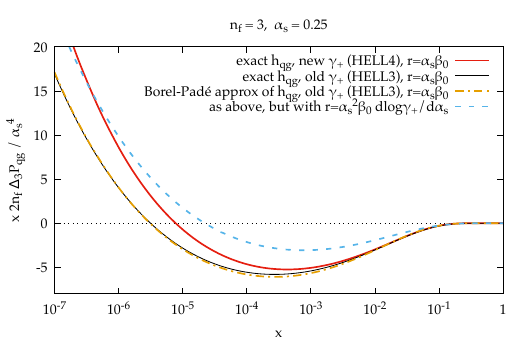}\\
  \includegraphics[width=0.495\textwidth,page=1]{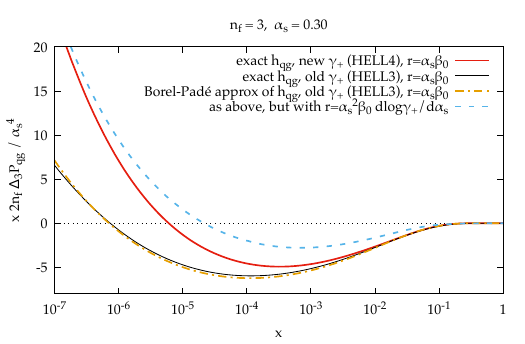}
  \includegraphics[width=0.495\textwidth,page=1]{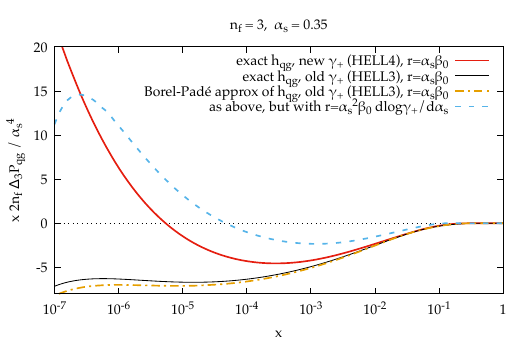}
  \caption{Comparison of $\Delta_3P_{qg}(\as,x)$ computed with different combinations of the ingredients and procedure.
    Each plot corresponds to a different value of $\as$, from 0.2 to 0.35 in steps of 0.05.}
  \label{fig:comparison}
\end{figure}
%%%%%%%%%%%%%%%%%%%%%%%%%%%%%%%%%%%%%%%%%%%%%%%%%%%%%%%%%%%%%%%%%%%%%

In fig.~\ref{fig:comparison} we show $2 n_f \Delta_3P_{qg}(\as,x)$,
multiplied by $x$ and divided by $\as^4$ for an improved readability, as a function of $x$
in a logarithmic scale.  We set $n_f=3$, which is the value relevant to small
scales, where $\as$ assumes large values.  The figure displays four panels,
corresponding to four different values of the coupling constant, namely
$\as=0.2$, $0.25$, $0.3$, and $0.35$ (left to right and top to bottom).
Currently, we do not consider larger $\as$ values since we also want to
present the results of \HELL version 3, which are limited to
$\as\lesssim0.35$. The plots in each panel have identical layouts,
and feature four curves, namely:
\begin{itemize}
\item our new best result (thick solid red), obtained according to the
procedure we have adoped in this work; in particular, we use the exact form
of $h_{qg}(\gamz)$, eq.~\eqref{hqgres10}, the modified $\gamma_+^{\NLLp}$ as
is described in sect.~\ref{sec:HELL-as}, and set $r(\as,N)=\as\beta_0$;
\item the same result as above (thin solid black), except for the fact
that we use here $\gamma_+^{\NLLp}$ extracted from \HELL version 3;
\item the old \HELL 3 result (dot-dashed orange), that uses a Borel-Pad\'e
procedure based on a finite number (16) of coefficients of the expansion of
$h_{qg}(\gamz)$;
\item the old \HELL 3 result (dashed blue), computed as the previous curve,
except that in this case the derivative form of $r$, namely
$r(\as,N)=\as^2\beta_0\frac{d}{d\as}\log\gamma_+(\as,N)$, is used (which was
considered the default result in \HELL 3).
\end{itemize}
Several comments are in order.

When comparing the first two curves, we immediately see that the results are
very similar at $\as=0.2$, and they are increasingly different as $\as$ gets
larger. This is in keeping with the genesis of the difference, namely the
modification in the construction of $\gamma_+^{\NLLp}$.  The largest part of
this effect comes from the modified resummation of the running coupling
contributions, as is described in appendix~\ref{sec:appHELL-RC}. Indeed, the
difference in the new construction amounts to a contribution of $\Ord(\as)$,
which is thus visible already at intermediate
values of $\as$, and becomes very significant as $\as$ gets larger.  The other
contribution to the difference between the two constructions, discussed in
appendix~\ref{sec:appHELL-as}, is instead designed to be a correction of
$\Ord(\as^2)$, and therefore it has a visible impact only at large $\as$.  We
also note that the new result is very stable across different $\as$ values,
while when using the old $\gamma_+^{\NLLp}$ there is a significant dependence on
$\as$, with the result bending down at small $x$ already when $\as=0.35$.  This
unexpected (and, we believe, unphysical) behaviour is induced by the spurious
pole discussed in appendix~\ref{sec:appHELL-RC}, and must be considered as a
failure of the previous construction at large $\as$ values, which is addressed
by the modifications we present in this work.

The comparison between the second and third curves is even more interesting.
The only difference between these curves is the use of the exact all-order
$h_{qg}$ expression in the former, and of its approximation based on the
Borel-Pad\'e procedure in the latter.
The striking feature of the comparison between these two curves is
that the agreement is remarkably good at all $\as$ values considered.  This
means that the Borel-Pad\'e procedure devised in ref.~\cite{Bonvini:2016wki}
works fairly (and even somewhat surprisingly) well, despite being based on
just 16 coefficients of the expansion of $h_{qg}$. Apart from considerations
on the procedure itself, this finding shows that results based on previous
versions of \HELL are fully reliable, if one limits oneself to employing
values of $\as$ compatible with a perturbative expansion.

Finally, the comparison between the last two curves shows the impact of
varying the approximation on which the construction of the ABF evolution
function of eq.~\eqref{eq:UABF} is based.  This difference is not considered
here for the first time, as it was already studied in
refs.~\cite{Bonvini:2016wki,Bonvini:2017ogt,Bonvini:2018xvt}, in the context
of the definition of an uncertainty band.  In the previous \HELL literature,
the last (dashed blue) curve has been considered as the ``central''
prediction, while the third curve (dot-dashed orange) was used as a variation
to construct the uncertainty band. In this work, instead, because of the
absence of an exact result obtained with
$r(\as,N)=\as^2\beta_0\frac{d}{d\as}\log\gamma_+(\as,N)$, we consider the
curve based on $r(\as,N)=\as\beta_0$ as the central value. This of course {\em
also} leads to differences with past results, which are in and of themselves
sizable. In part, such differences are mitigated by the uncertainty band
which we shall construct.
%%%%%%%%%%%%%%%%%%%%%%%%%%%%%%%%%%%%%%%%%%%%%%%%%%%%%%%%%%%%%%%%%%%%%%%%
\begin{figure}[t]
  \centering
  \includegraphics[width=0.495\textwidth,page=2]{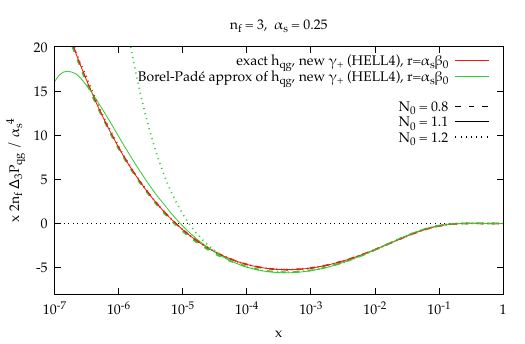}
  \includegraphics[width=0.495\textwidth,page=3]{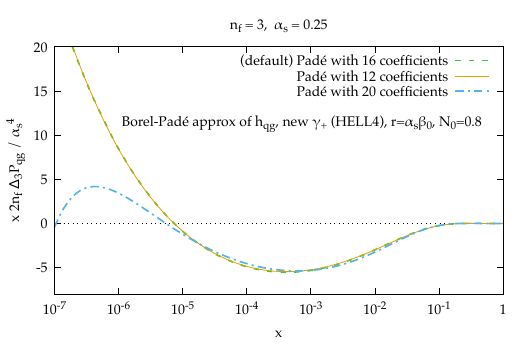}
  \caption{Variations of the Mellin inversion path (left-hand panel)
and of the order of the Pad\'e approximant (right-hand panel).
    The dashed green curve is the same in both plots.}
  \label{fig:stresstests}
\end{figure}
%%%%%%%%%%%%%%%%%%%%%%%%%%%%%%%%%%%%%%%%%%%%%%%%%%%%%%%%%%%%%%%%%%%%%%%%

The agreement between the second and third curves which appears in the
plots of fig.~\ref{fig:comparison} may raise the question of whether our
exact result for the function $h_{qg}$ is phenomenologically useful or not.
In order to (positively) answer this question, we consider some ``stress
tests'' of the two implementations of the resummed $P_{qg}$. While carrying
out those, we set the value of $\as$ equal to a sufficiently moderate value,
$\as=0.25$, in order to avoid potential problems due to large perturbative
corrections at larger $\as$ values, and consider the following two variations:
\begin{itemize}
\item a variation of the path used for Mellin inversion, which path we
parametrise as
\beq
\label{invNcontour0}
N(u) = N_0 + e^{i\theta}u, \qquad \frac\pi2 <\theta<\pi, \qquad u\in[0,\infty);
\eeq
\item a variation of the order of the Pad\'e approximant (obviously
applicable only to the ``old'' implementation).
\end{itemize}
The results are shown in fig.~\ref{fig:stresstests}.

In the plot on the left-hand panel, we show $2 n_f \Delta_3 P_{qg}(\as,x)$
multiplied by $x$ and divided by $\as^4$
(as in fig.~\ref{fig:comparison}) obtained with the exact
$h_{qg}$ (red curves) and with the Borel-Pad\'e procedure (green curves)
using three different paths for Mellin inversion. In particular, we change
the value of the parameter $N_0$\footnote{We do not show variations due to the
slope of the path, i.e.~the dependence upon $\theta$ of
eq.~(\ref{invNcontour0}), as it induces very little differences, mostly at
large $x$ where the damping washes out any resummation effects.} of
eq.~\eqref{invNcontour0}, from its default value $N_0=1.1$ (solid curves) to a
lower value $N_0=0.8$ (dashed curves) and to a higher value $N_0=1.2$ (dotted
curves). The solid red curve is identical to that of fig.~\ref{fig:comparison}.
In all cases, we use the new construction for $\gamma_+^{\NLLp}$.
We observe that the three red curves are in perfect agreement
among themselves: this proves that the implementation based on the exact
all-order $h_{qg}$ expression is very stable under variations of the path.
Conversely, the result based on the Borel-Pad\'e procedure depends
significantly on $N_0$: it is in good agreement with what is obtained with the
exact $h_{qg}$ for small $N_0$ values (we remark that $N_0=0.8$ was the value
used in \HELL 3), but it deviates from it in the small-$x$ region as $N_0$
gets larger.  This behaviour is likely due to the nature of the Pad\'e
approximant, which being a non-linear function may become overly sensitive to
small parameter variations.

In the plot on the right-hand panel, we change the order of the Pad\'e
approximant while fixing $N_0=0.8$, since the latter value underpins the best
agreement between the approximate and the exact result. Specifically, from the
default $[8/7]$ Pad\'e\footnote{We use the standard notation $[n/m]$ to
indicate a Pad\'e approximant with numerator of degree $n$ and denominator of
degree $m$.}  adopted in previous versions of \HELL, that uses 16 coefficients
of the expansion of $h_{qg}$, we consider a lower variation $[6/5]$ that uses
12 coefficients, as well as a higher variation $[10/9]$ that uses 20
coefficients.  The lower variation (solid yellow curve) is in perfect
agreement with the default result (dashed green), in spite of relying on
a smaller number of
coefficients: this means that either the result is dominated by the first
few terms of the expansion of $h_{qg}$, or the Pad\'e approximant
approximates very well the full result even with a small number of terms.
The higher variation (dot-dashed blue) predicts a splitting function which
is instead rather different w.r.t.~the other two. We point out that
this difference is not due to the incorrect values of the coefficients,
since up to order 20 (which is the maximum considered here)
they agree with the exact ones to better than permille
(see fig.~\ref{fig:ABFcoeff}); in fact, we have explicitly verified that by
using the exact coefficients in the Pad\'e approximant the result is the same
as before. Therefore, this test points to a failure of the procedure itself
when one tries and uses too many coefficients which is perhaps paradoxical,
but ultimately we interpret it to being due to the fact that a too-complicated
rational function becomes increasingly unreliable as a representation of a
non-rational function.

The conclusion of these studies is that the Borel-Pad\'e procedure is able to
provide one with an essentially correct result for the resummation of
$P_{qg}$, but only in a limited region of the space of parameters involved.
Finding this region is not straightforward (one looks for stability under
small parameter variations), and ultimately it can only be safely validated
by using the exact result, the availability of which of course renders the
whole exercise an academic one. Moreover, it is definitely possible that at
larger $\as$ value this region becomes smaller and smaller, thus making it
increasingly harder to pin it down. These considerations therefore clearly
highlight why our new exact result is crucial.

\subsection{Results for the singlet splitting function matrix}
\label{sec:results}

Having clarified, by using $P_{qg}$ as a prominent example whose conclusions
qualitatively apply to the other splitting functions as well,
in which aspects our new results are in keeping
with, or differ from, those obtained with the existing version of \HELL, we
are now in the position to present our predictions for all of the singlet
splitting functions, including at large values of $\as$.  We point out
that when constructing the resummed $P_{gg}$ according to
eq.~\eqref{Delta4gamma} we must take care of restoring the momentum
conservation condition,
\beq
\int_0^1 dx\, x \[P_{gg}(\as,x) + 2n_f P_{qg}(\as,x)\] = 0.
\eeq
This is in general violated by the large-$x$ damping discussed
before (which is applied to all of the resummed splitting functions).
The restoration is performed by adding a suitable term to the resummed
$P_{gg}$, as is discussed in detail in refs.~\cite{Bonvini:2017ogt}
(sect.~4.2.2 there). In view of the fact that we adopt the very same
procedure here, we do not comment on it any further.

%%%%%%%%%%%%%%%%%%%%%%%%%%%%%%%%%%%%%%%%%%%%%%%%%%%%%%%%%%%%%%%%%%%%%%%%%%%
\begin{figure}[t]
  \centering
  \includegraphics[width=0.495\textwidth,page=16]{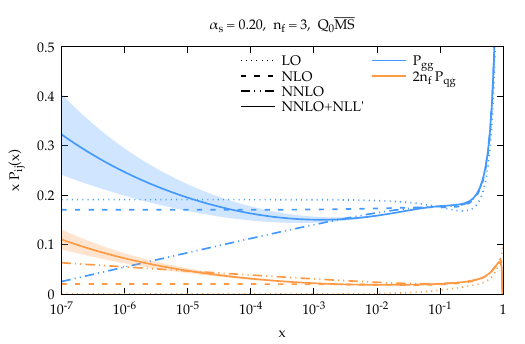}
  \includegraphics[width=0.495\textwidth,page=16]{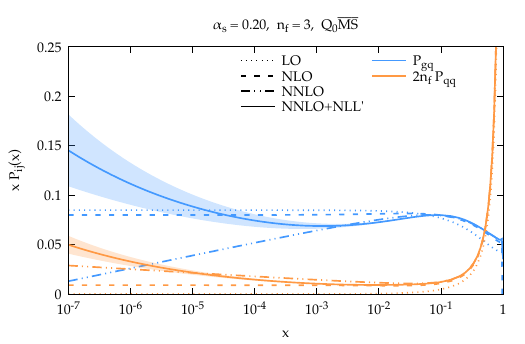}\\
  \includegraphics[width=0.495\textwidth,page=26]{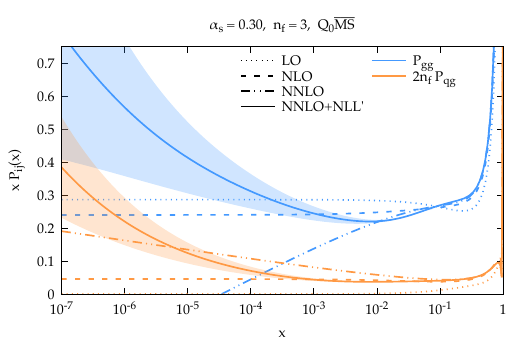}
  \includegraphics[width=0.495\textwidth,page=26]{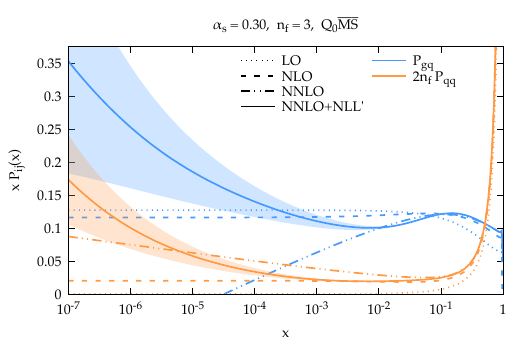}\\
  \includegraphics[width=0.495\textwidth,page=36]{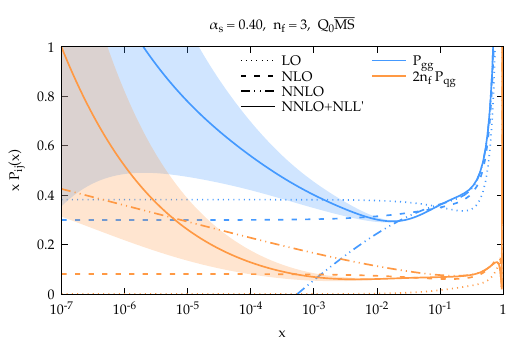}
  \includegraphics[width=0.495\textwidth,page=36]{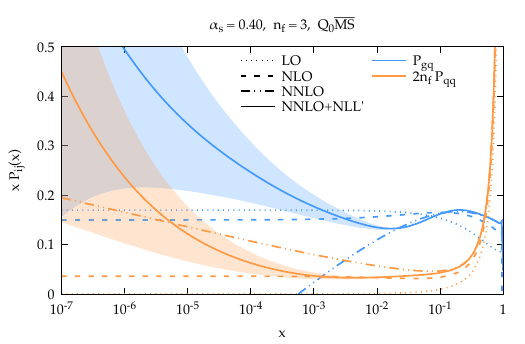}\\
  \includegraphics[width=0.495\textwidth,page=46]{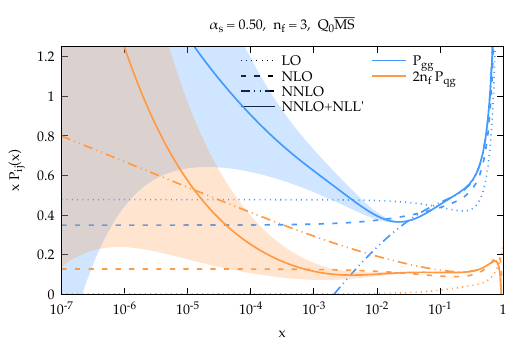}
  \includegraphics[width=0.495\textwidth,page=46]{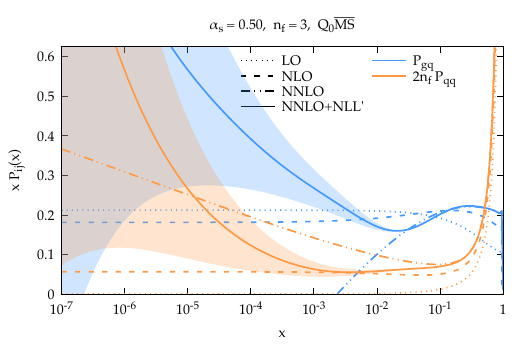}
  \caption{The fixed-order and resummed splitting functions as a function of $x$.
    Each row corresponds to a different values of $\as$, from 0.2 to 0.5 in steps of 0.1.
    In each row $P_{gg}$ (blue) and $P_{qg}$ (orange) are shown on the left,
    and $P_{gq}$ (blue) and $P_{qq}$ (orange) are shown on the right.}
  \label{fig:results}
\end{figure}
%%%%%%%%%%%%%%%%%%%%%%%%%%%%%%%%%%%%%%%%%%%%%%%%%%%%%%%%%%%%%%%%%%%%%%%%%%%

The resummed results matched to NNLO (solid), along with the fixed-order ones
(LO: dotted; NLO: dashed; NNLO: dot-dot-dashed), are shown in
fig.~\ref{fig:results}. The plots in the left column display the $P_{gg}$
(blue) and $2n_fP_{qg}$ (orange) splitting functions, while those in the right
column display the $P_{gq}$ (blue) and $2n_fP_{qq}$ (orange) splitting functions.
Each row corresponds to a value of $\as$, namely $\as=0.2$, $0.3$, $0.4$, and
$0.5$ (top to bottom).  In all cases we set $n_f=3$, which is the only value
of $n_f$ for which these large $\as$ values are relevant. The resummed results
are supplemented with an uncertainty band (shaded areas). At variance with
the previous \HELL 3 results, this band is obtained by considering a single
variation\footnote{The other variation used in ref.~\cite{Bonvini:2018xvt} is
the choice of the function $r(\as,N)$, that in our case is set to
$r(\as,N)=\as\beta_0$, as was previously discussed.} of the anomalous
dimension $\gamma_+^{\NLLp}$ which enters the construction of all of the
splitting functions. This variation has been described in
refs.~\cite{Bonvini:2017ogt,Bonvini:2018xvt}, and is related to the way the
resummation of running coupling effects mentioned in sect.~\ref{sec:gengampp}
is implemented (see appendix~\ref{sec:appHELL-RC} for further detail).  The
final band is obtained by symmetrising the difference of the variation with
respect to the central curve.

These results are {\em qualitatively} similar to those of the existing
literature.  For example, going from large to small $x$, the resummed NNLO+$\NLLp$
$P_{gg}$ and $P_{gq}$ follow initially the NNLO predictions before
dramatically steeping up at a certain $x$ value (which in turn depends on
$\as$), thus producing the well known and expected dip~\cite{Ciafaloni:2003kd,
Ciafaloni:2003rd,Altarelli:2005ni,Altarelli:1999vw}.  Also, the resummed
NNLO+$\NLLp$ $P_{qg}$ and $P_{qq}$ splitting functions tend to rise in a much
milder way w.r.t.~their NNLO counterparts, and are thus in a better agreement
with the corresponding NLO results in an intermediate range in $x$, as was
observed already in refs.~\cite{Bonvini:2017ogt,Bonvini:2018xvt}.

In addition to known qualitative behaviours, we can now investigate what
happens at large $\as$ values.  We first observe that when $\as$ grows the
curves of the resummed results at small $x$ become steeper, and the
corresponding rises begin at larger values of $x$ (w.r.t.~their small-$\as$
counterparts) for all of the splitting functions.  Moreover, the uncertainty
band of the resummed result becomes bigger, which is expected as the impact of
subleading terms becomes more important.  Keeping in mind that the way this
band is constructed is somewhat arbitrary, and it does not necessarily
faithfully represent the size of the subleading contributions, it is clear
that by going to large $\as$ values the resummed result becomes less reliable
towards small $x$'s --- the uncertainty band becomes huge.  A higher
logarithmic accuracy would be needed to tame the uncertainty and to obtain a
more reliable result, even though there is the concrete possibility that when
$\as$ is too large even the convergence of the resummed expansion is spoiled.
The next subsection will present a speculative discussion on this point,
and an alternative interpretation which questions its conclusions.

\subsection{When subleadingness ceases to be a useful concept\label{sec:subL}}

In order to qualitatively understand where we expect the resummed results to
be valid, we recall that we are resumming to all orders powers of
$\as\log(1/x)$. This means that the ordering for which this (re)organization
of the perturbative series is reliable is the one where
\beq
\as\log\frac1x \sim1, \qquad \as\ll1.
\eeq
The leftmost condition ensures that the organization of the series as
an expansion in powers of $\as$ of all-order functions of $\as\log(1/x)$
is meaningful, while the rightmost one ensures that formally-subleading
logarithmic corrections are {\em actually} subleading. Typically, we
consider that if the latter condition holds true while the former one
does not (i.e.~$\as\log(1/x)\gg1$), then the logarithmic-ordered
expansion of the resummed result is no longer appropriate, and the
results thus obtained become unreliable.

We shall argue later that this is not necessarily the case\footnote{In
this section, for the sake of simplicity the problem of the Landau pole
will be ignored, and $\as$ will be considered as well-defined everywhere.};
but before going into that, we shall briefly elaborate on such a standard
interpretation of the logarithmic ordering in the context of
small-$x$ physics.
%%%%%%%%%%%%%%%%%%%%%%%%%%%%%%%%%%%%%%%%%%%%%%%%%%%%%%%%%%%%%%%%%%%%%%%
\begin{figure}[t]
  \centering
  \includegraphics[width=0.6\textwidth, page=1]{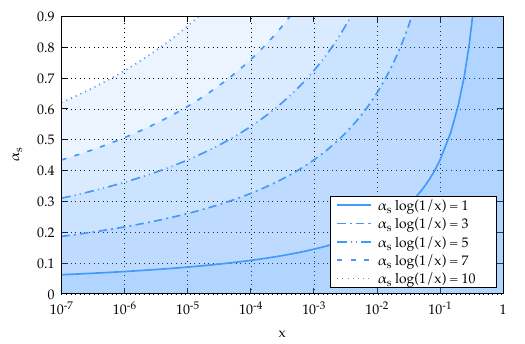}
  \caption{Curves with constant $\as\log(1/x)$ value. See the text for details.}
  \label{fig:validity}
\end{figure}
%%%%%%%%%%%%%%%%%%%%%%%%%%%%%%%%%%%%%%%%%%%%%%%%%%%%%%%%%%%%%%%%%%%%%%%
Clearly, the border between the regions \mbox{$\as\log(1/x)\sim1$} and
\mbox{$\as\log(1/x) \gg1$} is not sharp. We can have a rough idea of
the interplay between these regions by looking at figure~\ref{fig:validity}.
There, the curves represent the sets of $(x,\as)$ points
for which \mbox{$\as\log(1/x)$} assumes
a given value, which increases by moving southeast to northwest. Thus,
in the context of the standard interpretation of resummed results the
bottom right (upper left) corner is the region associated with the
most (least) reliable predictions. In particular, the region below and/or
to the right of the solid line (for which \mbox{$\as\log(1/x)=1$}) poses
no problem to the logarithmic ordering of the cross section. If anything,
when moving towards large-ish $x$'s, effects beyond small-$x$ resummation
also become important, and indeed the resummed results are damped there.
As one moves towards the non-solid curves, the impact of such a damping
becomes irrelevant, but at the same time the value of \mbox{$\as\log(1/x)$}
keeps on growing. Deciding where such a value is unconfortably large
is unfortunately an arbitrary choice. For example, if \mbox{$\as\log(1/x)=5$}
is deemed to be acceptable, then resummed predictions are reliable down to
$x\sim10^{-7}$ at $\as=0.3$, or $x\sim10^{-4}$ at $\as=0.5$ (clearly,
the larger the value of $\as$ the smaller the ``safe'' region in $x$).

In general, {\em if} we consider the bending down of the results in
fig.~\ref{fig:comparison} and fig.~\ref{fig:stresstests} as evidence
of the start of the breakdown of the logarithmic ordering which underpins
the resummed splitting functions, inspection of fig.~\ref{fig:validity}
shows that this indeed corresponds to \mbox{$\as\log(1/x)\sim5$}.

However, the very discussion given in the preceding sections implies
that the reason for that bending down may actually be more subtle,
namely driven in equal measure by the subleading nature of certain
contributions and by the way in which those are implemented in the
resummed results. In other words, certain classes of resummed results
might be such that formally-subleading terms may, or may not, become
numerically comparable to the leading ones, depending on the characteristics
of different approximations which are all at the same level from a
logarithmic standpoint. It is only in this case that the validity
of the logarithmic ordering of the resummed results becomes unreliable,
and further improvements are necessary to recover sensible predictions.

In order to further the previous point in an over-simplified context,
consider a model in which the perturbative expansion in a ``small''
parameter $a$ is potentially spoiled by large logarithms $L$ that emerge
order by order. Suppose also that the LL coefficients
associated with such logarithmic terms are well behaved, namely%
\footnote{All coefficients are understood to be divided by $1/k!$, as
is exemplified by eq.~\eqref{modelLL}.}
\beq
{\rm LL}\quad\longleftrightarrow\quad
\Big\{a^k L^k\Big\}_{k=0}^\infty
\qquad\Longrightarrow\qquad
\varsigma_{\rm LL}=\sum_{k=0}^\infty \frac{1}{k!}a^k L^k=e^{aL}\,,
\label{modelLL}
\eeq
with $\varsigma$ any quantity whose resummation is of interest
(i.e.~the splitting functions here, an observable, and so forth).
This simple formula captures the essence of resummation: when $L\sim 1/a$,
the coupling $a$ is not a good expansion parameter any longer, and one
must use $aL$ instead. By taking into account terms to all orders, one
obtains a well-defined expression $\varsigma_{\rm LL}$. Importantly,
$\varsigma_{\rm LL}$ is a perfectly valid result also in the region
where $L\gg 1/a$, i.e.~where the standard logarithmic-ordered counting
loses validity.
In other words, by means of
resummation one obtains results which are valid regardless of the value
assumed by the large logarithm; it is clear that this is strictly an
all-order property, and that the conclusion is universally valid only
because potentially subleading contributions have been ignored.

We now lift the latter restriction, i.e.~we consider the generic
NLL, NNLL, etc. contributions
\beq
{\rm N}^i{\rm LL}\quad\longleftrightarrow\quad
\Big\{c_i(k)a^{k+i} L^k\Big\}_{k=0}^\infty\,,\qquad
i=1,2,\ldots
\eeq
in the following two cases
\begin{align}
&c_1(k)=1\,,\;
c_2(k)=\frac{1}{2}\,,\;
c_3(k)=\frac{1}{6}\,,\;
\ldots
  &&\Longrightarrow\;\;
c_i(k)=\frac{1}{i!}\,,
\label{lincff}
\\*
&c_1(k)=k\,,\;
c_2(k)=\frac{k(k-1)}{2}\,,\;
c_3(k)=\frac{k(k-1)(k-2)}{6}\,,\;
\ldots
  &&\Longrightarrow\;\;
c_i(k)=\frac{\Gamma(k+1)}{i!\Gamma(k+1-i)}\,.
\label{geomcff}
\end{align}
Let us start with eq.~(\ref{lincff}). The resulting resummed
quantities are
\beq
\varsigma_{{\rm N}^i{\rm LL}}=
\sum_{k=0}^\infty \frac{1}{k!}c_i(k)a^{k+i} L^k=
\frac{1}{i!}a^i e^{aL}\,,
\eeq
and thence
\beq
\varsigma_{{\rm LL}}+\varsigma_{{\rm NLL}}+\varsigma_{{\rm N}^2{\rm LL}}
+\varsigma_{{\rm N}^3{\rm LL}}+\ldots=
\left(1+a+\frac{1}{2}a^2+\frac{1}{6}a^3+\ldots\right)
e^{aL}\,.
\label{modelNiLL}
\eeq
Therefore, here the subleadingness of the various contributions remains
a meaningful characteristic, {\em regardless} of the value of $L$, which
can be arbitrarily large --- the only necessary condition is that the
coupling constant remain in the perturbative domain. In other words,
one may truncate the resummed quantity at any logarithmic accuracy, and
be assured that the corresponding prediction is sensible, in the sense
that contributions of higher logarithmic accuracy will modify that
prediction in an increasingly minimal way.

Conversely, with eq.~(\ref{geomcff}) we obtain what follows,
\beq
\varsigma_{{\rm N}^i{\rm LL}}=
\sum_{k=0}^\infty \frac{1}{k!}c_i(k)a^{k+i} L^k=
\frac{1}{i!}a^i(aL)^i e^{aL}\,,
\eeq
so that
\beq
\varsigma_{{\rm LL}}+\varsigma_{{\rm NLL}}+\varsigma_{{\rm N}^2{\rm LL}}
+\varsigma_{{\rm N}^3{\rm LL}}+\ldots=
\left(1+a^2L+\frac{1}{2}a^4L^2+\frac{1}{6}a^6L^3+\ldots\right)
e^{aL}\,.
\label{modelNiLL2}
\eeq
Now, at variance with what happens with eq.~(\ref{modelNiLL}), if
eq.~(\ref{modelNiLL2}) truncated at a certain logarithmic order $i_0$
it does matter whether $aL\sim 1$ or $aL\gg 1$, since in the latter case
(in particular, when $aL\gg 1/a$) regardless of how many $i_0$ resummed
contributions one includes the result will not make any physical sense,
in spite of the valid definition of each of its components.

The only viable solution is therefore that of considering {\em all} of
the nominally subleading logarithmic terms, i.e.~to compute:
\beq
{\rm eq.}~(\protect\ref{modelNiLL2})
\;\;\;\;\longrightarrow\;\;\;\;
\sum_{i=0}^\infty\varsigma_{{\rm N}^i{\rm LL}}=
\sum_{i=0}^\infty \frac{1}{i!}a^i(aL)^i e^{aL}=
e^{a(1+a)L}\,,
\label{DRes}
\eeq
which is valid for any $L$, with the behaviour expected on the basis of
logarithmic dominance restored (since, regardless of the value of $L$,
\mbox{$aL>a^2L$} as long as $a$ is sufficiently small).
We point out that by playing the same game with eq.~(\ref{modelNiLL})
one does not obtain a result significantly different from a truncated one:
\beq
{\rm eq.}~(\protect\ref{modelNiLL})
\;\;\;\;\longrightarrow\;\;\;\;
\sum_{i=0}^\infty\varsigma_{{\rm N}^i{\rm LL}}=
\sum_{i=0}^\infty \frac{1}{i!}a^i e^{aL}=
e^{a(L+1)}\,.
\label{DRes2}
\eeq
The bottom line is that, depending on the characteristics of the
coefficients of the subleading contributions, such contributions may
become numerically comparable to or dominant over their leading counterparts,
thus spoiling the convergence properties of the resummed quantity
of interest, irrespective of the formal logarithmic accuracy which one
manages to achieve. In these cases the resummed prediction, if
truncated at any given logarithmic accuracy (however large), will
be unable to return a sensible physical result, and it must therefore be
further improved, for example by finding either an approximation
of the neglected subleading terms that can be effectively included
in those contributions which have been kept, or a pattern for the
neglected subleading terms which allows one to resum them\footnote{The
analytical inverse Mellin transform of the asymptotic $\xi\to 1$ form of
the weighted exponential of the function $\hh(s)$ of eq.~(\ref{sFsasyE1})
is a case of a resummation with $L=\log(1-\xi)$ where the coefficients
of a logarithmically ordered expansion behave quadratically (as in
eq.~(\ref{geomcff})). In spite of having figured out such a behaviour,
a double resummation was still not possible, since the coefficients
feature several contributions (as opposed to one in the example
above) whose number grows with the order, and of which only the first few
follow a clear pattern (e.g.~the leading ones which, as was just said,
behave quadratically.}.
The former strategy is clearly the least solid of the two, but
in practical cases it may be the only viable one\footnote{In the
context of the simple model discussed here, such a strategy would
amount to replacing $a$ with $a+a^2$ in the case of the coefficients
of eq.~(\ref{geomcff}), and $L$ with $L+1$ in the case of those of
eq.~(\ref{lincff}). The fact that these replacements allow one to
recover {\em exactly} the doubly-resummed quantities of eqs.~(\ref{DRes})
and~(\ref{DRes2}), respectively, is due to the simplicity of the model.}.

We conclude this section by stressing that there may be different
mechanisms which destroy the logarithmic ordering of resummed quantities,
and that we are not implying that small-$x$ resummation is underpinned
by geometrically-growing coefficients such as those of eq.~(\ref{geomcff}).
Nevertheless, we still stress that the commonly-assumed limitations on
the validity of resummed quantities may not necessarily be accurate,
and a case-by-case assessment is necessary. In our specific situation,
it is not unlikely that the improved approximations proposed in this
work imply an extended validity of the resummed splitting kernels.
Clearly, this improvement (being to some extent heuristic) goes hand-in-hand
with large theoretical systematics, and therefore the final evidence
of the validity of this approach can only come by means of a comparison
to data.

\section{Future directions: results for the  $qg$ Green function}
\label{sec:Gqgres}

While this is an aside in the present context, and does not play any role
in the numerical results obtained so far, it is worth mentioning the
fact that by means of the same techniques employed in sect.~\ref{sec:Pqg}
we have been able to obtain a closed, all-order form for the $\ep=0$ part of the
finite $\hat\G_{qg}$ Green function. Some of the details of the derivation
are given in appendix~\ref{sec:Gqgep0}; here, we limit ourselves to
reporting the results (see sect.~\ref{sec:hqgAO} for the normalisation
of the various quantities involved), which read
\begin{align}
\hat\G_{qg}(\ep\to0)&=
\cc_0(\hh(\gamma_s))+\cc_1(\gamma_s)
\exp\Bigg[
\half\int_0^1dy\,
\frac{\gamma_s\big(\chi_1(y\gamma_s)-\chi^\prime(y\gamma_s)\big)}
{\chi(y\gamma_s)}
+\gamma_s\psi_0(1)
\nonumber
\\*&\hspace{12em}
-\half\log\Gamma\big(1+\gamma_s\big)
+\half\log\Gamma\big(1-\gamma_s\big)
\nonumber
\\*&\hspace{12em}
-\half\log\Big(\gamma_s\chi(\gamma_s)\Big)
-\half\log\Big(-\gamma_s^2\chi^\prime(\gamma_s)\Big)\Bigg]\,,
\phantom{aaa}
\label{Fzfunres2}
\end{align}
with $\chi^\prime(z)=d\chi/dz$, $\gamma_s=\gamma_s(\bbas)$ and
\begin{align}
\cc_0(z)&=-\frac{1}{4}\left(3e^{2z}+e^{\frac{2}{3}z}\right),
\label{c0def}
\\
\cc_1(z)&=\frac14\[\frac3{1-z} + \frac{12}{2-z} - \frac{15}{3-z}\],
\label{c1def}
\end{align}
and where the function $\chi_1$ is defined as follows
\beq
\chi_1(z)=2\psi_1(1)-\psi_1(z)-\psi_1(1-z)\,.
\label{chi1fun}
\eeq
As it is also shown in appendix~\ref{sec:Gqgep0}, this result gives us
the possibility of determining a relationship between the finite
$\G_{qg}$ and $\G_{gg}$ Green functions. Specifically
(see appendix~\ref{sec:Gqgep0})
\beq\label{Fzfunres4}
\hat\G_{qg}(\ep\!\to\!0)=
\cc_0(\hh(\gamma_s))+\cc_1(\gamma_s)
\frac{\G_{gg}(\ep\!\to\!0)}{\gamma_s\chi(\gamma_s)}\,.
\eeq
Furthermore, from eqs.~(\ref{hqgres10}), (\ref{c0def}), \eqref{hqgdef}
and~\eqref{dualLO} we obtain:
\beq
\cc_0(\hh(\gamma_s))=-\frac{h_{qg}(\gamma_s)}{\gamma_s\chi(\gamma_s)}
\equiv
-\frac{\hat\gamma_{qg}}{\gamma_s}\,.
\label{c0vsgqg}
\eeq
Therefore, eq.~(\ref{Fzfunres4}) becomes
\beq
\gamma_s\,\hat\G_{qg}(\ep\!\to\!0)=-\hat\gamma_{qg}
+\bbas\,\cc_1(\gamma_s)\,\G_{gg}(\ep\!\to\!0)\,,
\label{Fzfunres5}
\eeq
which gives a direct relationship between the two renormalised Green
functions which only involves $\ep$-finite contributions. We can also
observe what follows: with eqs.~\eqref{Gqgdef}-\eqref{hhdef}, the
derivation of the master equation~(\ref{Gqgfact2}) w.r.t.~$\bbas$ gives
\beq
\frac{d\hat\G_{qg}^{(0)}}{d\bbas}=\frac{\gamma_s}{\ep\bbas}\,
\Gamma_{gg}\hat\G_{qg}+
\Gamma_{gg}\,\frac{d\hat\G_{qg}}{d\bbas}+
\frac{\hat\gamma_{qg}}{\ep\bbas}\,\Gamma_{gg}\,,
\label{dG0dabar0}
\eeq
whence, since $\G_{qg}$ is finite,
\beq
\gamma_s\,\hat\G_{qg}(\ep\!\to\!0)=-\hat\gamma_{qg}
+\bbas\lim_{\ep\to 0}
\left(\frac{\ep}{\Gamma_{gg}}\,\frac{d\hat\G_{qg}^{(0)}}{d\bbas}\right)\,.
\label{dG0dabar}
\eeq
By equating side-by-side this equation with eq.~(\ref{Fzfunres5})
we obtain
\beq
\lim_{\ep\to 0}
\left(\frac{\ep}{\Gamma_{gg}}\,\frac{d\hat\G_{qg}^{(0)}}{d\bbas}\right)=
\cc_1(\gamma_s)\,\G_{gg}(\ep\!\to\!0)\,.
\label{lim0G0}
\eeq
This is an identity whose correctness can be verified perturbatively,
by using the explicit results given elsewhere for the quantities that appear
in it.

The coefficients of the expansions of the finite Green functions in
terms of $\gamma_s$ can be obtained in a closed form. By defining
such coefficients as follows
\begin{align}
\G_{gg}(\ep\!\to\!0)&=\sum_{n=0}^\infty r_n\,\gamma_s^n\,,\label{eq:Gggcff}
\\
\hat\G_{qg}(\ep\!\to\!0)&=\sum_{n=0}^\infty s_n\,\gamma_s^n\,,\label{eq:Gqgcff}
\end{align}
they can be written in term of the complete exponential Bell polynomials,
\begin{align}
r_n&=\frac{1}{n!}\,B_n\left(\widehat{Z}_-\right),
\label{rncoeff}
\\
s_n&=\frac{1}{n!}\,B_n\left(\widehat{Z}_+ +\widehat{Z}_R\right)-
\frac{1}{n!}\left[
\frac{3}{4}\,B_n\left(\widehat{Z}_2^\prime\right)
+\frac{1}{4}\,B_n\left(\widehat{Z}_{\frac{2}{3}}^\prime\right)\right].
\label{sncoeff}
\end{align}
The arguments of the Bell polynomials are given in eq.~(\ref{hZpvecdef})
and as follows
\begin{align}
\widehat{Z}_\pm&=
\Big(0,0,\Big\{\widehat{Z}_\pm^{[i]}\Big\}_{i=3}^\infty\Big),
\\*
\widehat{Z}_R&=
\Big(\Big\{\frac{(i-1)!}{12^i}
\left(4^i+6^i-9^i+12^i\right)\Big\}_{i=1}^\infty\Big),
\\*
\widehat{Z}_\pm^{[i]}&=
-\half(1+(-)^{i+1})\psi_{i-1}(1)
+\half\sum_{j=1}^i(-)^j(j-1)!\Big(\pm B_{i,j}(V_1)+B_{i,j}(V_2)\Big)
\nonumber
\\*&\quad
+\frac{1}{i}\sum_{k=0}^{i-3}\binomial{i}{k}
   (-)^{i-k+1}(i-k-1)(i-k)\psi_{i-k-1}(1)
\left(\sum_{j=0}^k(-)^jj!B_{k,j}(V_1)\right)\,,
\end{align}
where in turn
\begin{align}
V_1&=
\Big(0,\Big\{-\(1-(-)^{k}\)k\psi_{k-1}(1)\Big\}_{k=2}^\infty\Big),
\\*
V_2&=
\Big(\Big\{\(1-(-)^k\)k(k-1)\psi_{k-1}(1)\Big\}_{k=1}^\infty\Big).
\end{align}
The numerical values of the first coefficients in eq.~\eqref{eq:Gggcff} are
\begin{subequations}
\begin{align}
  r_0 &= 1, \\
  r_1 &= 0, \\
  r_2 &= 0, \\
  r_3 &= \frac{8}{3}\zeta_3, \\
  r_4 &= -\frac{3}{4}\zeta_4,\\
  r_5 &= \frac{22}{5}\zeta_5,\\
  r_6 &= \frac{65}{9}\zeta_3^2 - \frac{5}{6}\zeta_6,\phantom{\hspace{16.5em}}
\end{align}
\end{subequations}
while the first coefficients of eq.~\eqref{eq:Gqgcff} are
\begin{subequations}
\begin{align}
  s_0 &= 0, \\
  s_1 &= -\frac{7}{12}, \\
  s_2 &= -\frac{41}{72}, \\
  s_3 &= -\frac{157}{1296} + \frac{2}{3}\zeta_3, \\
  s_4 &= \frac{2537}{7776} + \frac{29}{9}\zeta_3 -\frac{3}{4}\zeta_4, \\
  s_5 &= \frac{27595}{46656}+\frac{575}{108}\zeta_3
         -\frac{13}{16}\zeta_4 +\frac{12}{5}\zeta_5, \\
  s_6 &= \frac{2960431}{4199040}+\frac{3337}{648}\zeta_3+\frac{53}{9}\zeta_3^2
         -\frac{71}{96}\zeta_4+\frac{242}{45}\zeta_5-\frac{5}{6}\zeta_6.
\end{align}
\end{subequations}
These expressions allow one to obtain their analogues in the case of
an expansions in $\bbas$, as is explained in eqs.~(\ref{gggvsbbas1})
and~(\ref{gggvsbbas2}). We also observe that eq.~(\ref{sncoeff}) closely
mirrors eq.~(\ref{Fzfunres4}). More in details, $\widehat{Z}_-$ (which
gives rise to $\G_{gg}$) is turned into $\widehat{Z}_+$ by the factor
\mbox{$1/(\gamma_s\chi(\gamma_s))$}, while the rational part
\mbox{$\cc_1(\gamma_s)$} corresponds to $\widehat{Z}_R$. Finally,
the additive piece $\cc_0(\hh)$ of eq.~(\ref{Fzfunres4}) results in the
two terms in square brackets in eq.~(\ref{sncoeff}), as one expects
from eqs.~(\ref{gqgvsggg}) and~(\ref{qncoeff}).

\noindent
{\bf A tribute to ref.~\cite{Catani:1994sq}}

\noindent
In appendix~C of ref.~\cite{Catani:1994sq}, eq.~(C.13) gives the rational
part of the coefficients of the expansion of $\gamma_{qg}$; it is easy
to verify that these are what one obtains by expanding in $x$ the function
$\cc_0(x)$ defined in eq.~(\ref{c0def}) --- see eq.~(\ref{c0vsgqg}) for
the relationship between $\cc_0$ and $\gamma_{qg}$, and bear in mind
that, by ignoring Riemann $\zeta$'s, $\gamma_s=\bbas$. After presenting
that result, ref.~\cite{Catani:1994sq} proceeds to saying, verbatim:
``The derivation of the result (C.13) is left as an exercise for the
reader''. We therefore find it appropriate, after 30-plus years, to
ask the reader to show that the all-order (in $\ep$) rational part
of the finite $qg$ Green function reads as follows:
\begin{align}
  \G_{qg}(z)\Big|_{\substack{\zeta_k\to 0\\\pi^{2k}\to 0}}
  &=
-\frac{3e^{2z}}{4(1+2\ep)}-
\frac{3e^{2/3z}}{4(3+2\ep)}-
\frac{z^2}{4}
\label{Gqgfinrat2}
\\*&+
\frac{3}{4}\,\Bigg\{
(1+2z+2z^2)\sum_{n=0}^\infty\left(-2\ep\right)^n
\nonumber
\\*&\phantom{aaaaa}
+z^3\sum_{n=0}^\infty\ep^n\Bigg[
\frac{P_n(z)}{(1-z)^{2n+1}}
+2\sum_{i=1}^n (-2)^{i}
\frac{P_{n-i}(z)}{(1-z)^{2(n-i)+1}}\Bigg]\Bigg\}
\nonumber
\\*&
+\frac{3}{8}\,\Bigg\{
-z(2+z)\sum_{n=0}^\infty\left(-\frac{\ep}{2}\right)^n
\nonumber
\\*&\phantom{aaaaa}
+z^3\sum_{n=0}^\infty\left(2\ep\right)^n\Bigg[
\frac{P_n(z/2)}{(2-z)^{2n+1}}
-\sum_{i=1}^n \frac{(-)^i}{2^{2i}}
\frac{P_{n-i}(z/2)}{(2-z)^{2(n-i)+1}}\Bigg]\Bigg\}
\nonumber
\\*&
+\frac{9+6z+2z^2}{36}\sum_{n=0}^\infty\left(-\frac{2\ep}{3}\right)^n
+\frac{3z+z^2}{18}\sum_{n=0}^\infty\left(-\frac{\ep}{3}\right)^n
\nonumber
\\*&\phantom{aaaaa}
-\frac{5z^3}{36}\sum_{n=0}^\infty\left(3\ep\right)^n\Bigg[
\frac{P_n(z/3)}{(3-z)^{2n+1}}
-\frac{2}{5}\sum_{i=1}^n \frac{(-)^{i}(1+2^i)}{3^{2i}}
\frac{P_{n-i}(z/3)}{(3-z)^{2(n-i)+1}}\Bigg]\,,\phantom{aa}
\nonumber
\end{align}
with $P_n(z)$ the Eulerian polynomials of the second order.
Equation~(\ref{Gqgfinrat2}) is organised, with the exception
of the first two terms on its r.h.s., in terms of explicit
powers of $\ep$. This may or may not be convenient in view of
further manipulations. In the case it were not, eq.~(\ref{Gqgfinrat2})
can be rewritten as follows
\begin{align}
  \G_{qg}(z)\Big|_{\substack{\zeta_k\to 0\\\pi^{2k}\to 0}}
  &=
-\frac{3e^{2z}}{4(1+2\ep)}-
\frac{3e^{2/3z}}{4(3+2\ep)}
\label{Gqgfinrat}
\\*&+
\frac{3}{4}\,\frac{1-2\ep}{1+2\ep}\Bigg\{
1+\ep+z-z^2\,
\frac{e^{-z/\ep}}{\ep}\left(-\frac{z}{\ep}\right)^{1/\ep}
\left[\Gamma\left(-\frac{1}{\ep}\right)-
\Gamma\left(-\frac{1}{\ep},-\frac{z}{\ep}\right)\right]\Bigg\}
\nonumber
\\*&+
\frac{3}{2}\,\frac{1+\ep}{2+\ep}\Bigg\{
2+\ep+z-z^2\,
\frac{e^{-z/\ep}}{\ep}\left(-\frac{z}{\ep}\right)^{2/\ep}
\left[\Gamma\left(-\frac{2}{\ep}\right)-
\Gamma\left(-\frac{2}{\ep},-\frac{z}{\ep}\right)\right]\Bigg\}
\nonumber
\\*&
-\frac{3}{4}\,\frac{(1+\ep)(5+2\ep)}{(3+\ep)(3+2\ep)}\Bigg\{
3+\ep+z-z^2\,
\frac{e^{-z/\ep}}{\ep}\left(-\frac{z}{\ep}\right)^{3/\ep}
\left[\Gamma\left(-\frac{3}{\ep}\right)-
\Gamma\left(-\frac{3}{\ep},-\frac{z}{\ep}\right)\right]\Bigg\}\,,
\nonumber
\end{align}
in terms of the ordinary $\Gamma$ function and of its upper incomplete
counterpart. Among other things, the equality between eqs.~(\ref{Gqgfinrat2})
and~(\ref{Gqgfinrat}) is established by means of the identity
\beq
1+z\sum_{n=0}^\infty t^n\,\frac{P_n(z)}{(1-z)^{2n+1}}=
-\frac{e^{-z/t}}{t}\left(-\frac{z}{t}\right)^{1/t}\left[
\Gamma\left(-\frac{1}{t}\right)-
\Gamma\left(-\frac{1}{t},-\frac{z}{t}\right)\right].
\label{EulPGF}
\eeq

\section{Conclusions}
\label{sec:concl}

Driven by the original motivation of extending the numerical results
of small-$x$ resummation to cases characterised by $\as$
values of order one, which is necessary in the context of lepton-PDF
evolution that also accounts for coloured partons, we have reconsidered
the theoretical bases of such resummation, and their use in the
computer code \HELL.

While doing so, we have found a number of analytical results previously
unknown in the literature, that not only help extend the standard
small-$x$ treatments to large-$\as$ cases, but also
significantly improve the very basic ingredients of such treatments,
thus including those relevant to purely hadronic environments.

In this paper we have therefore focussed on two technical aspects of
our work, namely the derivation and presentation of the new analytical
results we have obtained, and the novel implementation strategies
that they have fostered, which have become the defaults in the new 4.0 version
of \HELL and which will be used for hadronic and leptonic PDFs alike;
phenomenological applications stemming from these improvements will
be presented in forthcoming papers.

Specifically, as far as new analytical results are concerned, we have started
with finding several all-order expressions for $\gamma_s$, the
leading-logarithmic divergent eigenvalue of the singlet anomalous-dimension
matrix. Although none of these is in a closed, explicit, non-integral
form given in terms of elementary functions, the various
expressions we have found offer complementary ways to study the impact of
$\gamma_s$ onto other quantities that employ it as an ingredient. The upshot
of this is that, in general, such quantities assume relatively simple forms if
expressed in terms of $\gamma_s$ rather than of $\as$; in other
words, $\gamma_s$ is an inherently complicated object, which helps capture
(most of) the complexity of the quantities it enters.

We have then managed to obtain a closed, all-order expression for the $qg$
anomalous dimensions, and closed forms for all of its coefficients relevant to
perturbative series in different expansion parameters.  Because of this, we
have further been able to obtain the first properly resummed expression of the
$P_{qg}$ splitting kernel --- what had been available up to present was an
expression stemming from a Borel-Pad\'e approximation of an asymptotic
expansion.  The difference between the exactly- and the Borel-Pad\'e-resummed
results for $P_{qg}$ are seemingly small if one compares the upcoming and
previous versions of \HELL which implement them, respectively, but this
conclusion is extremely misleading. In fact, we have shown how the
Borel-Pad\'e-resummed $P_{qg}$ depends very strongly on the integration path
employed to compute the inverse Mellin transform which underpins it --- the
previous implementation has simply fine-tuned the choice of such a path in
order to achieve maximum stability. Conversely, the exactly-resummed splitting
kernel is very stable upon path deformation. While studying the numerical
consequences of the availability of the latter splitting kernel, we have also
taken the opportunity to address some criticalities of the approximations
which were present in the \HELL implementation; we propose a new approach
which stems from a slightly different definition of $\gamma_+^{\NLLp}$, and
which is seen to behave in a more stable manner especially w.r.t.~to changes
in the value of $\as$.

As a final by-product of the analytical work on $\gamma_s$ and $\gamma_{qg}$
summarised above we have succeeded in determining the finite $\ord(\ep^0)$
$qg$ Green function to all orders in $\gamma_s$, as well as its relationship
with the all-order $\ord(\ep^0)$ $gg$ Green function. While these results have
no immediate consequences and serve no purpose in the upcoming \HELL
implementation, they help elucidate the involved $\MSb$ structure emerging
from small-$x$ dynamics, and may pave the way for future improved treatments
of contributions beyond leading logarithmic accuracy.

\acknowledgments{
We are grateful to Fabio Maltoni, Simone Marzani and Pier Monni for several
  discussions during the course of this work, and to Stefano Forte for a
  critical reading of the manuscript.  The work of MB is supported by the
  Italian Ministry of University and Research (MUR) grant PRIN 2022SNA23K
  funded by the European Union -- Next Generation EU, Mission 4, Component 2,
  CUP I53D23001410006.  GS acknowledges financial support from the EU Horizon
  Europe research and innovation programme under the Marie-Sk\l{}odowska Curie
  Action HIPFLAPP, grant agreement No.~101149076.  }

\appendix
\section{New analytical all-order results for $\gamma_s$}
\label{sec:gammagg}

For the algebraic manipulations carried out in this appendix, we consider
eq.~\eqref{dualLO} expressed in term of a generic complex number $a$, i.e.\ we
study the function $\gamma_s(a)$ defined as the solution of the equation
\beq
1=a\,\chi\big(\gamma_s(a)\big).
\label{mastergg}
\eeq
in the complex plane. The physical case corresponds to setting $a=\bbas$.

\subsection{Perturbative coefficients}
\label{sec:gggPert}
In the context of a perturbative approach, eq.~\eqref{mastergg} must be
solved for $\gggf$ order by order in $a$. We employ the ansatz
\begin{align}
\gggf(a)&=\sum_{j=0}^\infty g_j a^{j+1}=
a\left(1+\hgggf(a)\right)\,,\;\;\;\;\;\;\;\;
g_0=1\,,
\label{gggex}
\\*
\hgggf(a)&=\sum_{j=1}^\infty g_j a^j\,,
\label{hgggex}
\end{align}
and introduce the auxiliary quantities
\begin{align}
\chi_k&=2\zeta_{k+1}\delta_{{\rm mod}(k,2),0}\,,
\\*
f_{-1}(z)&=\frac{1}{z}\,,
\\*
f_\Sigma(z)&=2\psi_0(1)-\psi_0(1+z)-\psi_0(1-z)
\equiv
\sum_{k=1}^\infty 2\zeta_{2k+1}z^{2k}\,,
\end{align}
whence
\beq
\chi(z)=f_{-1}(z)+f_\Sigma(z)\,,
\eeq
so that eq.~(\ref{mastergg}) becomes
\beq
1=f_{-1}\big(1+\hgggf(a)\big)+
af_\Sigma\big(\gggf(a)\big)\,.
\eeq
We now expand in series of $a$ the r.h.s.~of this equation, by exploiting
the Fa\`a di Bruno formula for the composite derivatives of single-variable
functions, and obtain the identity,
\beq
1=1+\sum_{p=1}^\infty a^p\Bigg\{\frac{1}{p!}\sum_{i=1}^p
(-)^i\Gamma(i+1)B_{p,i}(G)
+\frac{\stepf(p\ge 2)}{(p-1)!}
\sum_{i=1}^{p-1} \chi_i\,\Gamma(i+1)B_{p-1,i}(\hat{G})
\Bigg\},\phantom{aa}
\label{ident12}
\eeq
where by $B_{n,k}$ we have denoted the incomplete exponential Bell
polynomials, to be computed with the following arguments:
\begin{align}
G&=\left(0,0,g_3 3!,\ldots,g_n n!,\ldots\right)\,.
\label{Gvec2}
\\*
\hat{G}&=\left(g_0 1!,g_1 2!,\ldots,g_n (n+1)!,\ldots\right)
\label{hGvec}
\\*&=
\left(1,0,0,g_3 4!,\ldots,g_n (n+1)!,\ldots\right)\,.
\end{align}
Equation~(\ref{ident12}) is solved by the following coefficients:
\beq
g_p=\sum_{i=1}^p\frac{1}{(p-i+1)!}\,B_{p,i}(Z)\,,
\label{gpsol}
\eeq
where
\begin{align}
Z&=\Big(0,\,0,\,2\zeta_3\,3!,\,0,\,2\zeta_5\,5!,0,\ldots
2\zeta_{2n+1}\,(2n+1)!,\ldots\Big)
\label{Zvec}
\\*&=
\left(\left\{\left.
\frac{d^k \hchi(z)}{dz^k}\right|_{z=0}\right\}_{k=1}^\infty\right).
\label{Zvec2}
\end{align}
In eq.~\eqref{Zvec2} we have introduced the function
\beq
\hchi(z)\equiv zf_\Sigma(z)=-z\Big[\psi_0(1+z)+\psi_0(1-z)+2\gE\Big],
\label{Gfun}
\eeq
which is ubiquitous in our approach, and which is such that
\beq
1+\hchi(z)=z\chi(z)\,.
\label{FPggg0}
\eeq
The sought perturbative result for $\gggf(a)$ is
given by eqs.~\eqref{gggex}, \eqref{hgggex}, and~\eqref{gpsol},
and reads
\beq
\gggf(a)=a\left(1+
\sum_{k=1}^\infty\sum_{i=1}^k \frac{a^k}{(k-i+1)!}\,B_{k,i}(Z)\right).
\label{hgggex2}
\eeq
We are unable to find an explicit closed-form, non-integral expression for
the series which appears in eq.~(\ref{hgggex2}) and, for the reasons
explained later (see in particular appendix~\ref{sec:gggFP}), we doubt
that such a form exists. Conversely, after some long and tedious
manipulations, we have found an {\em implicit} closed-form, non-integral
expression for $\gggf$, which reads
\beq \label{fk0LNLres}
\log\frac{\gggf(a)}{a}=
-\log(2a\psi_{0+}(a))
-\left.\frac{2a\psi_{0+}(a)}{2a\psi_{0+}(a)-\delta}\,
\log\left[\frac{1}{\delta}+
R\left(\frac{2a\psi_{0+}(a)(1+2a\psi_{0+}(a))}
{2a\psi_{0+}(a)-\delta}\right)\right]
\right|^{[\delta^0]}\,,
\eeq
where
\begin{align}
\psi_{k+}(z)&=\half\Big(\psi_k(1+z)+\psi_k(1-z)\Big)-
\psi_k(1)\stepf(k\le 1)\,,
\label{psipdef}
\\
\psi_{k-}(z)&=\half\Big(\psi_k(1+z)-\psi_k(1-z)\Big)\,,
\label{psimdef}
\end{align}
and
\beq
R(z)=\frac{\psi_{0+}\left(a-\frac{az}{1+2a\psi_{0+}(a)}\right)}
{\psi_{0+}(a)}\,.
\eeq
The notation $^{[\delta^0]}$ in eq.~(\ref{fk0LNLres}) implies that only
the $\ord(\delta^0)$ contribution of the expression preceding it must
be kept, which is what prevents one from obtaining an explicit result
for $\gggf$. Perturbatively in $a$, eqs.~(\ref{hgggex2})
and~(\ref{fk0LNLres}) are fully equivalent, but the structure of the
series expansion of the latter is peculiar; we shall return to
this point in appendix~\ref{sec:gggFP}.

Before moving to the study of the analytical properties of $\gggf(a)$,
we note that it is also possible to write $a$ as a function of
$\gamma_s = \gggf(a)$.
By means of the vector introduced in eq.~(\ref{Zvec}) and owing to
eq.~(\ref{Zvec2}), we have that
\beq
\log\left(1+\hchi(z)\right)=
\sum_{k=1}^\infty\frac{z^k}{k!}
\sum_{j=1}^k(-)^{j+1}(j-1)!B_{k,j}(Z)\,,
\label{log1pG}
\eeq
whence
\begin{align}
a^m&=\left(\chi(\gamma_s)\right)^{-m}
=\left(\frac{\gamma_s}{1+\hchi(\gamma_s)}\right)^{m}
\\*&=
\gamma_s^m\left[1+\sum_{k=1}^\infty\frac{\gamma_s^k}{k!}
\sum_{j=1}^k\left.\frac{d^j}{dz^j}\frac{1}{(1+z)^m}\right|_{z=0}
B_{k,j}(Z)\right]
\\*&=
\gamma_s^m\left[1+\sum_{k=1}^\infty\frac{\gamma_s^k}{k!}\,
A_k^{(m)}\right],
\label{atom}
\end{align}
where we have defined:
\beqn
A_k^{(m)}&=&\sum_{j=1}^k\frac{(-)^j(m-1+j)!}{(m-1)!}B_{k,j}(Z)\,.
\label{Akmdef}
\eeqn
Eqs.~\eqref{atom}--\eqref{Akmdef} can be conveniently used to turn a series in
$a$ into a series in $\gamma_s = \gggf(a)$.

\subsection{Integral representations\label{sec:gggIR}}
In order to study the analytical properties of $\gggf(a)$
one possibility, lacking a closed-form expression for its perturbative
series of eq.~\eqref{hgggex2}, is that of employing an integral
representation. This, we obtain as follows\footnote{Since the two are
trivially related to one another by eq.~\eqref{gggex}, we often deal
with $\hgggf$ rather than with $\gggf$.}.

Firstly, we use a Hankel-path representation of the reciprocal
of the $\Gamma$ function,
\beq
\frac{1}{(n-1-j)!}=\frac{1}{\Gamma(n-j)}=
\mp\frac{i}{2\pi}\int_{H^{\pm}} d\omega\,e^{\pm\omega}(\pm\omega)^{-n+j}\,.
\label{ooGamhank2}
\eeq
The path $H^-$ ($H^+$) is counterclockwise (clockwise),
starts and ends at \mbox{$(+\infty\pm i\ep)$} (\mbox{$(-\infty\mp i\ep)$}),
and crosses the real axis at some negative (positive) value. The
representation is valid for any integer, real, and complex argument,
and the paths can be freely deformed, provided the branch cut of the
integrand is not crossed. With eq.~(\ref{ooGamhank2}), eq.~(\ref{hgggex2})
leads to
\beq
\hgggf(a)=
\mp\frac{i}{2\pi}\int_{H^{\pm}}
\frac{d\omega}{(\pm\omega)^2}\,e^{\pm\omega}
\sum_{k=1}^\infty\sum_{i=1}^k
\left(\frac{a}{\pm\omega}\right)^k\left(\pm\omega\right)^i\,B_{k,i}(Z).
\label{hgggex3}
\eeq
Next, one Borel-transforms the series whose summation index is $k$,
and thus obtains an integrand whose structure is that of the l.h.s.~of
the following equation\footnote{Bar the $k=0$ term, which is absent,
and which will be thus subtracted by hand. This is straightforward,
since such a term is equal to one ($B_{0,i}=\delta_{i0}$).}
\beq
\exp\left(u\sum_{j=1}^\infty x_j\frac{t^j}{j!}\right)=
\sum_{n=0}^\infty\sum_{k=0}^n B_{n,k}(x_1,\ldots x_n)\,\frac{t^n}{n!}\,u^k\,,
\label{BellYgen}
\eeq
namely of the generating function of the incomplete exponential
Bell polynomials. This allows one to resum the series, and to
obtain the exponent on the r.h.s.~of eq.~(\ref{BellYgen}). In turn,
the series which appears there is easily resummed since, thanks
to eq.~(\ref{Zvec2}), it is nothing but the series expansion of
the function $\hchi(x)$ at $x=0$ (see eq.~\eqref{Gfun}). Therefore,
one ends up with
\beq
\hgggf(a)=
\mp\frac{i}{2\pi}\int_0^\infty dt\,e^{-t}\int_{H^{\pm}}
\frac{d\omega}{(\pm\omega)^2}\,e^{\pm\omega}\left\{
\exp\left[\pm\omega\,\hchi\!\left(\frac{a t}{\pm\omega}\right)\right]-1\right\}.
\label{hgggex4}
\eeq
In an analogous way, with a procedure that we refrain from reporting
here owing to its being fairly technical but without any insights
physics-wise, we obtain the following, equivalent, form:
\beq \label{hgggex10}
\hgggf(a)=
\mp\frac{i}{2\pi}\,\hchi(a)\int_0^\infty dt\,e^{-t}\int_{H^{\pm}}
\frac{d\omega}{(\pm\omega)^2}\,e^{\pm\omega}\,
\exp\left[\frac{\pm\omega}{\chi(a)}\left(
\hchi\!\left(\frac{a t}{\pm\omega}\hchi(a)+a\right)-\hchi(a)\right)\right].
\eeq
The integral representations of eqs.~(\ref{hgggex4}) and~(\ref{hgggex10})
are unusual, in that they feature both a Borel and a Hankel integral.
Conceptually, this is not problematic, but it may result in numerical
instabilities. It is therefore interesting to further manipulate
those results analytically.

Firstly, one notes that eqs.~(\ref{hgggex4}) and~(\ref{hgggex10}) have
in fact the same functional form. This can be made explicit by observing
that
\beq
\mp\frac{i}{2\pi}\int_0^\infty dt\,e^{-t}\int_{H^{\pm}}
\frac{d\omega}{(\pm\omega)^2}\,e^{\pm\omega}=1\,.
\label{HBeq1}
\eeq
Thus, by adding and subtracting $\hchi (a)$, and by exploiting
eq.~(\ref{HBeq1}) to move the $-\hchi (a)$ contribution under the
integral sign, eq.~(\ref{hgggex10}) becomes
\beq \label{hgggex10b}
\hgggf(a)=\hchi (a)\Bigg\{1
\mp\frac{i}{2\pi}\int_0^\infty dt\,e^{-t}\int_{H^{\pm}}
\frac{d\omega}{(\pm\omega)^2}\,e^{\pm\omega}\,
\left(\exp\left[\frac{\pm\omega}{\hchi (a)}\left(
\hchi\Big(\frac{a t}{\pm\omega}\hchi (a)+a\Big)-\hchi (a)\right)\right]
-1\right)\Bigg\},
\eeq
whereby the similarity with eq.~(\ref{hgggex4}) is manifest,
with the identification
\beq
\hchi\!\left(\frac{a}{\pm\homega}\right)\;\;\longleftrightarrow\;\;
\tilde{G}\!\left(\frac{a}{\pm\homega}\right)\equiv\frac{1}{\hchi (a)}
\hchi\!\left(\frac{a}{\pm\homega}\hchi (a)+a\right)-1\,.
\label{Gtildef0}
\eeq
By means of the change of variable
\beq
\omega=t\,\homega
\eeq
which preserves the Hankel path for any given $t>0$, eq.~\eqref{hgggex4}
reads
\beq
\hgggf(a)=
\mp\frac{i}{2\pi}\int_{H^{\pm}}
\frac{d\homega}{(\pm\homega)^2}
I_B\!\left(\mp\homega\mp\homega \hchi\!\left(\frac{a}{\pm\homega}\right),
\mp\homega\right)
\label{hgggex4b}
\eeq
with
\beq
I_B(\alpha,\beta)=
\int_0^\infty \frac{dt}{t}\,e^{-t}\left(e^{-\alpha t}-e^{-\beta t}\right)\,,
\label{IBint}
\eeq
while eq.~\eqref{hgggex10b} becomes (see eq.~\eqref{Gtildef0})
\beq\label{hgggex10c}
\hgggf(a)=\hchi (a)\Bigg\{1
\mp\frac{i}{2\pi}\int_{H^{\pm}}
\frac{d\homega}{(\pm\homega)^2}
I_B\!\left(\mp\homega\mp\homega \tilde{G}\!\left(\frac{a}{\pm\homega}\right),
\mp\homega\right)\Bigg\}.
\eeq
With an explicit computation we obtain
\beq
I_B(\alpha,\beta)=
\log\frac{1+\beta}{1+\alpha}\,,\qquad
\Re(\alpha)>-1\,,\qquad\Re(\beta)>-1\,.
\label{IBintres}
\eeq
While the constraints on $\alpha$ and $\beta$ in eq.~\eqref{IBintres} are
essential for the Borel integral to converge, the result is analytical,
and can therefore be employed with any values of $\alpha$ and $\beta$.
We shall therefore ignore those constraints in what follows, but observe
that we could not have done that if the Borel integral were computed only
numerically --- this is not irrelevant, since such contraints are
easily violated when increasing the value of $a$.

\begin{figure}[t]
  \begin{center}
  \includegraphics[scale=1.0,width=0.48\textwidth]{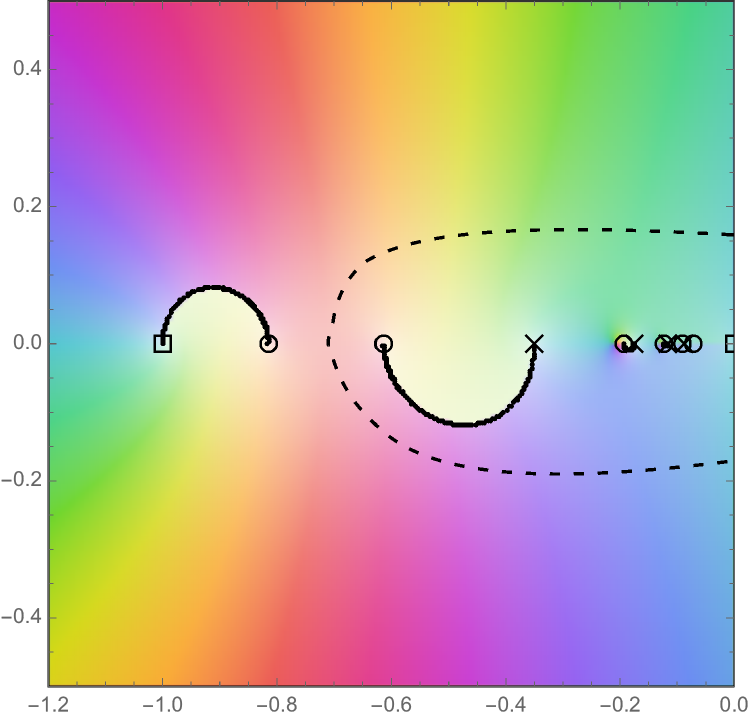}
~~
  \includegraphics[scale=1.0,width=0.48\textwidth]{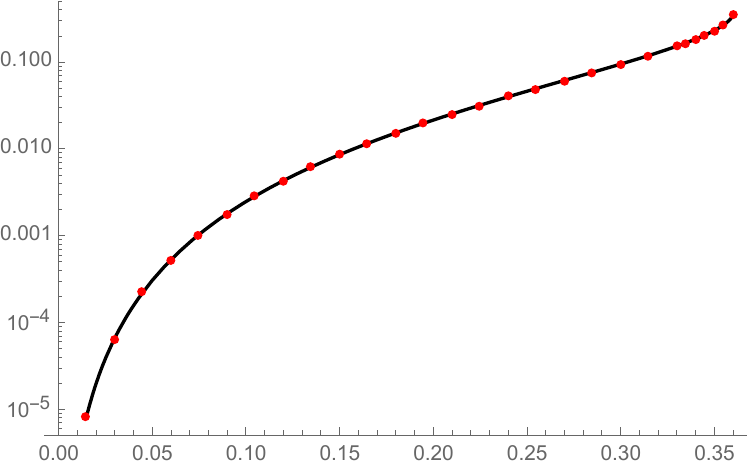}
  \caption{Integrand of eq.~(\ref{hgggex4c}) with $a=0.35$ (left-hand panel),
    and the result for $\hgggf(a)$ (right-hand panel). See the text for
    details. The plot on the left is obtained by means of the same
   function which has been employed in fig.~\ref{fig:hqg1s}.}
  \label{fig:a035}
  \end{center}
\end{figure}

In order to proceed in a definite way, let's start with eq.~(\ref{hgggex4b}),
and choose the lower signs there. Explicitly,
\beq
\hgggf(a)=
\frac{i}{2\pi}\int_{H^-}
\frac{d\homega}{\homega^2}\,
\log\frac{1+\homega}{1+\homega+\homega \hchi (-a/\homega)}\,.
\label{hgggex4c}
\eeq
We now have to bear in mind that the original reason for introducing
an Hankel integral was to represent the reciprocal of a factorial
(specifically, that in eq.~\eqref{hgggex2}). Therefore, the original
Hankel path could be chosen arbitrarily close to the real positive
axis\footnote{As an aside, one also observes that since one deals with
the reciprocal of $\Gamma$ functions with integer arguments, the Hankel
path can in fact be replaced by a closed path (counterclockwise, in the
case of $H^-$) around the origin, and again arbitrarily close to it.},
in view of the fact that each summand of the series that leads to
$\hgggf(a)$ has a pole at the origin\footnote{And a cut along the
positive real axis for non-integer $\Gamma$-function arguments,
which is not the case here.}. When the series is (partly) summed,
and one ends up with eq.~\eqref{hgggex4c}, (most of)
the poles move away from the origin. However, by assuming the series
to be uniformly convergent, they basically do so in a continuous way
with $a$, and one can therefore continuously deform the original
Hankel path to keep enclosing those poles. We point out that there may be
poles which do not fulfill this condition --- in other words, poles
that when $a\to 0$ do not move continuously towards the origin.
One example is given by the $\omega=-1$ pole which stems from the
numerator of the logarithm in eq.~(\ref{hgggex4c}). This makes sense,
since ultimately such a pole is associated with the $-1$ subtraction
term in eqs.~(\ref{hgggex4}) and~(\ref{hgggex10b}), which is always
present, and is independent of the value of $a$. The bottom line
is that, in situations analogous to the present one (where the Hankel
integral is associated with the reciprocal of factorials in a well-behaved
power series), we can formulate the following ansatz:
\begin{itemize}
\item The Hankel path must enclose all of the poles of the integrand which
tend continuously to the origin with decreasing values of the expansion
parameter of the series. This demands that all branch cuts to which those
poles belong be enclosed too, regardless of whether they also feature other
poles that do {\em not} fulfill the previous condition.
\end{itemize}
We note that there may be situations in which the path defined above
cannot be found; in such cases, the integral representation loses validity.
For the case at hand, we find that by following the prescription given
above the integrals are numerically very stable with an excellent
convergent speed. As an example, in the left panel of fig.~\ref{fig:a035}
we show as a dashed line the Hankel path that we have employed with $a=0.35$
--- the crosses, circles, and boxes are singularities of the integrand,
and the solid lines are branch cuts. Conversely, in the right panel of
the same figure we present the results obtained with eq.~(\ref{hgggex4c})
as solid red circles. These are superimposed over the values of
$\hgggf(a)$ (black line), obtained by inverting numerically
eq.~(\ref{mastergg}) (or, which is equivalent in this range of $a$,
by truncating the series of eq.~(\ref{hgggex2}) to its first sixty terms);
as one can see, we find an excellent agreement.

%%%%%%%%%%%%%%%%%%%%%%%%%%%%%%%%%%%%%%%%%%%%%%%%%%%%%%%%%%%%%%%%%%%
%%%%%%%%%%%%%%%%%%%%%%%%%%%%%%%%%%%%%%%%%%%%%%%%%%%%%%%%%%%%%%%%%%%

\begin{figure}[t]
  \begin{center}
  \includegraphics[scale=1.0,width=0.48\textwidth]{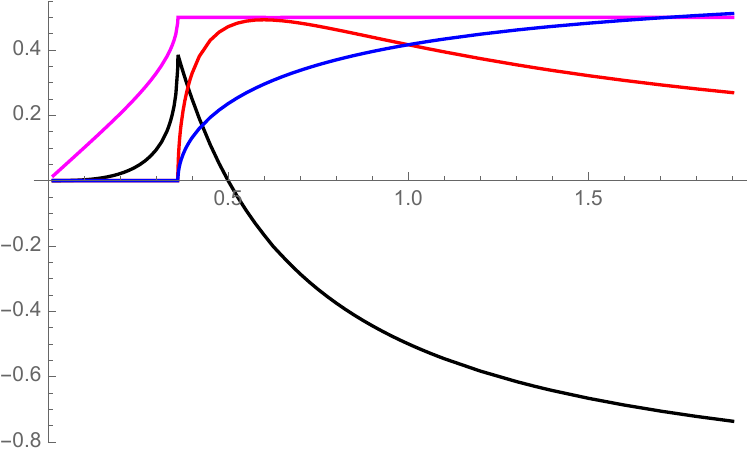}
~~
  \includegraphics[scale=1.0,width=0.48\textwidth]{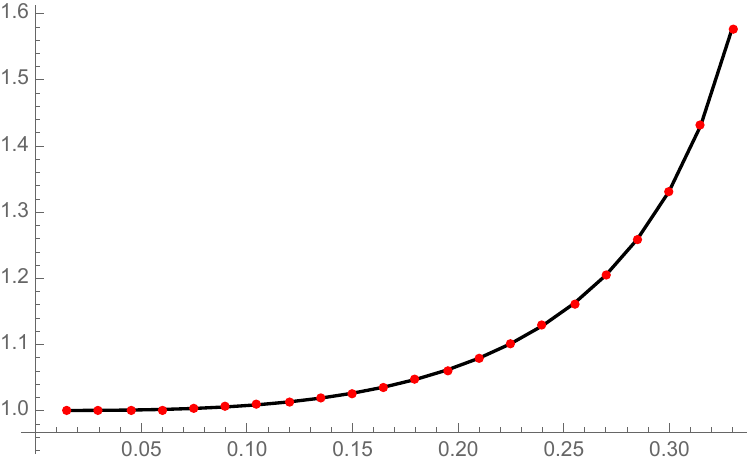}
  \caption{
Left-hand panel: integral-representation results for the real and imaginary
    parts of $\hgggf(a)$ (black and red) and of
    $\gggf(a)$ (magenta and blue), obtained with eq.~(\ref{hgggex4c}).
    Right-hand panel: ratio $\hgggf(a)/\hchi (a)$ computed with the results
    of the left-hand panel (red points), and with the integral representation
    of the ratio, eq.~(\ref{hgggex10c}).}
  \label{fig:ggg78}
  \end{center}
\end{figure}

While it does work in general, the integration-path prescription that has
led to the results shown in fig.~\ref{fig:a035} strictly depends on the
integrand: the simpler the latter, the more straightforward the procedure.
Therefore, it is convenient to push analytical simplifications as far
as one can. In the present case, an obvious simplification stems from
rewriting eq.~(\ref{hgggex4c}) as
\beq
\hgggf(a)=
\frac{i}{2\pi}\int_{H^-}
\frac{d\homega}{\homega^2}\,
\log(1+\homega)
-\frac{i}{2\pi}\int_{H^-}
\frac{d\homega}{\homega^2}\,
\log\Big(1+\homega+\homega \hchi (-a/\homega)\Big).
\label{hgggex4d}
\eeq
By choosing for the leftmost integral a contour around the origin that
leaves out the $(-1,0)$ singularity and the branch cut at $\homega<-1$
(which is in keeping with our prescription), one obtains
\beq
\frac{i}{2\pi}\int_{H^-}
\frac{d\homega}{\homega^2}\,
\log(1+\homega)=-1\,,
\eeq
whence, by employing eq.~(\ref{gggex}), one arrives at
\beq
\gggf(a)=
-a\,\frac{i}{2\pi}\int_{H^-}
\frac{d\homega}{\homega^2}\,
\log\Big(1+\homega+\homega \hchi (-a/\homega)\Big).
\label{gggex4d}
\eeq
Equation~(\ref{gggex4d}) can be expressed in two alternative ways which
may be more convenient for numerical evaluations and/or for the study
of its analytical behaviour. With a change of variable, and by exploiting
the identity in eq.~(\ref{FPggg0}) we obtain:
\beqn
\gggf(a)&=&
-\frac{i}{2\pi}\int_{H^-}
\frac{dz}{z^2}\,
\log\Big(1+az+az\, \hchi (-1/z)\Big)
\label{gggex4d2}
\\*&=&
-\frac{i}{2\pi}\int_{H^-}
\frac{dz}{z^2}\,
\log\Big(1-a\,\chi(-1/z)\Big).
\label{gggex4d3}
\eeqn
In this way, having succeeded in rendering the functions $\hchi$ and $\chi$
independent of $a$, one can then easily compute the integral which is
relevant to the definition of $\hh$, eq.~\eqref{hhdef}. Explicitly,
\beq
\hh(\gamma_s(\bar a)) = \int_0^{\bar{a}}\frac{da}{a}\,\gggf(a)=
\frac{i}{2\pi}\int_{H^-}
\frac{dz}{z^2}\,
{\rm Li}_2\Big(\bar{a}\,\chi(-1/z)\Big).
\label{gggint}
\eeq
In conclusion, we remind the reader what we have observed at the
beginning of this section, namely that the two representations given by
eqs.~(\ref{hgggex4}) and~(\ref{hgggex10}) are functionally identical.
Therefore, the manipulations carried out here starting from the former
equation can also be applied to the latter one.
We have indeed done so, as a further check of the mutual consistency
of our results. One interesting aspect of this is presented in
fig.~\ref{fig:ggg78}. The left panel there shows both the real and the
imaginary parts of $\hgggf(a)$ and of $\gggf(a)$,
obtained with eq.~(\ref{hgggex4c}). The results for the former are
then employed to compute the ratio $\hgggf(a)/\hchi (a)$, with
$\hchi (a)$ in the denominator evaluated analytically --- this operation
results in the red circles displayed in the right panel of the figure.
There, they are compared with the values of the same ratio, this time
obtained directly from eq.~(\ref{hgggex10c}), i.e.~with the alternative
integral representation of $\hgggf$. We see that there is
a perfect agreement between the two types of predictions.

%%%%%%%%%%%%%%%%%%%%%%%%%%%%%%%%%%%%%%%%%%%%%%%%%%%%%%%%%%%%%%%%%%%
%%%%%%%%%%%%%%%%%%%%%%%%%%%%%%%%%%%%%%%%%%%%%%%%%%%%%%%%%%%%%%%%%%%

\subsection{Fixed-point representation\label{sec:gggFP}}
Fig.~\ref{fig:ggg78}, and specifically its right panel, helps us make
the following observation: $\hchi (a)$ is a reasonable approximation for
$\hgggf(a)$, precisely as eq.~(\ref{hgggex10c}) suggests. In fact,
the structure of that equation, and of eq.~(\ref{Gtildef0}) in particular,
implies that the following sequence is of interest:
\beq
\big\{\gggf^{(n)}(a)\big\}_{n=0}^\infty\,,\qquad
\gggf^{(0)}(a)=a\,,
\label{gseq}
\eeq
where the elements $n\ge 1$ are defined recursively:
\begin{align}
\gggf^{(n+1)}(a)&=a\big[1+\hchi\big(\gggf^{(n)}(a)\big)\big]
\label{gseqrec}
\\*&=
a\big[1+\hchi\big(a\big(1+\hchi\big(a\big(1+\hchi\big(\ldots\big)
\big)\big)\big)\big)\big].
\label{gseqrec2}
\end{align}
Equation~(\ref{gseqrec}) is in the form of a fixed-point sequence;
namely, if it converges to $\gggf^{(\infty)}(a)$, this is
a fixed point of the function $\Phi_a(x)$, with
\beq
\gggf^{(\infty)}(a)=\Phi_a\left(\gggf^{(\infty)}(a)\right)\,,
\;\;\;\;\;\;\;\;\;
\Phi_a(x)=a\big(1+\hchi (x)\big)\equiv a\,x\chi(x)\,.
\label{FPggg}
\eeq
having exploited the identity of eq.~(\ref{FPggg0}).
As the notation suggests, in terms of the fixed-point problem, $a$ is
an external parameter. If the convergence point $\gggf^{(\infty)}(a)$
exists, it coincides with the sought solution $\gggf(a)$,
as one can see from the fact that eq.~(\ref{FPggg}) becomes an
identity upon exploiting eq.~(\ref{mastergg}).
%%%%%%%%%%%%%%%%%%%%%%%%%%%%%%%%%%%%%%%%%%%%%%%%%%%%%%%%%%%%%%%%%%%
\begin{figure}[t]
  \begin{center}
  \includegraphics[scale=1.0,width=0.48\textwidth]{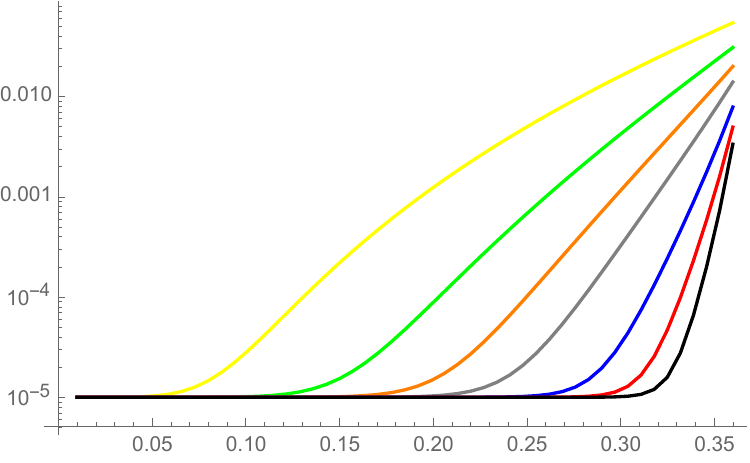}
  \caption{Checks of eq.~(\ref{mastergg}) with the elements of the sequence
    of eq.~(\ref{gseq}).}
  \label{fig:ggg9}
  \end{center}
\end{figure}
%%%%%%%%%%%%%%%%%%%%%%%%%%%%%%%%%%%%%%%%%%%%%%%%%%%%%%%%%%%%%%%%%%%
There is therefore nothing particularly new here, except for the fixed-point
observation, and for the related fact that the sequence of eq.~(\ref{gseq})
gives one another approximant of $\gggf(a)$. The quality of such an
approximant is shown in fig.~\ref{fig:ggg9}, where we plot\footnote{The
$10^{-5}$ factor is there in order for us to use a logarithmic scale
in the plot.}
\beq
a\,\chi\big(\gggf^{(n)}(a)\big)-1+10^{-5}
\eeq
for $n$ equal to $1$ (yellow), $2$ (green), $3$ (orange), $4$ (gray),
$6$ (blue), $8$ (red), and $10$ (black). Therefore, at least for values
not too close to $a=1/\log(16)$, the convergence appears to be much faster
than that of the original series in eq.~(\ref{gggex}).

We hasten to stress that the positive feature of the fixed-point
solution we have just mentioned applies only to real values of $a$.
In the complex plane (and also for larger real values of $a$) the
situation is very different, and the fixed-point recursive sequence
becomes very unstable. Indeed, a modern trend in mathematics defines
fractal structures by means of the speed of convergence (or lack thereof)
of fixed-point sequences associated with given functions.
%%%%%%%%%%%%%%%%%%%%%%%%%%%%%%%%%%%%%%%%%%%%%%%%%%%%%%%%%%%%%%%%%%%
\begin{figure}[t]
  \begin{center}
  \includegraphics[scale=1.0,width=0.48\textwidth]{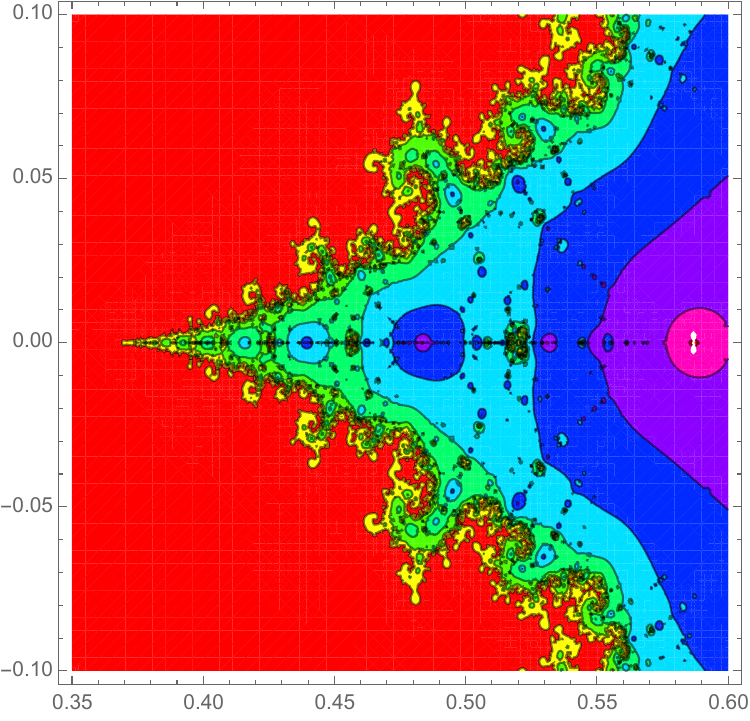}
~~
  \includegraphics[scale=1.0,width=0.48\textwidth]{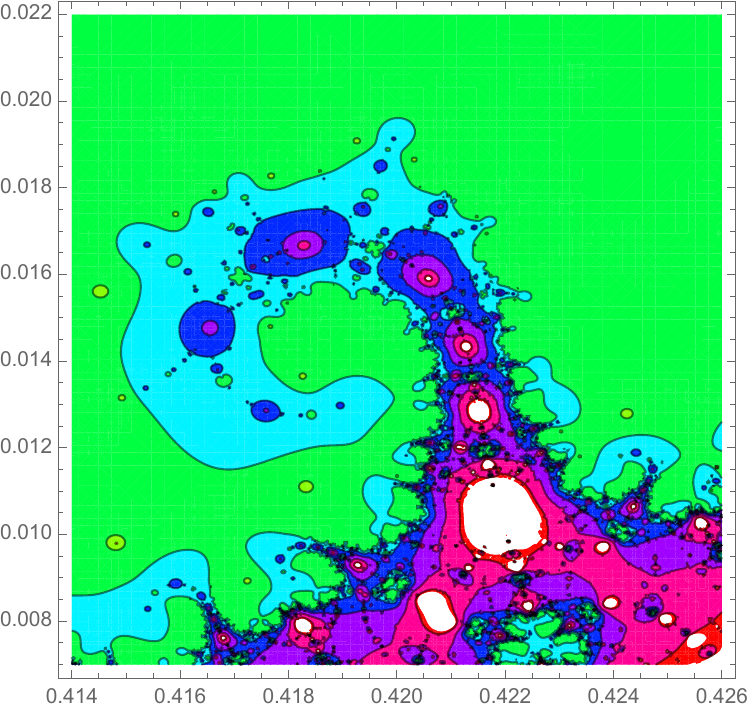}
  \caption{Fractal structures emerging from the recursive fixed-point sequence
    of eq.~(\ref{gseqrec}). The plot on the right-hand panel is a zoomed-in
    version of that on the left-hand panel, with different colour choices.
    Both plots have been obtained by using the {\tt Mathematica}~\cite{Mathematica}
    functions {\tt ContourPlot} and {\tt FixedPoint}. The colours
    are controlled by {\tt Mathematica}, and the same remarks made
    in fig.~\ref{fig:hqg1s} apply here.}
  \label{fig:fractal}
  \end{center}
\end{figure}
%%%%%%%%%%%%%%%%%%%%%%%%%%%%%%%%%%%%%%%%%%%%%%%%%%%%%%%%%%%%%%%%%%%
We are indeed able to obtain fractal-type results from
eq.~(\ref{gseqrec}), as is shown in fig.~\ref{fig:fractal}, with
the uniformly-coloured regions associated with fastly-convergent
and stable solutions; unstable, non-converging solution make up the
rest of the complex plane.

We conclude this section with the following heuristic observation:
the structure of the perturbative expansion of eq.~(\ref{gseqrec2})
is analogous to that of eq.~(\ref{fk0LNLres}) (in terms of the functions
$\hchi$ and $R$, respectively; neither expansion is shown explicitly here).
We interpret this as a further evidence that all non-integral-representation
results for $\gggf(a)$ can be cast as recursive solutions, with the features
that we have outlined before.

\section{Solving equations for the resummation of $\gamma_{qg}$}
\label{sec:appqg}

In this appendix, we report the results for some of the quantities of
interest in the determination of $h_{qg}$ and of the green function $\G_{qg}$.

\subsection{Determining the function $\hh$}\label{sec:hhexpr}

Here we study in more detail the function $\hh$ introduced in
eq.~\eqref{hhdef}, which is used in the definition of the transition function
$\Gamma_{gg}$, eq.~\eqref{Gggdef}.

We start by performing the change of integration variable $a=1/\chi(y)$,
so that the function $\hh(\gamz)$ eq.~\eqref{hhdef} becomes
\begin{align}\label{eq:hhgs}
  \hh(\gamz)
  &\equiv \int_0^{1/\chi(\gamz)} da\,\frac{\gggf(a)}{a}
  = -\int_0^\gamz dy\,y \frac{\chi^\prime(y)}{\chi(y)}
\end{align}
where we have used eq.~\eqref{mastergg} and we have exploited
\beq
\frac{da}{dy}=-\frac{\chi^\prime(y)}{\chi^2(y)}.
\label{masterggder}
\eeq
As $\gamz$ is in general a complex variable, this integral has to be
performed in the complex plane by choosing an integration path.
To simplify this aspect, we further change integration variable as $y=w\gamz$,
so that eq.~\eqref{eq:hhgs} becomes
\beq\label{eq:hhv1}
\hh(\gamz) = -\gamz^2 \int_0^1 dw\,\frac{w\chi^\prime(w\gamz)}{\chi(w\gamz)},
\eeq
where now the integration variable $w$ is real.
We then perform the following manipulations:
\begin{align}\label{eq:hhv2}
  \hh(\gamz)
  &= -\gamz \int_0^1 dw\,w\frac{d}{dw}\log\chi(w\gamz) \nonumber\\
  &= -\gamz \log\chi(\gamz) + \gamz \int_0^1 dw\,\log\chi(w\gamz) \nonumber\\
  &= -\gamz \log\chi(\gamz) + \gamz \int_0^1 dw\,\log\frac{1+\hchi(w\gamz)}{w\gamz} \nonumber\\
  &= \gamz -\gamz \log\[1+\hchi(\gamz)\] + \gamz \int_0^1 dw\,\log\[1+\hchi(w\gamz)\]
\end{align}
where we have carried out an integration by parts in the second step and we
have made use in the last two steps of the function $\hchi$ introduced in
eq.~\eqref{Gfun}, which is related to the LO BFKL kernel function $\chi$ as is
shown in eq.~\eqref{FPggg0}.  We observe that the integral representations in
eq.~\eqref{eq:hhv1} and eq.~\eqref{eq:hhv2} are suitable for numerical
implementation for generic complex values of the argument $\gamz$, and we
indeed use them in the \HELL code.  Unfortunately we did not manage to compute
the integrals analytically in terms of known functions.

By exploiting eq.~\eqref{log1pG} to rewrite the basic quantity which appears in
eq.~\eqref{eq:hhv2}, we have
\beq\label{hhvsgampXX}
  \hh(\gamz) = \sum_{k=0}^\infty q_k \gamz^{k+1}
  = \gamz-\frac{3}{2}\zeta_3\gamz^4-\frac{5}{3}\zeta_5\gamz^6
  +\frac{12}{7}\zeta_3^2\gamz^7-\frac{7}{4}\zeta_7\gamz^8
  +\frac{32}{9}\zeta_3\zeta_5\gamz^9+\ldots\,,
\eeq
where the form of the coefficients is
\beq
q_0 = 1\,,\quad
q_k = \frac{k}{(k+1)!}\,\sum_{j=1}^k(-)^j (j-1)!B_{k,j}(Z)\,,
\label{cffomvsgam}
\eeq
where $Z$ has been defined in eq.~\eqref{Zvec}.
Furthermore, by using the Fa\`a di Bruno formula, it is possible to write any
power of $\hh(\gamz)$ as a linear combination of powers of $\gamz$:
\beq
\hh^n(\gamz)=
\sum_{k=n}^\infty\frac{n!}{k!}\,
B_{k,n}\left(\Big\{\!l!q_{l-1}\Big\}_{l=1}^{k-n+1}\right)\,\gamz^k\,.
\label{hhton}
\eeq
The results above can be used to obtain an explicit expression for $\gamz$ in
terms of $\hh(\gamz)$.
Indeed, by exploiting known results for the reversion of series, we have
\beq
\gamz = \sum_{k=0}^\infty p_k \hh^{k+1}(\gamz) =
\hh+\frac{3}{2}\zeta_3\hh^4+\frac{5}{3}\zeta_5\hh^6
+\frac{51}{7}\zeta_3^2\hh^7+\frac{7}{4}\zeta_7\hh^8
+\frac{193}{9}\zeta_3\zeta_5\hh^9+\ldots\,.
\label{gamvshh1XX}
\eeq
The closed form of the coefficients is
\beq
p_0 = 1\,,\quad
p_k = \frac{1}{(k+1)!}\,
\sum_{j=1}^{k} (-1)^j \frac{(k+j)!}{k!}
B_{k,j}\left(\Big\{\! l! q_{l} \Big\}_{l=1}^{k-j+1}\right)\,,
\eeq
and the generic power of $z$ reads thus
\beq
\gamz^n =
\sum_{k=n}^\infty\frac{n!}{k!}\,
B_{k,n}\left(\Big\{\! \;l!p_{l-1}\Big\}_{l=1}^{k-n+1}\right)\,\hh^k(\gamz)\,.
\label{gamvshh2XX}
\eeq

As is discussed immediately after eq.~\eqref{eq:PqgNLL-ktfact}, the
most straightforward way to obtain the resummed $qg$ splitting kernel
is through the Mellin inversion of its anomalous-dimension counterpart. In
turn, owing to eq.~\eqref{hqgres10}, this operation requires a complex-plane
line integration that involves the function $\hh(z)$. This is problematic,
owing to the behaviour of $\hh(z)$ at large $\abs{z}$. In order to
see this, we choose for simplicity the following integration path,
which may be used to carry out the inverse Mellin transform:
\beq
s=c+\rho\left(\cos\theta +i\sin\theta\right)=
c+\rho\,e^{i\theta}\,.
\label{spath}
\eeq
We limit ourselves to considering it in the upper half-plane of
the complex domain, i.e.~for
\beq
c>0\,,\qquad
\rho\ge 0\,,\qquad
0<\theta\le \pi\,,
\label{URquad}
\eeq
since the arguments that follow can be repeated essentially unchanged in
the case of negative imaginary parts. A direct calculation leads us to
the following asymptotic behaviour:
\beq
\hh(s)\Big|_{s=c+\rho\,e^{i\theta}}\stackrel{\rho\to\infty}{\longrightarrow}
-\half\log\left(\gE-i\theta+\log s\right)
-(s-1)\sum_{n=1}^{n_{\rm max}}\frac{e^{i\hth}\Gamma\left(n,i\hth\right)}
{(\gE-i\theta+\log s)^n}
-(f_R+if_I)\,,
\label{sFsasy}
\eeq
where are $f_R$, $f_I$ two constants (whose forms are irrelevant here),
$n_{\rm max}$ is an integer parameter, and we have defined
\beq
\hth=\frac{\pi}{2}-\theta\,.
\label{hthdef}
\eeq
While the asymptotic behaviour of the inverse Mellin transform of the
exponent of eq.~(\ref{sFsasy}) (times a numerical prefactor equal to either
$2$ or $2/3$, per eq.~(\ref{hqgres10})), can be computed analytically, in
itself this is not particularly useful since, even for large $\abs{s}$,
the agreement between the value of $\hh(s)$ computed with eq.~(\ref{sFsasy})
and the one obtained numerically is poor, unless $n_{\rm max}$ is relatively
large ($\gtrsim 8$). Given that the sought configuration-space result is
then the convolution of $n_{\rm max}$ individual inverse-Mellin transforms,
it is clear that one ends up with a quantity which does not behave
well numerically. A possible solution stems from exploiting the identity
\beq
-\sum_{n=1}^\infty\frac{1}{\log^n\rhob}\,
e^{i\hth}\Gamma\left(n,i\hth\right)
\;\;\longrightarrow\;\;
-\frac{e^{i\hth}}{\rhob}\left(\Eof(\log\rhob-i\hth)+i\pi\right)\,,
\;\;\;\;\;\;
\rhob\;=\;\rho\,e^{\gE}\,,
\label{E1replcmp}
\eeq
by means of which one arrives at the following alternative expression
\begin{align}
\hh(s)\Big|_{s=c+\rho\,e^{i\theta}}\stackrel{\rho\to\infty}{\longrightarrow}&
-\half\log\left(\gE-i\theta+\log s\right)
\nonumber \\*&
-e^{-\gE}e^{i(\hth+\theta)}
\left[\Eof\Big(\log s +\gE-i(\hth+\theta)\Big)+i\pi\right]
-(f_R+if_I)
\nonumber \\*
\stackrel{\rho\to\infty}{\longrightarrow}&
-\half\log\Big(\gE-i{\rm Arg}(s)+\log s\Big)
-i e^{-\gE}
\Eof\Big(\log s +\gE-i\frac{\pi}{2}\Big)\,.
\label{sFsasyE1}
\end{align}
Here, $\Eof$ denotes the exponential integral defined by\footnote{Note that this function concides with the {\tt ExpIntegralEi} module in Mathematica.}
\beq
\Eof(z)=-\int_{-z}^\infty dt\,\frac{e^{-t}}{t}
=
-i\pi\floor*{\frac{\pi+{\rm Arg}(z)}{2\pi}}+\gE+\log z
+\sum_{k=1}^\infty\frac{z^k}{kk!}\,.
\label{Eofdef}
\eeq
Unfortunately, this manipulation trades one problem for another:
eq.~(\ref{sFsasyE1}) is a fairly good approximation of the exact
$\hh(s)$, but the inverse Mellin transform of its (weighted) exponential
turns out to be very difficult to compute analytically\footnote{More
precisely: such a transform is the sum of two distributions, one of
which is a $\delta(1-\xi)$ with an easy-to-compute coefficient. The
other contribution is a generalised plus distribution for which we
were unable to obtain a closed form (among other things, it requires
a double summation of $\log(1-\xi)$ terms, since the nominally subleading
contributions have coefficients which scale in a manner that (over)compensates
the subleadingness --- see also sect.~\ref{sec:subL}).}.

\subsection{Coefficients of the bare quark Green function
$\hat\G_{qg}^{(0)}$}\label{sec:G0qg}

We now compute the bare quark Green function $\hat\G_{qg}^{(0)}$ as is
introduced in eq.~\eqref{Gqgfact}, after having factored out some constant
terms as in eq.~\eqref{Gqg0hatdef}. In particular, we are interested in
the expansions in series of $\ep$ of the coefficients of the
$\ord(\as^{k+1})$ terms. The starting point is eq.~(4.10) of
ref.~\cite{Catani:1994sq}, which we recast as\footnote
{Henceforth, without loss of generality, we set $\mu=Q$.}
\begin{align}
  \hat\G_{qg}^{(0)}
  &=\frac{\bbas}{\ep}
    \left(\rho_0(\ep)+\sum_{k=1}^\infty \frac{\rho_k(\ep)}{\ep^k}
    \bbas^k\right)
    \label{Gzqgdef}
  \\*&\equiv
       \frac{\bbas}{\ep}
       \sum_{k=0}^\infty\sum_{i=0}^\infty \frac{\rho_{k,i}}{\ep^{k-i}}
       \bbas^k\,,
       \label{Gqgzex}
\end{align}
where we have introduced the shorthand notation
\beq
\rho_{k,i}=\frac{1}{i!}\,\frac{d^i\rho_k}{d\ep^i}(0)\,.
\label{rhokidef}
\eeq
The coefficients $\rho_k$ are defined as follows:
\begin{align}
\rho_0(\ep)&=\frac{e^{\ep\psi_0(1)}}{\Gamma(1+\ep)}\,
\frac{3(4+\ep)}{2(2+\ep)(3+\ep)}\,,
\label{rho0def}
\\
\rho_k(\ep)&=\rho_0(\ep)\left(\frac{e^{\ep\psi_0(1)}}{\Gamma(1+\ep)}\right)^k
\frac{\hat{d}_k(\ep)}{k}\,,
\label{rhokdef}
\end{align}
where (note that $\cc_1(\ep)=1$):
\begin{align}
\hat{d}_k(\ep)&=\frac{1}{k+1}\,
\frac{(4+(1-3k)\ep)\prod_{i=1}^3(i+\ep)}
{(4+\ep)\prod_{j=1}^3(j+(1-k)\ep)}
\nonumber\\*&\quad\times\;
\frac{\Gamma(1+\ep)^2\Gamma(1-k\ep)\Gamma(1+k\ep)}
{\Gamma(1+(1-k)\ep)\Gamma(1+(1+k)\ep)}\,\hat{c}_k(\ep)\,,
\label{hatddef}
\\
\hat{c}_k(\ep)&=
\hat{c}_{k-1}(\ep)\hat{I}_{k-1}(\ep)
\equiv
\hat{c}_{k-1}(\ep)
\left(\hat{I}_{k-1}^{(A)}(\ep)-\hat{I}^{(B)}(\ep)\right)\,,
\label{hatcdef}
\\
\hat{I}_{k}^{(A)}(\ep)&=I_F(\ep)\,\hat{J}_{k}^{(A)}(\ep)\equiv
I_F(\ep)\,\frac{\Gamma(1+2\ep)\Gamma(k\ep)\Gamma(1-k\ep)}
{\Gamma((1+k)\ep)\Gamma(1+(1-k)\ep)}\,,
\label{hatIAdef}
\\
\hat{I}^{(B)}(\ep)&=I_F(\ep)\,\hat{J}^{(B)}(\ep)\equiv
I_F(\ep)\,\Gamma(1+\ep)\Gamma(1-\ep)\,,
\label{hatIBdef}
\\
I_F(\ep)&=\frac{\Gamma(1+\ep)^2}{\Gamma(1+2\ep)}\,.
\label{IFdef}
\end{align}
The recursive definition in eq.~(\ref{hatcdef}) can be made explicit,
\beq
\hat{c}_k(\ep)=I_F^{k-1}(\ep)\sum_{q=0}^{k-1}(-)^{k-1-q}
\left(\hat{J}^{(B)}(\ep)\right)^{k-1-q}
\sum_{\{j_1\ldots j_q\}}\hat{J}_{j_1}^{(A)}(\ep)\ldots
\hat{J}_{j_q}^{(A)}(\ep)\,.
\label{chatex}
\eeq
Here
\beq
\big\{j_1\ldots j_q\big\}\;\subseteq\;\big\{1,\ldots k-1\big\}
\label{qset}
\eeq
is a $q$-element unordered set, and the rightmost sum in eq.~(\ref{chatex})
is understood to run over all possible such sets --- by convention,
if $q=0$ the argument of the sum is set equal to one.

By using eq.~(\ref{chatex}), the coefficients $\rho_k(\ep)$ will be given
as sums of terms, each of which has a multiplicative structure. Denoting
by $f(\ep)$ any such term, we shall use the identity\footnote{Clearly,
no manipulation is necessary for terms that are already in an exponential
form, of which there is only one here, namely $\exp(\ep\psi_0(1))$.}
\beq
f(\ep)=\exp\big(\log f(\ep)\big)\,,
\eeq
so that:
\beq
\frac{d^n f}{d\ep^n}(0)=f(0)\,B_n\left(\frac{d\log f}{d\ep}(0),
\ldots,\frac{d^n\log f}{d\ep^n}(0)\right),
\label{tmp1}
\eeq
with $B_n$ the complete exponential Bell polynomials. Thus, each element
of the product that defines $f(\ep)$ will correspond to a set of inputs
to the Bell polynomials, with the set relevant to $f(\ep)$ the sum of
such sets. The $\ep$-dependent terms in eq.~\eqref{rho0def}
and~\eqref{rhokdef}, bar $\hat{d}_k(\ep)$, lead us to define
\begin{align}
Z_{56}(k)&=\Big(\ldots ,\,
(-)^{m-1}(m-1)!(1/4^m-1/3^m-1/2^m)\,,
\ldots\Big)_{m=1}^n
\nonumber\\*&\quad+
(k+1)\Big(\ldots ,\,
(\delta_{1m}-1)\psi_{m-1}(1)\,,
\ldots\Big)_{m=1}^n.
\label{Z56vec}
\end{align}
The $\ep$-dependent terms in eq.~\eqref{hatddef}, bar $\hat{c}_k(\ep)$,
gives instead:
\begin{align}\label{Z34vec}
  Z_{34}(k)
  &=\Big(\ldots ,\, (-)^{m-1}(m-1)!( (1-3k)^m-1 )/4^m, \ldots\Big)_{m=1}^n \nonumber \\*
  &\quad -\Big(\ldots ,\, (-)^{m-1}(m-1)!( 1+1/2^m+1/3^m )( (1-k)^m-1 ), \ldots\Big)_{m=1}^n \nonumber\\*
  &\quad + \Big(\ldots ,\, (2+(1+(-1)^m)k^m-(1+k)^m-(1-k)^m)\psi_{m-1}(1)\,, \ldots\Big)_{m=1}^n.
\end{align}
Equations~\eqref{hatIAdef}--\eqref{IFdef} give
\begin{align}
  Z_A(k) &= \Big(\ldots ,\, (2^m+(1+(-1)^m)k^m-(1-k)^m-(1+k)^m)\psi_{m-1}(1), \ldots\Big)_{m=1}^n, \label{ZAvec} \\
  Z_B &= \Big(\ldots ,\,(1+(-1)^m)\psi_{m-1}(1), \ldots\Big)_{m=1}^n, \label{ZBvec} \\
  Z_F &= \Big(\ldots ,\, (2-2^m)\psi_{m-1}(1), \ldots\Big)_{m=1}^n, \label{ZFvec}
\end{align}
respectively. With these, eq.~\eqref{tmp1} leads to the sought
results\footnote{Which are actually valid also for $n=0$.\label{ft:rhok}}
\begin{align}
\frac{d^n\rho_0}{d\ep^n}(0)&=B_n\Big(Z_{56}(0)\Big)\,,
\label{rho0res}
\\
\frac{d^n\rho_1}{d\ep^n}(0)&=\half\,
B_n\Big(Z_{34}(1)+Z_{56}(1)\Big)\,,
\label{rho1res}
\\
\frac{d^n\rho_k}{d\ep^n}(0)&=
\frac{1}{k(1+k)}
\sum_{q=0}^{k-1}(-)^{k-1-q}
\sum_{\{j_1\ldots j_q\}}\hat{J}_{j_1}^{(A)}(0)\ldots
\hat{J}_{j_q}^{(A)}(0)
\label{rhokres}
\\*&\quad\times
B_n\Big(Z_{34}(k)+Z_{56}(k)+(k-1)Z_F+(k-1-q)Z_B
+Z_A(j_1)+\ldots +Z_A(j_q)\Big)\,.\nonumber
\end{align}
Note that
\beq
\hat{J}_{j}^{(A)}(0)=\frac{1+j}{j}\,,
\eeq
and that the arguments of the Bell polynomials in
eqs.~\eqref{rho0res}--\eqref{rhokres} do not feature $\psi_0(1)$,
i.e.~the results are independent of the Euler-Mascheroni constant.
In these arguments, the only irrational numbers are $\psi_i(1)$
with $i\ge 1$, i.e.~Riemann's $\zeta_{i+1}$ of either parity.
We expect significant cancellations in the sums on the right-hand side of
eq.~\eqref{rhokres}; bear in mind that this structure stems from
the difference on the rightmost side in eq.~\eqref{hatcdef}, which
leads to a proliferation of terms upon recursion.

An alternative result can be obtained by starting from the observation
that the $n^{\rm th}$ coefficient of the series expansion in $\ep$ of the
product of $m+1$ functions $f_q(\ep)$ reads
\beq
\frac{1}{n!}\,\frac{d^n f_0\ldots f_m}{d\ep^n}(0)=
\frac{1}{n!}\!\!\sum_{\stackrel{j_0\ldots j_m=0}{j_0+\ldots j_m=n}}^n
\!\!\binomial{n}{j_0\ldots j_m}
\frac{d^{j_0} f_0}{d\ep^{j_0}}(0)\ldots
\frac{d^{j_m} f_m}{d\ep^{j_m}}(0)\,,
\label{df0fm}
\eeq
with the usual definition of the multinomial coefficient:
\beq
\binomial{n}{j_0\ldots j_m}=
\frac{n!}{j_0!\ldots j_m!}=
\binomial{j_0}{j_0}\binomial{j_0+j_1}{j_1}\ldots
\binomial{j_0+\ldots +j_m}{j_m}\,.
\label{multcff}
\eeq
By applying eq.~(\ref{df0fm}) to the computation of the derivatives
of $\rho_k(\ep)$ ($k\ge 2$), with $m=k-1$ and
\begin{align}
f_0&=\frac{\rho_k(\ep)}{\hat{c}_k(\ep)}\,I_F^{k-1}(\ep)\,,
\\*
f_q&=\hat{J}_{q}^{(A)}(\ep)-\hat{J}^{(B)}(\ep)\,,\qquad
1\le q\le k-1\,,
\end{align}
and by bearing in mind the definitions in eqs.~(\ref{Z56vec})--(\ref{ZFvec}),
we arrive at:
\begin{align}
  \frac{d^n\rho_k}{d\ep^n}(0)
  &= \frac{1}{k(1+k)} \sum_{\stackrel{j_0\ldots j_{k-1}=0}{j_0+\ldots j_{k-1}=n}}^n
    \binomial{n}{j_0\ldots j_{k-1}}
    B_{j_0}\big(Z_{34}(k)+Z_{56}(k)+(k-1)Z_F\big)
\nonumber\\*&\qquad\qquad\qquad\qquad\qquad\times
\prod_{q=1}^{k-1} \Big[\hat{J}_{q}^{(A)}(0)B_{j_q}\big(Z_A(q)\big)-B_{j_q}\big(Z_B\big)\Big]\,.
\label{rhokres2}
\end{align}
The equivalence of eqs.~\eqref{rhokres} and~\eqref{rhokres2} can be shown by
using the rightmost expression of eq.~\eqref{multcff} in eq.~\eqref{rhokres2},
and by repeatedly applying the Binomial identity
\beq
B_n(x_1+y_1,\ldots x_n+y_n)=\sum_{i=0}^n\binomial{n}{i}
B_{n-i}(x_1,\ldots x_{n-i})B_i(y_1,\ldots y_i)
\label{BellCcomb}
\eeq
of the complete exponential Bell polynomials.

\subsection{Extraction of the $h_k$ coefficients and construction of
the function $h_{qg}$}
\label{sec:hqg2}
In the literature, the all-order result for $h_{qg}$ we have presented in
sect.~\ref{sec:hqgAO} was unknown, and therefore the determination of its
perturbative coefficients necessarily followed a different procedure with
respect to we have adopted in sect.~\ref{sec:hqgPert}. Here, we derive the
coefficients employing a more traditional approach; still, we stress that the
emerging results, at variance with those in the literature, are entirely
analytical {\em and} not recursive (in other words, the perturbative-index
dependence is completely parametrical).

The relevant quantity is $\hat\gamma_{qg}$, which enters the integral in
eq.~\eqref{Gqgdef}, and whose perturbative expansion in terms of $\bbas$ is
given by eq.~\eqref{gqgvsggg}, that we report here for convenience:
\begin{equation}
\hat\gamma_{qg}(\bbas)=\sum_{k=0}^\infty t_k \bbas^{k+1}\,.
\label{gqgdef}
\end{equation}
The $t_k$ will be determined order-by-order by means of an expansion of
eq.~\eqref{masterqg}, bearing in mind that the left-hand side there is finite.

One starts by computing $\Gamma_{gg}$, according to its definition in
eq.~\eqref{Gggdef}, in a way which emphasises the role of $\bbas$ as the expansion
parameter. With eq.~(\ref{gggex}) we have
\begin{equation}
  \Gamma_{gg}(\bbas,\ep) = \exp\left[\frac{1}{\ep} \sum_{j=0}^\infty g_j \,\int_0^{\bbas}\frac{da}{a}\,a^{j+1} \right]
  = \exp\left[\frac{1}{\ep} \sum_{k=1}^\infty \frac{g_{k-1}}{k} \bbas^{k} \right].
\end{equation}
By employing the properties of the Bell polynomials (in particular,
their generating function and the homogeneous behaviour of the incomplete
ones), we arrive at the following expression:
\begin{align}
\Gamma_{gg}(\bbas,\ep)&=\sum_{n=0}^\infty\Gamma_{gg}^{(n)}(\ep)\,\bbas^n\,,
\label{Gggres}
\\
\Gamma_{gg}^{(n)}(\ep)&=\frac{1}{n!}\,
\sum_{k=0}^n\frac{1}{\ep^k}B_{n,k}(G_0)\,,
\label{Gggcff}
\end{align}
where
\beq
G_0=\left(g_0 0!,\,g_1 1!,\,g_2 2!,\ldots,g_n n!,\ldots\right).
\label{G0vec}
\eeq
By replacing eqs.~\eqref{Gggres} and~\eqref{gqgdef} into eq.~\eqref{Gqgdef},
after performing the integration and changing summation indices we arrive at
\beq
\hat\Gamma_{qg}(\bbas,\ep)=\frac{1}{\ep}
\sum_{n=0}^\infty\frac{\bbas^{n+1}}{n+1}
\sum_{k=0}^n t_{n-k}\Gamma_{gg}^{(k)}(\ep)\,.
\label{Gqgres}
\eeq
Thus eq.~(\ref{masterqg}), thanks to eqs.~(\ref{Gqgzex}), (\ref{Gqgres}),
and~(\ref{Gggres}), becomes
\begin{align}\label{tmp2}
\hat\G_{qg}(\bbas,\ep)&=\frac{1}{\ep}
\sum_{n=0}^\infty\bbas^{n+1}\sum_{k=0}^n
\frac{1}{(n-k)!}\Bigg[
\left(\sum_{i=0}^\infty\frac{\rho_{k,i}}{\ep^{k-i}}\right)
\left(\sum_{p=0}^{n-k}\frac{(-)^p}{\ep^p}\,B_{n-k,p}(G_0)\right)
\\*&\qquad\qquad\qquad\qquad\qquad\quad
-\frac{1}{k+1}
\sum_{j=0}^k \frac{t_{k-j}}{j!}
\left(\sum_{i=0}^j\frac{1}{\ep^i}B_{j,i}(G_0)\right)
\left(\sum_{p=0}^{n-k}\frac{(-)^p}{\ep^p}\,B_{n-k,p}(G_0)\right)
\Bigg].
\nonumber
\end{align}
The singular structure of the two terms in squared brackets in
eq.~(\ref{tmp2}) is manifest. For a given $\ord(\bbas^n)$ term,
and ignoring the $1/\ep$ prefactor, the singular terms are
\begin{align}
&1/\ep^{k+p-i}\,,&&\hspace{-6em} -\infty<k+p-i\le n\,,
\\
&1/\ep^{i+p}\,,&&\hspace{-6em} 0\le i+p\le n\,.
\end{align}
Since $\hat\G_{qg}$ is a finite quantity, this implies that at $\ord(\bbas^n)$
we shall have to impose that the residues of the $n$ poles and the constant
term (owing to the $1/\ep$ prefactor) vanish; the corresponding equations
must be solved for the $t_i$ coefficients. With tedious but straightforward
manipulations of the sums, eq.~(\ref{tmp2}) can be recast in the following
form:
\begin{align}
\hat\G_{qg}&=\frac{1}{\ep}
\sum_{n=0}^\infty\bbas^{n+1}
\!\!\sum_{m=-\infty}^n\frac{1}{\ep^m}
\sum_{k=0}^n
\frac{1}{(n-k)!}
\label{Gqgres2}
\Bigg[\sum_{p=0}^{n-k}(-)^p\stepf(p\ge m-k)
\rho_{k,k+p-m}B_{n-k,p}(G_0)
\\*&\qquad\qquad\qquad
-\frac{1}{k+1}
\sum_{i=0}^m \sum_{j=i}^k (-)^{m-i}\stepf(m-n+k\le i\le k)
 \frac{t_{k-j}}{j!} B_{j,i}(G_0)B_{n-k,m-i}(G_0) \Bigg].
\nonumber
\end{align}
Since at $\ord(\bbas^n)$ the absence of divergencies in eq.~(\ref{Gqgres2})
gives $(n+1)$ conditions, there is a lot of redundancy as far as the
determination of the $t_i$ coefficients is concerned. It turns out, for
example, that the $(n+1)$ conditions at a given $n$ feature all $t_j$
coefficients with $0\le j\le n$. However, it is actually more convenient
to consider the $m=0$ contributions associated with different $n$'s.
This is because eq.~(\ref{Gqgres2}) in that case leads to
\beq
m=0\;\;\Longrightarrow\;\;
\ldots
\sum_{k=0}^n
\frac{1}{(n-k)!}
\Bigg[\sum_{p=0}^{n-k}(-)^p\rho_{k,k+p}B_{n-k,p}(G_0) -
\frac{1}{k+1}\sum_{j=0}^k\frac{t_{k-j}}{j!}
B_{j,0}(G_0)B_{n-k,0}(G_0)
\Bigg]\phantom{aa}
\label{Gqgresm0}
\eeq
whence
\begin{align}
\frac{t_n}{n+1}&=
\sum_{k=0}^n\frac{1}{(n-k)!}
\sum_{p=0}^{n-k}(-)^p\,\rho_{k,k+p}\,B_{n-k,p}(G_0)
\nonumber\\*&=
\sum_{i=0}^n\sum_{k=0}^i\frac{(-)^{i-k}}{(n-k)!}\,
\rho_{k,i}\,B_{n-k,i-k}(G_0)\,.
\label{Gqgresm02}
\end{align}
Equation~(\ref{Gqgresm02})
is remarkable, because it tells us that it is not even necessary to
solve (recursively or otherwise) a system of linear equations in order
to arrive at the $t_i$ coefficients, since these are already in a closed
form.

We now focus on an alternative determination of $h_{qg}$, which we remind
the reader to be implicitly defined by
\beq
\hat\gamma_{qg}(\bbas) = \bbas\,h_{qg}(\gamma_s(\bbas))\,,
\eeq
with expansion
\beq
h_{qg}(\gamz)=\sum_{n=0}^\infty h_n \gamz^n\,.
\label{hqggam}
\eeq
The coefficients $h_n$ are obtained by exploiting the variable replacement of
eq.~\eqref{atom} inside eq.~\eqref{gqgdef} and by using the results of
eq.~\eqref{Gqgresm02}. After having computed in this way a few of the $h_n$
coefficients, we trade $\gamz$ for $\hh \equiv \hh(\gamz)$ by employing
the relationships in eqs.~\eqref{gamvshh1XX}--\eqref{gamvshh2XX},
whence
\begin{subequations}
\begin{align}
h_{qg}(\gamz)&=1+\frac{5}{3}\hh+\frac{14}{9}\hh^2+\frac{82}{81}\hh^3+
\frac{122}{243}\hh^4+\frac{146}{729}\hh^5+\ord(\zeta_i)+\ldots
\label{hqgres10a}
\\*&=
1+\sum_{k=1}^\infty\frac{2^{k-2}}{k!}\left(3+\frac{1}{3^k}\right)\hh^k
+\ord(\zeta_i)
\label{hqgres10b}
\\*&=
\frac{1}{4}\left(3e^{2\hh}+e^{\frac{2}{3}\hh}\right)+\ord(\zeta_i)
\label{hqgres10c}
\\*&=
\frac{1+\hchi(\gamz)}{4}\left(3e^{2\hh}+e^{\frac{2}{3}\hh}\right),
\label{hqgres102}
\end{align}
\end{subequations}
which coincides with eq.~(\ref{hqgres10}).
In eq.~(\ref{hqgres10a}), by $\ord(\zeta_i)$ we have denoted all of the
terms with coefficients which are not rational numbers, in that they feature
(also) Riemann's $\zeta$'s. In the same equations, the dots indicate
terms of higher powers of $\hh$, of which we have computed a few,
although obviously not all of them. However, by induction we have
established that their general form is that reported in eq.~(\ref{hqgres10b}).
The resulting series can be easily summed, leading to eq.~(\ref{hqgres10c}).
Then, crucially, eq.~(\ref{hqgres102}) shows that all terms with non-rational
coefficients can in fact be accounted for by simply including a
$\gamz$-dependent prefactor, while keeping the same exponential structure.
The key message to take home here is that most of the inherent complications
of the function $h_{qg}$ are captured by replacing the dependence on $\bbas$
in its rational part with that on $\hh(\gamma_s(\bbas))$ --- essentially,
$\hh$ collects
the non-rational behaviour of the result we sought, up to some residual
non-rational dependence which, in turn, is completely accounted for
by an almost trivial $\gamma_s$-dependent prefactor (which confirms, once
again, the role of $\gamma_s$ in the significant simplifications of the
analytical results with respect to their $\bbas$-expanded counterparts).

\subsection{Alternative closed-form expression for $t_n$ coefficients}
\label{sec:exprtn2}

If one assumes to know the coefficients $h_n$ as in
eq.~(\ref{hncoeff}), one can obtain a different, more compact,
closed-form expressions for the coefficients $t_n$ eq.~\eqref{Gqgresm02}.
This is done by realizing that eq.~\eqref{hqggam} and eq.~\eqref{gqgdef}
imply that
\beq
\sum_{k=0}^\infty t_k \bbas^k = \sum_{n=0}^\infty h_n \gamma_s^n(\bbas)\,,
\eeq
and then
\beq
t_k=\left.\frac{1}{k!}\frac{d^k}{d\bbas^k}h_{qg}\right|_{\bbas=0}=
\frac{1}{k!}\sum_{n=0}^k h_n \left.
\frac{d^k\gamma_s^n}{d\bbas^k}\right|_{\bbas=0}\,,
\eeq
The derivative on the right-hand side of this equation is computed by
employing the Fa\`a di Bruno formula, whence
\beq
\left.\frac{d^k\gamma_s^n}{d\bbas^k}\right|_{\bbas=0}=
n!B_{k,n}\big(\hat{G}\big)\,,
\eeq
with the argument of the Bell polynomial defined in eq.~(\ref{hGvec})
(note that this is the list of the derivatives of $\gamma_s$ computed
at $\bbas=0$). Therefore
\beq
t_k=\frac{1}{k!}\sum_{n=0}^k h_n n!B_{k,n}\big(\hat{G}\big)=
\frac{1}{k!}\sum_{n=0}^k \left(\frac{3}{4}\,B_n\left(\widehat{Z}_2\right)
+\frac{1}{4}\,B_n\left(\widehat{Z}_{\frac{2}{3}}\right)\right)
B_{k,n}\big(\hat{G}\big)\,,
\label{tnclosed}
\eeq
having used eq.~(\ref{hncoeff}) and the fact that
\mbox{$B_{k,n}()=0$} for any $n>k$, which allows one to set the upper
bound of the summation over $n$ equal to $k$.

Eq.~(\ref{tnclosed}) is in fact a blueprint for expressing the coefficients
of the expansion in $\bbas$ of a quantity whose coefficients of the
expansion in $\gamma_s$ are known. Explicitly, if
\beq
\sum_{n=0}^\infty c_n\gamma_s^n(\bbas)=
\sum_{k=0}^\infty c_k^\prime\bbas^n
\label{gggvsbbas1}
\eeq
then
\beq
c_k^\prime=\frac{1}{k!}\sum_{n=0}^k c_n n!B_{k,n}\big(\hat{G}\big).
\label{gggvsbbas2}
\eeq

\subsection{Construction of finite quark Green function $\G_{qg}$}
\label{sec:Gqgep0}

The method employed in sect.~\ref{sec:hqgAO} for the determination
of $h_{qg}$ can be used to compute the other non-divergent component
of eq.~(\ref{masterqg}), namely $\hat\G_{qg}$ (or, more precisely, its
\mbox{$\ep\to 0$} limit). Note that, according to eqs.~(\ref{G0master}),
(\ref{BkCkpos}), and~(\ref{Fhsplit}), we have
\beq
\lim_{\ep\to 0}\hat\G_{qg}=A_0=B_0=F^0\,.
\label{GqgvsFz}
\eeq
As a power series, $F^0$ can be directly determined by taking the
$\ord(\ep^0)$ term of $F(\hh)$ from its definition in eq.~(\ref{masterqg2}).
In order to do this, one needs to exploit the series expansion of
\mbox{$\hat\G_{qg}^{(0)}$}, eq.~\eqref{Gqgzex}, and of $\Gamma_{gg}^{-1}$
(eq.~(\ref{Gggres}), with \mbox{$\ep\to -\ep$} there). The former is
a series in $\bbas$, which is turned into a series in $\gamma_s(\bbas)$
by means of eq.~(\ref{atom}). As far as \mbox{$\Gamma_{gg}^{-1}$} is
concerned, we can simply employ eq.~(\ref{hhton}).

From its definition in eqs.~(\ref{masterqg2}) and~(\ref{Fhsplit}),
and the results for $\hat\G_{qg}^{(0)}$ and $\Gamma_{gg}$, we have
\beq
F^0\equiv
\lim_{\ep\to 0}\hat\G_{qg}^{(0)}\Gamma_{gg}^{-1}=
\lim_{\ep\to 0} \sum_{k=0}^\infty\sum_{i=0}^\infty
\sum_{j=0}^\infty\frac{\ep^i}{\ep^j\ep^{k+1}}\,\rho_{k,i}\,\bbas^{k+1}\,
\frac{(-\hh)^j}{j!}\\,.
\eeq
The $\ep\to 0$ limit then is simply taken by selecting the
$\ord(\ep^0)$ terms on the right-hand side of this equation by constraining
\beq
i-j-k-1=0\,.
\label{ijkconstr}
\eeq
Owing to eqs.~(\ref{atom}) and~(\ref{hhton}), this implies that the
$i^{\rm th}$ $\ep$-derivative of a $\rho_k$ coefficient must be computed only
up to \mbox{$\ord(\gamma_s^{j+k+1})$}. Therefore, if the overall accuracy
one is interested in is some $\ord(\gamma_s^M)$, then
\beq
F^0=\sum_{k=0}^{M-1}\sum_{j=0}^{M-1-k}\rho_{k,j+k+1}\,
\bbas^{k+1}\,\frac{(-\hh)^j}{j!}\,.
\label{Fzsums2}
\eeq
Before proceeding, we point out that in view of the attempt to find a
functional expression for $F^0$, it is crucially important that the
derivatives of polygamma functions be treated symbolically, and not
converted into their equivalent real-number expressions. Specifically,
the problem affects the derivatives with odd indices, since for example
\mbox{$\pi^4\propto\psi_3(1)$} but also \mbox{$\pi^4\propto\psi_1(1)^2$}.
In any case, following what has been done in other, simpler, cases, the
first step is the determination of the rational part of $F^0$. The
explicit calculation of the right-hand side of eq.~\eqref{Fzsums2} leads to
\beq
F^0 = -\frac{7}{12}\,\gamma_s-\frac{41}{72}\,\gamma_s^2-
\frac{157}{1296}\,\gamma_s^3+\frac{2537}{7776}\,\gamma_s^4+
\frac{27595}{46656}\,\gamma_s^5+\ldots
+\ord(\psi_2(1)\gamma_s^3)+\ord(\psi_3(1)\gamma_s^4)\,.
\eeq
The arguments of the $\ord()$ symbols on the right-hand side denote the
instances of the Riemann $\zeta$ at the lowest order in $\gamma_s$ where they
start to appear --- only the first with even and odd indices are reported.
This implies that $F^0$ features neither $\psi_0(1)=-\gE$
nor $\psi_1(1)=\pi^2/6$ (but higher powers of $\pi$ will be present).
By computing a sufficiently large number of terms of the series,
one finds that the sequence of rational coefficients is
\beq
\left\{-\frac{7}{12},\,-\frac{41}{72},\,-\frac{157}{1296},\,
  \frac{2537}{7776}, \ldots\right\}
=
\left\{-\frac{2^{k-2}}{k!}\left(3+\frac{1}{3^k}\right)\right\}_{k=1}^\infty
+ \left\{\frac{3}{4}+\frac{3}{2^{k+1}}-
\frac{5}{4\mydot 3^k}\right\}_{k=1}^\infty\,.
\label{Fzratseq}
\eeq
We note that the two sequences on the right-hand side of eq.~\eqref{Fzratseq}
computed in $k=0$ would give $-1$ and $1$ respectively, thus summing up to zero.
For subsequent manipulations, it will turn out to be convenient to include
such harmless $k=0$ contributions in the definitions of the following
functions, already reported in the main text as eqs.~(\ref{c0def})
and~(\ref{c1def}):
\begin{align}
\cc_0(z)&\equiv
-\sum_{k=0}^\infty\frac{2^{k-2}}{k!}\left(3+\frac{1}{3^k}\right)z^k
\nonumber\\*
&= -\frac{1}{4}\left(3e^{2z}+e^{\frac{2}{3}z}\right) \,,
\label{c0ser}
\\
\cc_1(z)&\equiv \sum_{k=0}^\infty\left( \frac{3}{4}+\frac{3}{2^{k+1}}-
\frac{5}{4\mydot 3^k} \right)z^k
\nonumber\\
  &= \frac14\(\frac3{1-z} + \frac{12}{2-z} - \frac{15}{3-z}\)\,,
\label{c1ser}
\end{align}
so that
\beq
F^0=\cc_0(\gamma_s)+\cc_1(\gamma_s)+
\ord(\psi_2(1)\gamma_s^3)+\ord(\psi_3(1)\gamma_s^4)\,.
\eeq
Then, one finds a crucial property of $F^0$ by assuming, in keeping
with eq.~(\ref{hqgres10b}), that the natural argument of the $\cc_0$
function is $\hh(\gamma_s)$ rather than simply $\gamma_s$. This results
in the following identity:
\beq
\frac{F^0-\cc_0(\hh)}{\cc_1(\gamma_s)}=1+\delta_1F^0+\delta_2F^0+\ldots\,.
\label{Fzmrat}
\eeq
The defining characteristic of the functions $\delta_iF^0$ is that
they feature monomials that contain exactly $i$ instances of Riemann's
$\zeta$, each {\em uniquely} associated with a power of $\gamma_s$.
In other words, if one finds some term $\psi_a(1)\gamma_s^b$ in
$\delta_1F^0$, then $\psi_a(1)$ will not appear anywhere else
in $\delta_1F^0$. This simplifies significantly the task of finding
the structures of the functions $\delta_iF^0$, an operation that
requires their explicit computations for the lowest values of $i$ in
the attempt to find a pattern for larger $i$ values. Obviously,
one starts with $i=1$, and by trial and errors one finds
\beqn
\delta_1F^0&=&d_1^e(\gamma_s)+d_1^o(\gamma_s)\,,
\\*
d_1^e(\gamma_s)&=&
-2\gamma_s\int_0^1 dy\,\psi_{0+}(y\gamma_s)
+2\gamma_s\,\psi_{0+}(\gamma_s)
-\gamma_s^2\,\psi_{1-}(\gamma_s)\,,
\label{d1e2}
\\*
d_1^o(\gamma_s)&=&
-\gamma_s^2\int_0^1 dy\,y\,\psi_{1+}(y\gamma_s)\,,
\label{d1o2}
\eeqn
where we have used the definitions eqs.~\eqref{psipdef}--\eqref{psimdef}.
Then, eq.~\eqref{Fzmrat} becomes
\beqn
\frac{F^0-\cc_0(\hh)}{\cc_1(\gamma_s)}
&=&1+d_1^e(\gamma_s)+d_1^o(\gamma_s)+\delta_2F^0+\ldots
\label{Fzmrat00}
\\*
&\equiv&
\exp\Big[d_1^e(\gamma_s)+d_1^o(\gamma_s)\Big]
+\bar{\delta}_2F^0+\ldots\,.
\label{Fzmrat2}
\eeqn
In other words, while eq.~(\ref{Fzmrat00}) is simply equal to our
starting point, eq.~(\ref{Fzmrat}), in eq.~(\ref{Fzmrat2}) we have
made the further crucial assumption that the contributions of $d_1^e$
and $d_1^o$ exponentiate: this does not change the structure of the
single-Riemann-$\zeta$ terms in $F^0$ (which is why this operation is
legitimate), but it does change that of the terms with multiple $\zeta$'s,
hence the change of notation \mbox{$\delta_2F^0\to \bar{\delta}_2F^0$}.
Ultimately, whether this exponentiation is appropriate or not is
determined by the capability of finding a pattern in the (iteratively
modified) $\bar{\delta}_iF^0$ sequence. This indeed turns out to be
the case --- by computing exactly the contributions up to and including
those that feature monomial with four Riemann's $\zeta$, we could establish
the existence of a clear pattern, and thence to resum the resulting
series. Eq.~(\ref{Fzmrat2}) becomes
\beq
\frac{F^0-\cc_0(\hh)}{\cc_1(\gamma_s)}=
\exp\Big[\sum_{n=1}^\infty \Big(d_n^e(\gamma_s)+d_n^o(\gamma_s)\Big)\Big]\,,
\label{Fzmrat4}
\eeq
while eqs.~(\ref{d1e2}) and~(\ref{d1o2}) generalise as follows
\begin{align}
\sum_{n=1}^\infty d_n^e(\gamma_s)&=
-\int_0^1dy\,\frac{\gamma_s\psi_{0+}(y\gamma_s)}
{1-2y\gamma_s\psi_{0+}(y\gamma_s)}
-\gamma_s\int_0^1dy\,\psi_{0+}(y\gamma_s)
\nonumber
\\*&\quad
-\log\Big(1-2\gamma_s\psi_{0+}(\gamma_s)\Big)
-\half\log\Big(1+2\gamma_s^2\psi_{1-}(\gamma_s)\Big)\,,
\label{dnesum}
\\
\sum_{n=1}^\infty d_n^o(\gamma_s)&=
-\int_0^1dy\,\frac{y\gamma_s^2\psi_{1+}(y\gamma_s)}
{1-2y\gamma_s\psi_{0+}(y\gamma_s)}\,,
\label{dnosum}
\end{align}
where the quantities defined in eqs.~(\ref{psipdef}) and~(\ref{psimdef})
emerge in a natural manner.

By computing analytically the second integral on the right-hand side of
eq.~(\ref{dnesum}), and by exploiting some obvious identities among
the $\psi_{k\pm}(z)$ functions, eq.~(\ref{Fzmrat4}) leads us to the
sought result
\begin{align}
F^0&=\cc_0(\hh)+\cc_1(\gamma_s)
\label{Fzfunres}
\exp\Bigg[
\int_0^1dy\,\frac{y\gamma_s^2\big(\psi_1(1)-\psi_1(1-y\gamma_s)\big)}
{1-2y\gamma_s\psi_{0+}(y\gamma_s)}
+\gamma_s\psi_0(1)
\nonumber
\\*&\qquad\qquad\qquad\qquad\qquad
-\half\log\Gamma\big(1+\gamma_s\big)
+\half\log\Gamma\big(1-\gamma_s\big)
\nonumber
\\*&\qquad\qquad\qquad\qquad\qquad
-\half\log\Big(1-2\gamma_s\psi_{0+}(\gamma_s)\Big)
-\half\log\Big(1+2\gamma_s^2\psi_{1-}(\gamma_s)\Big)\Bigg]\,.
\end{align}
This expression can also be rewritten in terms of quantities which
have already played a prominent role. One starts by observing that
(see eq.~(\ref{FPggg0}))
\beq
1+\hchi (\gamma_s)\equiv
\gamma_s\chi(\gamma_s)=
1-2\gamma_s\psi_{0+}(\gamma_s)\,,
\label{Gfid1}
\eeq
and that
\beq
1+\hchi (\gamma_s)-\gamma_s \hchi^\prime(\gamma_s)\equiv
-\gamma_s^2\chi^\prime(\gamma_s)=
1+2\gamma_s^2\psi_{1-}(\gamma_s)\,,
\label{Gfid2}
\eeq
having denoted by a prime the first derivative of a function. Since
$\hchi (z)$ is an odd function of $z$, its series expansion features only
$\psi_i(1)$ terms with even $i$, and therefore not all terms in
eq.~(\ref{Fzfunres}) can be expressed by means of it. We can therefore
employ a function which we define by analogy to eq.~\eqref{eq:chidef}, namely:
\beq
\chi_1(z)=2\psi_1(1)-\psi_1(z)-\psi_1(1-z)\,.
\eeq
We then have
\beq
\half\Big(\chi_1(z)-\chi^\prime(z)\Big)=\psi_1(1)-\psi_1(1-z)\,.
\label{chichip}
\eeq
By means of these relationships we obtain the final form reported in the
main text, eq.~(\ref{Fzfunres2}).

At this point, one notes that the structure of the exponential in
eq.~(\ref{Fzfunres}) is quite similar (although not identical) to
that of the function $R_N$ in eq.~(3.17) of ref.~\cite{Catani:1994sq};
this is relevant, since $R_N$ is the $\ord(\ep^0)$ term of the
renormalised $\G_{gg}$ Green function, in keeping with $F^0$ being the
$\ord(\ep^0)$ term of the renormalised $\hat\G_{qg}$ Green function,
eq.~\eqref{GqgvsFz}. As a matter of fact, one can manipulate
the formulae so that indeed $R_N$ emerges, since owing to eq.~\eqref{chichip}:
\begin{align}
R_N(\text{eq.~(3.17) of~\cite{Catani:1994sq}}) &=
\exp\Bigg[
\half \gamma_s\int_0^1dy\,
\frac{\chi_1(y\gamma_s)-\chi^\prime(y\gamma_s)}
{\chi(y\gamma_s)}
+\gamma_s\psi_0(1)
\nonumber \\*
&\phantom{\exp\Bigg[\quad\quad}
-\half\log\Gamma\big(1+\gamma_s\big)
+\half\log\Gamma\big(1-\gamma_s\big)
\nonumber \\*
&\phantom{\exp\Bigg[\quad\quad}
+\half\log\Big(\gamma_s\chi(\gamma_s)\Big)
-\half\log\Big(-\gamma_s^2\chi^\prime(\gamma_s)\Big)\Bigg].
\phantom{aaaaa}
\label{RNCH}
\end{align}
It is then a matter of trivial algebra to arrive at eq.~(\ref{Fzfunres4}).

\section{Details on numerical implementation}
\label{sec:appHELL}

\subsection{Numerical integrations for $P_{qg}$}
\label{sec:appHELL-Pqg}

The numerical implementation of the resummed $P_{qg}$ requires the computation
of the inverse Mellin transform of eq.~\eqref{gqgABFint3}, which is in turn
given by an integral in the complex $z$ plane.  We now review the strategy
adopted in the \HELL code.

\begin{figure}[t]
  \centering
  \includegraphics[width=0.5\textwidth]{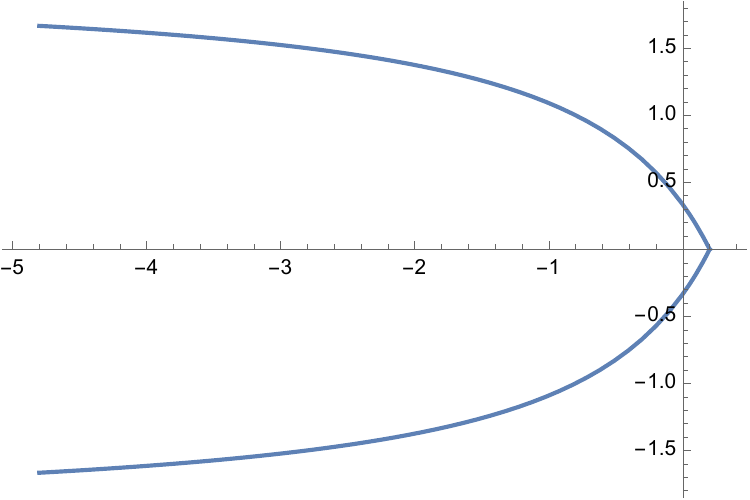}
  \caption{Contour used for the computation of the $z$ integral.}
  \label{fig:contour}
\end{figure}

The integration in $z$, at fixed $N$, must be performed along a Hankel-like
contour enclosing the negative real axis.  We find it convenient to
parametrise the contour as
\beq\label{eq:zcontour}
z(t) = A - t + iB\(1-\frac1{1+t}\), \qquad t\in[0,\infty),
\eeq
for the upper part of the contour, while the lower part is given by complex
conjugation, $\bar z(t)$, to be traveled in the opposite direction.  Suitable
values for the parameters are $A=0.2$, $B=2$. In particular the latter ensures
that we travel sufficiently far from the negative real axis, otherwise
oscillations may show up.  Conversely, $A$ must be small in order to avoid the
singularities of $h_{qg}$.  The contour is shown in fig.~\ref{fig:contour}.
%%%%%%%%%%%%%%%%%%%%%%%%%%%%%%%%%%%%%%%%%%%%%%%%%%%%%%%%%%%%%%%%%%%%%
%%%%%%%%%%%%%%%%%%%%%%%%%%%%%%%%%%%%%%%%%%%%%%%%%%%%%%%%%%%%%%%%%%%%%

Then, we also need to compute the inverse Mellin transform.
For this, we can use a typical Mellin inversion path parametrised as
\beq\label{invNcontour}
N(u) = N_0 + e^{i\theta}u, \qquad \frac\pi2 <\theta<\pi, \qquad u\in[0,\infty),
\eeq
for the upper part, and its complex conjugate (in the opposite direction) for
the lower part. Here, $N_0=1$ and $\theta=\frac23\pi$ are reasonable values.
Now, defining\footnote{We remind the reader that the quantity $r(N,\as)$ has
been introduced in sect.~\ref{sec:gengamqg} in the context of the definition
of the approximation of the anomalous dimension, eq.~\eqref{gamPABF},
used for the resummation of the running coupling contributions.}
\beq
H(z,N) \equiv \Gamma\(1+\frac{\gamma_+(N,\as)}{r(N,\as)}\) \frac1z \, e^{\frac z{r(N,\as)}}\(\frac{r(N,\as)}z\)^{\frac{\gamma_+(N,\as)}{r(N,\as)}} h_{qg}(z),
\eeq
we can rewrite eq.~\eqref{gqgABFint3} as
\begin{align}
\gamma_{qg}^{\NLLp}(\as,N)
  &= \frac{\as}{3\pi}\TR\,
  \frac1{2\pi i} \oint dz\, H(z,N) \nonumber\\
  &= \frac{\as}{3\pi}\TR\, \frac1{2\pi i} \int_0^\infty dt\[\frac{dz}{dt} H(z(t),N) - \frac{d\bar z}{dt} H(\bar z(t),N)\]
\end{align}
and thus its inverse Mellin transform as (the bar denotes complex conjugation)
\begin{align}
P_{qg}^{\NLLp} (\as,x)
  &= \frac{\as}{3\pi}\TR\,
    \frac1{2\pi i} \int_0^\infty du \Bigg\{
    \frac{dN}{du} x^{-N(u)} \frac1{2\pi i} \int_0^\infty dt\[\frac{dz}{dt} H(z(t),N(u)) - \frac{d\bar z}{dt} H(\bar z(t),N(u))\]\nonumber\\
  &\hspace{7.3em}-\frac{d\bar N}{du} x^{-\bar N(u)} \frac1{2\pi i} \int_0^\infty dt\[\frac{dz}{dt} H(z(t),\bar N(u)) - \frac{d\bar z}{dt} H(\bar z(t),\bar N(u))\]
    \Bigg\}\nonumber\\
  &= -\frac{\as}{3\pi}\TR\,\frac1{4\pi^2} \int_0^\infty du \int_0^\infty dt \Bigg\{
    \frac{dN}{du} x^{-N(u)} \[\frac{dz}{dt} H(z(t),N(u)) - \frac{d\bar z}{dt} H(\bar z(t),N(u))\]\nonumber\\
  &\hspace{11.3em}+\frac{d\bar N}{du} x^{-\bar N(u)} \[\frac{d\bar z}{dt} H(\bar z(t),\bar N(u)) - \frac{dz}{dt} H(z(t),\bar N(u))\]
    \Bigg\}\nonumber\\
  &= -\frac{\as}{3\pi}\TR\,\frac1{2\pi^2} \int_0^\infty du \int_0^\infty dt\, \Re\Bigg\{
    \frac{dN}{du} x^{-N(u)} \[\frac{dz}{dt} H(z(t),N(u)) - \frac{d\bar z}{dt} H(\bar z(t),N(u))\] \Bigg\},
\end{align}
which gives a simple and stable numerical implementation.

One practical aspect that we have to consider is that the function
$h_{qg}(z)$, eq.~\eqref{hqgres10}, depends on the function $\hh(z)$,
eq.~\eqref{hhdef}, which is given by an integral appearing at the exponent.
Unfortunately we have not been able to compute the integral analytically, so
by computing it during integration in $z,N$ one would end up with nested
integrals, which are very slow to evaluate.  In order to achieve a
sufficiently fast numerical implementation, our strategy consists in
precomputing the function $\hh(z)$ along the $z$ contour,
eq.~\eqref{eq:zcontour}, so that $h_{qg}(z)$ can be quickly evaluated by
performing a simple interpolation.  For the computation of $\hh(z)$ along the
contour we use one of the expressions derived in appendix~\ref{sec:hhexpr},
specifically eq.~\eqref{eq:hhv1} or eq.~\eqref{eq:hhv2}, which have similar
performances.

\subsection{Approximate fixed-order anomalous dimension at large $\as$}
\label{sec:appHELL-as}

As was mentioned in section~\ref{sec:HELL-as}, the \HELL construction of the
resummed anomalous dimension $\gamma_+^{\NorLLp}$ requires an approximation of the
fixed-order anomalous dimension used to resum the collinear singularities of
the BFKL kernel.  The inverse of this fixed-order anomalous dimension (called
$\chi_s$ in ref.~\cite{Bonvini:2017ogt}) must be added to the fixed-order BFKL
kernel, after subtracting the double counting.  This procedures removes the
singular behaviour at $\gamma=0$ of the kernel $\chi_0(\gamma)$,
eq.~\eqref{eq:BFKLkerLO}.  However, the kernel has an additional, subleading,
singularity at $\gamma=-1$.  When computing the inverse to obtain the resummed
anomalous dimension (after the symmetrisation procedure) this singularity
would dominate the asymptotic behaviour at large $N$, in disagreement with the
logarithmically divergent behaviour of the actual anomalous dimension.

Therefore, in order for this procedure not to spoil the original large-$N$
behaviour, in ref.~\cite{Bonvini:2017ogt} it has been suggested to replace
the exact fixed-order anomalous dimension with an approximation that,
while preserving the physical small-$N$ features,
behaves asymptotically as a (negative) constant.
This is helpful because, when this constant is larger than $-1$,
the dual of the approximate anomalous dimension dominates over the subleading
pole of the BFKL kernel, thus determining the asymptotic form of the resummed
anomalous dimension, which then coincides with the one we started from.
At this point, in order to restore the physical logarithmic large-$N$
behaviour, one can simply subtract the approximate fixed-order prediction
from the resummed anomalous dimension (which have the same asymptotic
form), and add back the exact one.

For this procedure to work the constant must be greater than $-1$,
but its value depends on $\as$.
Let us review the construction of the approximation.
The LO and NLO anomalous dimensions close to $N=0$ behave as
\begin{align}
  \gamma_0(N) &= \frac{a_{11}}{N} + a_{10} +\Ord(N)\,,
\nonumber\\*
  \gamma_1(N) &= \frac{a_{22}}{N^2} + \frac{a_{21}}{N} + a_{20} +\Ord(N)\,,
\end{align}
where $a_{22}=0$ (accidental zero).
The approximation proposed in ref.~\cite{Bonvini:2017ogt} is
\beq\label{BestgammaApprox}
\gamma^{\rm approx}(N) = \frac{a_1}N + a_0 - (a_1+a_0)\frac{2N}{N+1},
\eeq
which is valid both at the LO and the NLO with appropriate coefficients.
At the NLO these are given by
\begin{align}
  a_1 &= \as a_{11} + \as^2 a_{21}, \nonumber\\
  a_0 &= \as a_{10} + \as^2 a_{20},
\end{align}
and at the LO one simply neglects the $\Ord(\as^2)$ terms; the coefficients
are thus given by
\begin{align}
  a_{11} &= \frac{C_A}\pi, \nonumber\\
  a_{21} &= n_f\frac{26C_F-23C_A}{36\pi^2}, \nonumber\\
  a_{10} &= -\frac{11C_A + 2n_f(1-2C_FC_A+4C_F^2)}{12\pi}, \nonumber\\
  a_{20} &= \frac1{\pi^2}\[\frac{1643}{24} - \frac{33}2 \zeta_2 - 18 \zeta_3 + n_f\(\frac49\zeta_2-\frac{68}{81}\) + n_f^2\frac{13}{2187}\].
\end{align}
The last term of eq.~\eqref{BestgammaApprox} is needed to restore the constraint
\mbox{$\gamma^{\rm approx}(N=1)=0$} due to momentum conservation, and it is thus
largely arbitrary. The choice of the specific functional form of this term
is dictated by simplicity, as one additional requirement of the approximation
is the possibility of computing the inverse function analytically.

We propose here to modify this function so that the functional form
remains simple but the asymptotic value at large $N$ is larger.
We thus generalize eq.~\eqref{BestgammaApprox} as
\beq\label{BestgammaApprox2}
\gamma^{\rm approx}(N) = \frac{a_1}N + a_0 - (a_1+a_0)\frac{(1+\nfac)N}{N+\nfac},
\eeq
where $0< \nfac\leq1$ is a parameter that controls the asymptotic behaviour.
At large $N$, eq.~\eqref{BestgammaApprox2} behaves as a constant,
\beq\label{eq:gammaapproxasy}
\gamma(N) \overset{N\to\infty}{\to} -(1+\nfac)a_1 -\nfac a_0.
\eeq
By modifying the original value ($\nfac=1$) into a smaller value we see that
the asymptotic limit of $\gamma(N)$ becomes larger, as desired.

In practice, rather than choosing a specific value of $\nfac$, we have decided
to promote it to a function of $\as$. A simple form that gives satisfactory
results is
\beq\label{eq:cnew}
\nfac = \frac1{1+6\as^2}.
\eeq
In this way at small $\as$ this reproduces the original value $\nfac=1$,
where it is sufficient, but as $\as$ increases $\nfac$ decreases,
reaching e.g.\ $\nfac\sim1/5$ at $\as=0.8$.

The inverse (i.e.\ the dual) of eq.~\eqref{BestgammaApprox2} can be
computed analytically
\begin{align}\label{chisanalytic}
\chi_s(M)
&= \frac{a_1+\nfac a_0-\nfac M + \sqrt{(\nfac M+a_1-\nfac a_0)^2 + 4\nfac (1+\nfac) a_1(a_1+a_0)}}{2[(1+\nfac)a_1+\nfac a_0+M]},
\end{align}
which of course reduces to the result of ref.~\cite{Bonvini:2017ogt} for
$\nfac=1$.

\subsection{A modified approximation for the resummation of running
coupling effects in $\gamma_+$}
\label{sec:appHELL-RC}

The resummation of running-coupling effects to the duality relation between
the BFKL kernel and the DGLAP anomalous dimension is obtained by solving the
BFKL evolution equation with running coupling, and then by deriving from its
solution the desired anomalous
dimension~\cite{Thorne:1999sg,Thorne:1999rb,Thorne:2001nr,Altarelli:2005ni}.
A convenient way of performing this task is that of considering an
approximation of the kernel such that the sought solution can be computed
analytically, simplifying the numerical implementation of this procedure.
Originally, an approximation was proposed in ref.~\cite{Altarelli:2005ni},
which was later used in the first version of \HELL~\cite{Bonvini:2016wki}.
Then, a different form of the approximation has been introduced in
ref.~\cite{Bonvini:2017ogt}, which has solved some problems of the original
approximation, leading to a more reliable result which has been implemented in
\HELL from version 2 onwards.  In this appendix, we shall briefly review this
approximation, and propose a simple modification of it which allows us to
avoid an undesired effect showing up at large $\as$.

Without giving a thorough explanation (for more detail, see section~3 of
ref.~\cite{Bonvini:2017ogt}), we recall that the approximation of the BFKL
kernel $\chi(\as,\gamma)$ which we consider here has two aspects: an
approximation of the $\as$ dependence, and an approximation of the
$\gamma$ dependence.  As far as the former is concerned, a linear
approximation in $\as$ is employed,
\beq\label{eq:chi_linearapprox}
\chi(\as,\gamma)
= \chi(\alpha_0,\gamma) + (\as-\alpha_0) \chi'(\alpha_0,\gamma),
\eeq
where the prime denotes a derivative with respect to $\as$.
In the equation above, $\alpha_0$ is a reference value for $\as$ around
which the linear approximation is computed, and it is typically given by
$\as$ computed at the hard scale. A simpler approximation is also considered,
\beq\label{eq:chi_linearapprox2}
\chi(\as,\gamma)
= \as \frac{\chi(\alpha_0,\gamma)}{\alpha_0},
\eeq
which is used as an alternative implementation for the resummation of these
running coupling effects, to construct an uncertainty band on the result, see
sect.~\ref{sec:results}.

Then, the $\gamma$ dependence of the kernel is approximated as
\begin{align}\label{eq:chi_coll_approx}
  \chi(\as,\gamma)
  &\simeq \chi_{\rm coll}(\as,\gamma) \nonumber\\
  &= c(\as) + \frac{\Mmin^3(\as) \kappa(\as)}4 \[\frac1{\gamma}+\frac1{2\Mmin(\as)-\gamma} - \frac2{\Mmin(\as)}\] \nonumber\\
  &= c(\as) + \frac{\kappa(\as)}2 \(\gamma-\Mmin(\as)\)^2  + \Ord\((\gamma-\Mmin)^3\),
\end{align}
which is a symmetric form with respect to $\gamma=\Mmin(\as)$ based on two
poles such that its expansion around the symmetry point coincides with the one
of the double-leading kernel up to second order.  Indeed, the parameters
$c(\as)$ and $\kappa(\as)$, as well as the position of the minimum
$\Mmin(\as)$, are computed from the double-leading kernel, and used in the
construction of the approximation.  This is the key property of this
approximation, as the region around the minimum encodes the leading behaviour
at small $x$, which then must be correctly reproduced.

For reasons upon which we shall comment in a moment, we propose here a
simple modification of this approximation, where the two poles are shifted
by some amount $\eta$ in mutually opposite directions to preserve the symmetry:
\begin{align}\label{eq:chi_coll_approx_new}
  \chi(\as,\gamma)
  &\simeq \chi_{\rm coll,new}(\as,\gamma) \nonumber\\
  &= c(\as) + \frac{\(\Mmin(\as)+\eta\)^3 \kappa(\as)}4 \[\frac1{\gamma+\eta}+\frac1{2\Mmin(\as)-\gamma+\eta} - \frac2{\Mmin(\as)+\eta}\].
\end{align}
The parameters of the approximation have been adapted such that its expansion
around the minimum coincides with the previous approximation,
eq.~\eqref{eq:chi_coll_approx}.  Therefore, this approximation has all the
properties of the previous one, but the poles are located in different
places.  The solution to the running-coupling duality equation within
this new approximation is
\beq\label{eq:gamma_RC}
\gamma_{\rm rc}(\as,N) = \Mmin + \beta_0\asb
\[ z \frac{k_{\nu}'(z)}{k_{\nu} (z)} -1 \],
\eeq
with (omitting the dependence on $\as$ for improved readability)
\begin{subequations}\label{eq:RCfunctions}
\begin{align}
  \frac1{\asb} &= \frac1{\as} + \frac{\kappa'-2c'/(\Mmin+\eta)^2}{\kappa-\as\kappa'+2(N-c+\as c')/(\Mmin+\eta)^2}\\
  z &= \frac1{\beta_0\asb}\sqrt{\frac{N-c+\as c'}{(\kappa-\as\kappa')/2 +(N-c+\as c')/(\Mmin+\eta)^2}}\\
  \nu &= \(\frac{c'}{N-c+\as c'} + \frac{\kappa'-2c'/(\Mmin+\eta)^2}{\kappa-\as\kappa'+2(N-c+\as c')/(\Mmin+\eta)^2}\) \asb z.
\end{align}
\end{subequations}
and where $k_\nu(z)$ is a Bateman function.  This solution tends to the one of
ref.~\cite{Bonvini:2017ogt} in the limit $\eta\to 0$.  We also clearly see that
the overall effect of having introduced the parameter $\eta$ amounts to a
global\footnote{Except in the first term of eq.~\eqref{eq:gamma_RC}, which
remains $\Mmin$.  This is however irrelevant, as this term is subtracted as a
double-counting contribution when combining this solution with the
double-leading result.}  shift in $\Mmin\to\Mmin+\eta$, which is thus very
easy to implement numerically.

\begin{figure}[t]
  \centering
  \includegraphics[width=0.495\textwidth]{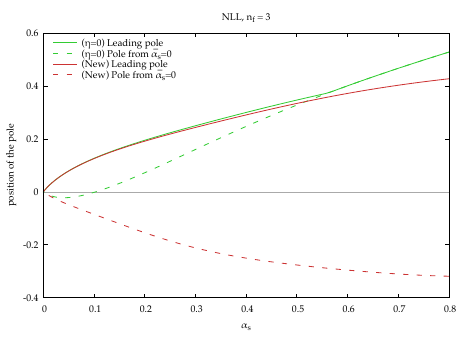}
  \includegraphics[width=0.495\textwidth]{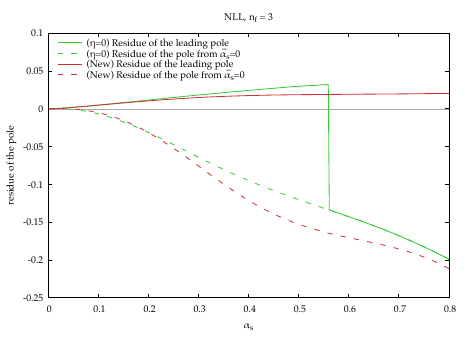}
\caption{The left-hand side panel shows the leading (i.e.\ rightmost) pole of
    $\gamma_{\rm rc}(\as,N)$ (solid lines) and the pole due to $\asb=0$,
    eq.~\eqref{eq:poleasb} (dashed lines), in the original configuration with
    $\eta=0$ (green curves) and with the new suggested value of $\eta$,
    eq.~\eqref{eq:etachoice} (red curves).  The right-hand side panel shows
    the corresponding residues using curves with the same styles.  In the
    $\eta=0$ case the pole due to $\asb=0$ becomes the leading one at
    $\as\sim0.55$, thus inducing a jump in the residue of the leading pole;
    with the new choice for $\eta$ the pole due to $\asb=0$ is always
    negative.}
  \label{fig:poles}
\end{figure}

The reason for the new approximation resides in the singularities of this
solution.  When $\as$ is sufficiently small, the leading (rightmost)
singularity in $N$ is dictated by the rightmost zero of the Bateman function
$k_\nu(z)$ appearing in the denominator of eq.~\eqref{eq:gamma_RC}.  However,
there are other poles not due to the Bateman function but to the
factors $\asb$ and $z$.  The former in particular gives a pole at
\beq\label{eq:poleasb}
N = N_-(\as)\equiv c(\as) - \frac{\kappa(\as)}2 \(\Mmin(\as)+\eta\)^2
\eeq
with residue
\beq\label{eq:residueasb}
r_-(\as) \equiv -\as\beta_0\[ \as c'(\as) - \frac{\as\kappa'(\as)}2 \(\Mmin(\as)+\eta\)^2 \].
\eeq
This residue is typically negative, and typically larger in absolute value
than the positive residue due to the Bateman pole.  Moreover, in the original
$\eta=0$ case, this pole becomes the leading one at sufficiently large $\as$,
see fig.~\ref{fig:poles}.
%%%%%%%%%%%%%%%%%%%%%%%%%%%%%%%%%%%%%%%%%%%%%%%%%%%%%%%%%%%%%%%%%%%%%%%%%
%%%%%%%%%%%%%%%%%%%%%%%%%%%%%%%%%%%%%%%%%%%%%%%%%%%%%%%%%%%%%%%%%%%%%%%%%

Having a pole with a large negative residue which becomes even the dominant
one as $\as$ increases has the effect of producing splitting functions which
decrease with decreasing $x$ at small $x$'s.  This is unexpected, based on the
features seen at the LL and at smaller values of $\as$, as well as from
general Regge theory considerations.  In fact, the presence of this
singularity is strongly related to the specific functional form that we have
used to construct the approximation of the kernel, and as such it does not
necessarily represent a true physical feature.

This is where the $\eta$ parameter comes to help. Owing to the positivity of
$\kappa(\as)$, a positive value of $\eta$ reduces the value of $N_-$,
eq.~\eqref{eq:poleasb}, pushing the pole to the left and thus making it less
dangerous.  A positive parameter $\eta$ is also natural. Indeed, the actual
double-leading kernel does no longer have the collinear pole at $N=0$, because
it has been resummed by duality with the (approximate) fixed-order anomalous
dimension discussed in sect.~\ref{sec:appHELL-as}.  Rather, after resummation
the leading pole to the left of the minimum is positioned by construction in
the asymptotic value of the approximate anomalous dimension at large $N$,
eq.~\eqref{eq:gammaapproxasy}.  The shape of the actual double-leading kernel
is thus better reproduced by setting
\beq\label{eq:etachoice}
\eta = (1-\nfac) a_1 + \nfac a_0,
\eeq
in terms of the parameters introduced in sect.~\ref{sec:appHELL-as}.  In this
way, $\eta$ becomes a function of $\as$, which begins at $\Ord(\as)$.  As
such, this change with respect to the previous implementation is
``perturbative'', and thus it has a smaller effect at small $\as$.  At the
large values of $\as$ that we have considered in this work, the introduction
of $\eta$ with the form eq.~\eqref{eq:etachoice} solves the issue found at
$\eta=0$.  In particular, the leading pole is always the one coming from the
Bateman function, while for the considered $\as$ values the spurious pole
$N_-$, eq.~\eqref{eq:poleasb}, is always negative
(see figure~\ref{fig:poles}).

In particular, as one can see from the figure, the leading pole is basically
unchanged after the introduction of the $\eta$ parameter, up until the point
where the spurious pole from $\asb=0$ dominates in the $\eta=0$ case.  This
confirms that the rightmost pole coming from the Bateman function is stable
upon changes of unphysical parameters, and represents a physical property of
the result.  On the contrary, the spurious pole from $\asb=0$ is strongly
dependent on the choice of $\eta$, which confirms its unphysical nature. The
use of $\eta$ from eq.~\eqref{eq:etachoice} strongly reduces the impact of
this unphysical pole, and thus represents a significant improvement over the
previous $\eta=0$ implementation, which is necessary to deal reliably with
the region of large $\as$.

\subsection{Solution to the branch-change problem in $\gamma_+$}
\label{sec:appHELL-branch}

We give here some detail on how we solved the branch-change problem discussed
in section~\ref{sec:HELL-as}.  First of all, we want to clarify how the
duality is solved along the inverse Mellin contour.  This contour is typically
given by eq.~\eqref{invNcontour} in the upper half-plane $\Im(N)\geq0$, and by
its complex conjugate path in the lower half-plane.  As the function
$\gamma_+(\as,N)$ is real, namely
\mbox{$\gamma_+(\as,\bar N)=\overline{\gamma_+(\as,N)}$}, it is sufficient
to consider it on the upper half-plane.

The original \HELL approach consisted in starting from the computation of
$\gamma_+(\as,N_0)$, with $N_0$ lying on the real axis where we can use robust
numerical zero-finding methods, and then moving along the contour in small
steps, using every time the value of $\gamma_+$ from the previous step as an
initial guess for the Secant method, which is a variant of the well known
Newton method where the derivative of the function is replaced by a
numerically computed difference quotient.  The only change we have made to
this procedure is related to the choice of the starting point.  Any of these
zero-finding algorithms require a reasonably good guess of the result, and it
is much easier to find it if $N$ is real and large.  Indeed, in this limit,
the solution of the duality should, by construction, correspond to the
fixed-order anomalous dimension that was used to resum the kernel, i.e.\ the
one discussed in appendix~\ref{sec:appHELL-as}. In order to exploit this
property, we take a starting real value of $N$ which is sufficiently large (we
currently use $N=15$ in the code), and from there we move towards $N_0$ in
small steps along the real axis, using the result at the previous step as
an initial guess. For this procedure to be even more robust, along this path on
the real axis we perform for each value of $N$ a scan of the (real) function
to be set to zero in order to make sure that there is a single sign flip in
the region under consideration, and then apply the Regula Falsi
method\footnote{If there is a single zero in the considered range, the Regula
Falsi method finds it in a very robust and fast way.}  in that region.  Note
that this path along the real axis is ignored in the end --- it is just used to
provide the correct starting point $N_0$ for the original procedure.\footnote
{This is important at large $\as$, because at $N_0\sim1$ the resummed
anomalous dimension is no longer so similar to the fixed-order one.}

As was mentioned, it may happen at large values of $|\Im(N)|$ that for two
values of $\as$ even very close to one another the duality condition selects
two different branches of the anomalous dimension along the $N$ contour.  This
is normal because we expect this function to have branch cuts, but undesirable
because it produces different large-$x$ behaviours after Mellin inversion.
One possible workaround is to modify the contour such that the duality always
selects the same branch. This works up to a point: indeed, in order to avoid
switching branch, the contour must become very steep as $\as$ gets larger,
leading to numerical instabilities in the Mellin inversion.  Therefore, we
have decided to keep the contour fixed (we employ $N_0=1.1$ and
$\theta=\frac23\pi$ in eq.~\eqref{invNcontour}) and to adopt the procedure
discussed in section~\ref{sec:HELL-as}, which is articulated in a number of
steps:
\begin{itemize}
\item Firstly, the resummed contribution to the anomalous dimension (the
 actual resummed result minus its fixed-order expansion), that we denote here
 by $\Delta\gamma_+(\as,N)$, is computed along the Mellin inversion contour.
\item Then, its inverse Mellin transform, denoted by $\Delta P_+(\as,x)$, is
 numerically computed ignoring the issue of the change-of-branch.
\item Since the large-$x$ behaviour of such $\Delta P_+(\as,x)$ is unstable
 (due to the changing of the branch) and is in any case inaccurate (as we are
 resumming small-$x$ logarithms without control on the large-$x$ region), we
 replace it with a cubic extrapolation (in $\log x$) for $x>x_{\rm extr}$
 based on the values of $x$ below this boundary of the large-$x$ region.  In
 the code we currently use $x_{\rm extr}=0.02$. We call the result of this
 procedure $\Delta P_+^{\rm extr}(\as,x)$.
\item The extrapolation destroys momentum conservation, which we have to impose
 back by adding a suitable contribution.  In order to do so, we need to know the
 integral of \mbox{$x \Delta P_+^{\rm extr} (\as,x)$} after the extrapolation is
 performed.  This cannot be done analytically, so we have to revert to
 numerical methods.  Rather than simply computing the integral numerically, we
 find it more convenient to approximate the entire function with a suitable
 polynomial basis, so that the integral (and in general, the Mellin transform)
 can be computed analytically.  Therefore, the fourth step consists in
 performing this polynomial approximation.
\item Once this projection onto polynomials is performed, not only we can
 easily restore momentum conservation, but we also have a simple
 parametrisation available for the Mellin transform of the patched
 \mbox{$\Delta P_+$}.  We thus use it as a substitute of the original
 $\gamma_+$ for all the remaining tasks in \HELL, e.g., for the computation of
 the resummed $\gamma_{qg}$ from eq.~\eqref{gqgABFint3}.
\end{itemize}

It is instructive to give some extra detail on the last two steps.  First of
all, we recall that the dominant behaviour of the function $\Delta P_+$ at low $x$ is
given by $x^{-1-N_B}$, where $N_B=N_B(\as)>0$ is the position of the leading
pole of eq.~\eqref{eq:gamma_RC}.  We thus find it more convenient to apply the
approximation procedure to the function
\beq\label{fdP}
f(\as,x) = x^{1+N_B(\as)+p} \Delta P_+^{\rm extr}(\as,x),
\eeq
which has a flatter behaviour at small $x$ when $p\sim0$.  We introduced $p$
to be able to slightly modify the asymptotic small-$x$ behaviour, in order to
have a handle to possibly improve the polynomial approximation.  For such an
approximation, we considered Chebyshev polynomials of the first kind, which are very convenient
from a numerical point of view, as they provide accurate approximations with
fast routines.  It is important however to choose a proper functional form
of $x$ to be used as the variable in
these polynomials.  The simplest option is to consider polynomials in
$\log x$, which would allow us to express, after the projection onto the
basis and further manipulations,
\beq\label{fdPcheblog}
f^{\rm cheb, log}(\as,x) = \sum_{k=0}^n c_k \log^kx,
\eeq
whose Mellin transform is easily computed,
\beq\label{fdPcheblogMell}
\tilde f^{\rm cheb, log}(\as,N) = \sum_{k=0}^n c_k \frac{(-)^k k!}{(N+1)^{k+1}}.
\eeq
Here, $n$ is the order of the polynomial approximation.
In practice, we found it more convenient to use another functional form
for the argument of the polynomials, namely a fractional power of $x$, so that
\beq\label{fdPchebpow}
f^{\rm cheb, pow}(\as,x) = \sum_{k=0}^n c_k x^{k/q},
\eeq
whose Mellin transform is even simpler than that of eq.~(\ref{fdPcheblogMell}),
\beq\label{fdPchebpowMell}
\tilde f^{\rm cheb, pow}(\as,N) = \sum_{k=0}^n \frac{c_k}{N+1+k/q}.
\eeq
In both cases, we perform the projection onto the Chebyshev polynomials in
the range \mbox{$x_{\rm min}\leq x\leq1$}, with $x_{\rm min}=10^{-7}$.
We observe that the extrapolation below $x_{\rm min}$ is better behaved
when using a power, so we adopt this choice, with $q=6$, $n=12$ and $p=-0.1$.
The Mellin transform of $\Delta P_+^{\rm extr}(\as,x)$ is thus approximated by
\begin{align}
  \int_0^1 dx\, x^N\,\Delta P_+^{\rm extr}(\as,x)
  &= \int_0^1 dx\, x^{N-1-N_B-p}\,f(\as,x) \nonumber\\
  &\simeq \tilde f^{\rm cheb, pow}(\as,N-1-N_B-p) \nonumber\\
  &= \sum_{k=0}^n \frac{c_k}{N-N_B-p+k/q}.
\end{align}
We can now restore momentum conservation by defining
\beq
\Delta P_+^{\rm new}(\as,x) \equiv \frac{f^{\rm cheb, pow}(\as,x)}{x^{1+N_B+p}} -c_{\rm mc}
\eeq
with
\beq
c_{\rm mc} = 2 \tilde f^{\rm cheb, pow}(\as,-N_B-p)
\eeq
such that
\beq
\int_0^1 dx\, x\,\Delta P_+^{\rm new}(\as,x) = 0.
\eeq
The corresponding anomalous dimension is
\begin{align}
  \Delta\gamma_+^{\rm new}(\as,N)
  &= \tilde f^{\rm cheb, pow}(\as,N-1-N_B-p) - \frac{c_{\rm mc}}{N+1} \nonumber\\
  &= \sum_{k=0}^n c_k\(\frac1{N-N_B-p+k/q} - \frac1{(N+1)(-N_B-p+k/q)}\).
\end{align}
This anomalous dimension is now used in all of the subsequent manipulations
in \HELL. This is very convenient, as with a limited number of parameters (13
coefficients for $n=12$) we can reconstruct the anomalous dimension at any $N$
in a very fast way, as opposed to the previous implementation that required
the sampling of the anomalous dimension on a specific $N$-space path with many
($\sim2000$) points.

\phantomsection
\addcontentsline{toc}{section}{References}
\bibliographystyle{JHEP}
\bibliography{smallx-upgrade}

\end{document}